\documentclass[11pt,english]{article}
\usepackage{amssymb}
\usepackage{babel}
\usepackage{latexsym}
\usepackage{graphics}
\usepackage{amsmath}
\usepackage{amsfonts}
\usepackage{hyperref}
\makeatletter
\@addtoreset{equation}{section}
\makeatother

\def\dalemb#1#2{{\vbox{\hrule height .#2pt
        \hbox{\vrule width.#2pt height#1pt \kern#1pt
                \vrule width.#2pt}
        \hrule height.#2pt}}}

\def\cA{{\cal A}}

\def\cM{{\cal M}}

\def\0{{\sst{(0)}}}
\def\1{{\sst{(1)}}}
\def\2{{\sst{(2)}}}
\def\3{{\sst{(3)}}}
\def\4{{\sst{(4)}}}
\def\5{{\sst{(5)}}}
\def\6{{\sst{(6)}}}
\def\7{{\sst{(7)}}}
\def\8{{\sst{(8)}}}
\def\n{{\sst{(n)}}}

\def\ep{\epsilon}
\def\td{\tilde}

\def\half{{\textstyle{1\over2}}}
\def\hp{ \frac{1}{2}}
\def\qu{\frac{1}{4}}

\let\a=\alpha \let\b=\beta \let\g=\gamma \let\d=\delta \let\e=\epsilon
  \let\q=\theta  \let\k=\kappa
\let\l=\lambda \let\m=\mu \let\n=\nu  \let\r=\rho
\let\s=\sigma \let\t=\tau  \let\f=\phi  
\let\w=\omega  \let\D=\Delta  \let\L=\Lambda
    
    \let\G=\Gamma

\def\nn{\nonumber} \def\bd{\begin{document}} \def\ed{\end{document}}
\def\ds{\documentstyle} \let\fr=\frac \let\bl=\bigl \let\br=\bigr
\let\Br=\Bigr \let\Bl=\Bigl
\let\bm=\bibitem
\let\na=\nabla
\let\pa=\partial \let\ov=\overline
\newcommand{\be}{\begin{equation}}
\newcommand{\ee}{\end{equation}}
\def\ba{\begin{array}}
\def\ea{\end{array}}
\def\ft#1#2{{\textstyle{{\scriptstyle #1}\over {\scriptstyle #2}}}}
\def\fft#1#2{{#1 \over #2}}
\def\del{\partial}
\def\sst#1{{\scriptscriptstyle #1}}
 \def\oneone{\rlap 1\mkern4mu{\rm l}}
\def\ie{{\it i.e.\ }}
\def\via{{\it via}}
\def\semi{{\ltimes}}
\def\str{{\rm str}}
\def\Dm{{{D_{\sst{max}}}}}
\def\vac{ \left | 0 \right \rangle }
\def\kvac{ \left | k \right \rangle }

\def\sp{\; \; \;}

\def\bol{ \left | B (p^+) \right \rangle}
\def\bo1{ \left | B^0 (p^+) \right \rangle}

\def\bolt{ \left | B (p^+) \right \rangle_{\t}}

\def\boxl{ \left | B (x^-) \right \rangle}

\def\<{ \langle }
\def\>{ \rangle }

\def\vf{\varphi}

\def\ls{{(l,0)}}
\def\lv{{(l,\pm1)}}
\def\lt{{(l,\pm2)}}

\def\lse#1{{(l_{#1},0)}}
\def\lve#1{{(l_{#1},\pm1)}}
\def\lte#1{{(l_{#1},\pm2)}}

\def\lsg#1{{5(l_{#1},0)}}
\def\lvg#1{{5(l_{#1},\pm1)}}
\def\ltg#1{{5(l_{#1},\pm2)}}

\def\lsi#1{{5{(#1,0)}}}
\def\lvi#1{{5{(#1,\pm1)}}}
\def\lti#1{{5{(#1,\pm2)}}}

\def\lsr#1{{1{(#1,0)}}}
\def\lvr#1{{1{(#1,\pm1)}}}
\def\ltr#1{{1{(#1,\pm2)}}}

\def\cn{{\cal N}}
\def\cao{{\cal O}}
\def\cD{{\cal D}}
\def\cE{{\cal E}}
\def\cF{{\cal F}}
\def\cG{{\cal G}}
\def\cH{{\cal H}}
\def\cK{{\cal K}}
\def\cO{{\cal O}}
\def\cP{{\cal P}}
\def\cQ{{\cal Q}}
\def\cR{{\cal R}}
\def\cS{{\cal S}}
\def\cT{{\cal T}}
\def\cU{{\cal U}}
\def\cV{{\cal V}}
\def\cW{{\cal W}}
\newcommand{\nono}{\nonumber}
\newcommand{\dtilde}[1]{\tilde{\tilde{#1}}}
\newcommand{\hatb}[1]{\hat{\ov{#1}}}
\newcommand{\hatt}[1]{\hat{\tilde{#1}}}
\newcommand{\emnr}{{e_\m}^{\n\r}}

\newcommand{\hsp}{\hspace{0.5cm}}

\newcommand{\ho}[1]{$\, ^{#1}$}
\newcommand{\hoch}[1]{$\, ^{#1}$}
\newcommand{\bea}{\begin{eqnarray}}
\newcommand{\eea}{\end{eqnarray}}
\newcommand{\ra}{\rightarrow}
\newcommand{\lra}{\longrightarrow}
\newcommand{\Lra}{\Leftrightarrow}
\newcommand{\ap}{\alpha^\prime}
\newcommand{\bp}{\tilde \beta^\prime}
\newcommand{\tr}{{\rm tr} }
\newcommand{\Tr}{{\rm Tr} }
\newcommand{\NP}{Nucl. Phys. }

\newcommand{\ams}{{\it Institute for Theoretical Physics,
University of Amsterdam, \\
Valckenierstraat 65, 1018XE Amsterdam, The Netherlands} \\
{\tt skenderi, taylor@science.uva.nl}}
\newcommand{\auth}{Kostas Skenderis and Marika Taylor}

\def\red{\color{red}}

\vspace{10pt}

\begin{document}

\begin{flushright}
\hfill{ITFA-2008-12}
\end{flushright}

\vspace{25pt}

\begin{center}

{\Large \bf The fuzzball proposal for black holes}

\vspace{20pt}

\auth

\vspace{15pt}

\vspace{8pt}

{\ams}

\vspace{20pt}

\underline{ABSTRACT}
\end{center}

The fuzzball proposal states that associated with a black hole of
entropy $S$ there are
$\exp S$ {\it horizon-free non-singular} solutions that asymptotically look
like the black hole but generically differ from the black hole up to the
horizon scale. These solutions, the fuzzballs, are considered to be the
black hole microstates while the original black hole represents the
average description of the system. The purpose of this report is to
review current evidence for the fuzzball proposal, emphasizing the use
of AdS/CFT methods in developing and testing the proposal. In
particular, we discuss the status of the proposal for 2 and 3 charge
black holes in the D1-D5 system, presenting new derivations and
streamlining the discussion of their properties. Results to date support
the fuzzball proposal but further progress is likely to require going beyond
the supergravity approximation and sharpening the definition of a
``stringy fuzzball''.  We outline how the fuzzball proposal could
resolve longstanding issues in black hole physics, such as Hawking radiation
and information loss. Our emphasis throughout is on connecting different
developments and identifying open  problems and directions for future research.

\pagebreak

\tableofcontents
\addtocontents{toc}{\protect\setcounter{tocdepth}{2}}

\section{Introduction}

The physics of black holes has been at the center stage of
theoretical physics for over thirty years, as black holes
raise puzzles that directly challenge many cherished fundamental
physical properties such as as unitarity and locality.
Defining questions in this area have been
\begin{itemize}
\item Why does a black hole have entropy proportional to its horizon area?
\item Is there information loss because of black holes?
\item How does one resolve spacetime singularities, such as those inside black
holes or in Big Bang cosmologies?
\end{itemize}
The fact that black holes appear to have entropy is puzzling on
many counts. Typically in a quantum system the correspondence
principle relates the quantum states to the classical phase space
and the entropy to the volume of this phase space in Planck units.
Black holes  however are uniquely fixed in terms of the conserved
charges they carry (black holes have ``no hair'' in classical general 
relativity) so the
classical phase space is zero dimensional\footnote{Strictly speaking
such uniqueness theorems have only been proven for very special
systems and in general black holes do carry hair. However the
volume of the corresponding classical phase space is still far too
small to account for black hole entropy upon quantization.}.
One could argue that perhaps the states counted by the Bekenstein-Hawking
formula are purely quantum mechanical with no classical limit,
but even in this case one is faced with the puzzle that
the entropy is proportional to the area of the horizon, rather than
the volume enclosed in it, as one might have anticipated based on the fact that the entropy is an
extensive quantity. This has been taken as a hint of a fundamental
property of quantum gravitational theories, namely that they
are {\it holographic} \cite{'t Hooft:1993gx}: any $(d{+}1)$-dimensional gravitational theory should have a
description in terms of $d$-dimensional quantum field theory without
gravity with one degree of freedom per Planck area.
We will discuss holography and its realization in the AdS/CFT
correspondence extensively in this report.

Classically black holes are completely black, but
semi-classically they thermally radiate \cite{Hawking:1974sw}. This fact considerably
strengthens the case for taking seriously the analogy between the black hole
laws and thermodynamics and searching for an underlying statistical
explanation of this thermodynamic behavior. More importantly, the thermal nature of the radiation
has led to one of the biggest recent conundrums in theoretical
physics: the information loss paradox \cite{Hawking:1976ra}.
Matter in a pure state may be thrown into black hole
but only thermal radiation comes out. So it would appear as if a pure state
has evolved into a mixed state,
thus violating unitarity. This issue has been debated vigorously over the
years and has been taken as an indication that well accepted physical
principles, such as unitarity or locality, may have to be abandoned,
see \cite{ILreviews} for reviews. As we will review later,
recent progress based on the AdS/CFT correspondence implies a
unitary evolution and the holographic nature of the correspondence
also implies that spacetime locality is only approximate.

Black holes have a curvature singularity hidden behind their
horizon. Near these singularities Einstein gravity breaks down
and a fundamental question is how (and when) the quantum theory of
gravity resolves these singularities and what effect the
resolution has. When considering black holes
with macroscopic horizons, one might anticipate that semi-classical
computations, such as those implying Hawking radiation,
would be applicable. This conclusion has also been challenged in the
literature, see \cite{tHooft,Giddings:2006sj,Mazur:2001fv,Corley:1996ar} for
a (small) sample of works in this direction.
Here recent developments also offers a new perspective: non-singular
spacetimes would generically differ from the black hole
background up to the horizon scale, rather than only in the
neighborhood of the singularity.

Over the last 15 years great progress has been achieved in
string theory in addressing black hole issues, as we now briefly
review. Firstly, for
a class of supersymmetric black holes the black hole entropy
was understood using D-branes. The basic idea is simple:
supersymmetric states (generically) exist for all values
of the parameters of the underlying theory.
Changing the gravitational strength one can interpolate
between the description of the system as a black hole
and the description in terms of bound states
of D-branes.  Thus one can compute the degeneracy in the
D-brane description and thanks to supersymmetry extrapolate this
result to the black hole regime. Starting from \cite{Strominger:1996sh}
such computations were done for a class
of black holes and exact agreement was found with the Bekenstein-Hawking
entropy formula. In more recent times, the agreement was extended to subleading
orders where on the gravitational side one takes into account the effect
of higher derivative terms and on the D-brane side one computes subleading
terms in the large charge limit of the degeneracy formulas, see \cite{Sen:2007qy} for a
review. These developments show that the gravitational entropy indeed
has a statistical origin, but they do not directly address any of the
other black hole issues since the computations involve in an essential way
an extrapolation from weak to strong coupling.

Further progress was achieved with the advent of the AdS/CFT correspondence.
The asymptotically flat black holes under consideration have a
near-horizon region that contains an AdS factor, so AdS/CFT
is applicable. The black hole microstates
can then be understood as certain supersymmetric states of the dual CFT.
Since the AdS/CFT duality is (conjectured to be) an exact equivalence
and the boundary theory is unitary, the dynamics of the black hole
microstates is unitary. Although this in principle shows that
there is no information loss it still does not
explain what is the gravitational nature of the black hole microstates,
nor does it show where Hawking's original argument goes wrong.

The AdS/CFT correspondence however implies more: for every stable
state of the CFT, there should exist a corresponding regular
asymptotically AdS geometry that encodes in its asymptotics the vevs
of gauge invariant operators in that state. Thus, for every CFT
state that one counts to account for the black hole entropy there
should exist a corresponding asymptotically AdS geometry. These solutions
will generically be stringy in the interior although some will be well
described in the (super)gravity limit.

Now since these solutions have the same behavior near the AdS
boundary as the near-horizon limit of the original asymptotically
flat black hole, one can attach the asymptotically flat
region as in the original black hole. We thus find that associated with the original black hole
there is an exponential number of solutions that look like the black
hole up to the horizon scale but differ from it in the interior; the
interior region is replaced by the asymptotically AdS solutions just discussed and
there is one such solution per microstate. The
ones that have a good supergravity description everywhere
should not have horizons, since for those the semiclassical
arguments that relate horizons to entropy should be applicable and
each of these solutions should correspond to a pure state. This is the
fuzzball proposal for black holes, formulated in works of Mathur and
collaborators in \cite{Lunin:2001jy,Lunin:2002qf,Mathur:2002ie,Lunin:2002bj}.

The purpose of this report is to review and
make a critical appraisal of the fuzzball proposal. Other reviews on
this subject include
\cite{Mathur:2005zp,Mathur:2005ai,Mathur:2008wi,Bena:2007kg}.
In the next section we introduce the fuzzball proposal, discuss
its relation with the AdS/CFT correspondence, explaining more fully the argument in the preceding
paragraph, and sketch why this proposal would resolve the black hole puzzles.
In section 3 we explain how holography works. In particular, we discuss
how the vevs of gauge invariant operators are encoded in supergravity
solutions. Then in section 4 we discuss the best understood and
simplest example, namely
the two charge D1-D5 system where one can explicitly find all fuzzball
geometries (visible in supergravity)
and test the general arguments.

To make further progress with the
fuzzball proposal, one would like to understand black holes which have
macroscopic horizons, such as the 3-charge D1-D5-P system.
Section 5 contains a discussion of what is
known about the 3-charge system, in particular the candidate fuzzball
geometries which have been constructed. Much of the work in the
current literature has been focused on constructing explicit examples
of fuzzball geometries, with a view to reproducing the entropy of the
black hole. Many open questions remain as to how the fuzzball
proposal would address black hole puzzles, and we discuss these issues
in section 6. Our emphasis throughout is on connecting the different
developments and emphasizing open problems and directions for further research
rather than reviewing exhaustively the literature.

\section{Generalities}

\subsection{What is the fuzzball proposal?}

{ Consider a black hole solution with associated gravitational entropy $S$.
According to the fuzzball proposal associated with this black hole there are
$\exp S$ {\it horizon-free non-singular} solutions that asymptotically look
like the black hole but generically differ from the black hole up to the
horizon scale. These solutions, the fuzzballs, are considered to be the
black hole microstates while the original black hole represents the
average description of the system.}

To complete this definition we should specify
which class of theories we consider and what precisely we
mean by ``solutions''. We will work within the framework of
string theory; black holes are then solutions of the
corresponding low energy effective action. Lower dimensional
black holes, such as for example the 4d Schwarzchild black hole,
are viewed as resulting from a corresponding 10d solution upon
compactification. Black hole solutions
typically involve only a very small number of the low energy fields,
e.g. only the metric for the Schwarzchild black hole, or the metric plus
a gauge field for the Reissner-Nordstr\"{o}m solution.

The corresponding fuzzball solutions would however in general
be solutions of the full theory, not just its low
energy approximation. Only a subset of fuzzballs would solve the
low energy field equations
and in general these solutions would involve all low energy fields, not
just the fields participating in the black hole solution.
In fact, as it will become clear later on, it is crucial that
the fuzzball solutions involve many other fields. This explains in part
why people have not stumbled upon  such an exponential number
of regular solutions that resemble black holes.

Fuzzballs that involve string scale physics
would in general only have a sigma model description or, if
string field theory were adequately developed, they
would be non-singular solutions of the string field equations.
Some of these string solutions, however, may have an
extrapolation to low energies, i.e. there would exist a
corresponding supergravity solution but it would contain
small regions of high curvature. Furthermore, there would also
be cases where the differences between fuzzball supergravity solutions
are comparable to the corrections coming from
the leading higher derivative corrections to the string theory effective
action. In such cases these solutions will not be reliably distinguishable
within supergravity.

To properly define the dynamics of a system on a non-compact
spacetime we have to specify boundary conditions for
all fields. We will loosely refer to the boundary conditions
as non-normalizable modes and we will refer to the leading part of the asymptotics
which is affected by dynamics as normalizable modes.
All fuzzball solutions would share the same non-normalizable
modes with the original black hole spacetime but they would differ in
the normalizable modes. The precise notion of
normalizable and non-normalizable modes
depends on the asymptotics under consideration, and will be clear
in the specific examples of interest later.

\subsection{Fuzzballs and black hole puzzles}

The fuzzball proposal has the potential of resolving all black hole
related puzzles, although more work is required in order to demonstrate this
with sufficient precision.

Firstly, the black hole entropy becomes of a standard statistical origin:
there is a corresponding solution for every black hole microstate.
In other words, the entropy is related to the volume of the classical
phase space. The fact that the entropy grows like an area will be
seen to be a direct consequence of holography, at least in the examples
where AdS/CFT is applicable.

Since the geometries have no horizons, there is no
information loss either. Matter coming from infinity
will escape back to infinity at late times. A typical
fuzzball geometry is expected to look like the black hole
asymptotically, but it would differ from it up to
the horizon scale (although this has not been demonstrated
to date for solutions with macroscopic non-extremal horizons).
One might anticipate that this difference in the ``inner horizon
region'' is responsible for obtaining a different answer than
in Hawking's original computation.

Boundary conditions and regularity in the interior are
expected to fix the fuzzball solutions and so the resolution of the
black hole singularity is already built into this proposal.
Notice also that the ``size'' of the fuzzball is determined
dynamically from these requirements.  We will discuss these issues
in more detail in section \ref{open},
after reviewing the current results and literature on the fuzzball
proposal.

\subsection{AdS/CFT and the fuzzball proposal}

The black holes whose entropy we best understand microscopically
have an near-horizon region that contains an AdS factor.
For these black holes one can use the AdS/CFT correspondence and
for this reason the general discussion in the previous section
can be made much more precise.

Let us consider for example the 3-charge D1-D5-P system which was the
first to be understood quantitatively. The near-horizon region is
$AdS_3 \times S^3 \times X_4$, with $X_4$ either $T^4$ or K3 (more properly,
the near-horizon region is $BTZ \times S^3 \times X_4$ \cite{skenderis_sfetsos}) .
The entropy of this system was originally computed in \cite{Strominger:1996sh}
by finding the degeneracy of the D1-D5-P bound states at weak coupling and then
extrapolating the result to the black hole phase. It was later realized that
this computation is  part of the AdS/CFT duality: what one counts is
the degeneracy of
certain supersymmetric states of the dual CFT.

Using gravity/gauge theory duality however one can say more. Given a
state in (a deformation of) the CFT, the duality implies that there
is a corresponding asymptotically AdS spacetime with non-trivial
matter fields capturing the parameters of deformation and the vevs
of gauge invariant operators in the given state. The detailed
correspondence will be discussed in the next section, but for the
current argument one only needs the existence of such a
correspondence. Now consider one of the supersymmetric states
counted in accounting for the entropy of, say, the Strominger-Vafa
black hole. Associated with this state there should exist a regular
asymptotically AdS solution. We thus arrive at the conclusion that
there should exist $\exp S$ regular solutions which asymptotically
look like the near-horizon region of the original black hole. These
solutions share the same non-normalizable modes as the near-horizon
limit of the original black hole solution, since we are considering
states and not deformations of the CFT, but differ in their
normalizable modes, which capture the non-trivial vevs in the state
under consideration. One can now attach back\footnote{As emphasized
the fuzzball solutions and near-horizon region of the black hole
have the same leading behavior near the AdS boundary (since they
have the same non-normalizable modes) so the gluing is the same in
both cases.} the asymptotically flat region to arrive at the
conclusion that there should exist $\exp S$ regular solutions that
look like the original black hole up to the horizon scale but differ
in the interior; the interior has been replaced by the asymptotically
AdS solution corresponding to each state. These are the fuzzball
solutions.  The place where each solution starts to
differ from the black hole is controlled by the vev of the lowest
dimension operator in this state and solutions corresponding to
different states are distinguished by the vevs of higher
dimension operators  \cite{Skenderis:2006ah, Kanitscheider:2006zf,
Kanitscheider:2007wq}.

What this argument emphatically does {\it not} imply is that the
solutions would be supergravity solutions and indeed the majority of
fuzzball solutions will not be, although some will be well
described by supergravity.
For states where operators dual to supergravity fields acquire large
vevs the solution will differ appreciably from the black hole
solution already at the supergravity level (but still look like the
black hole at the asymptotically flat infinity). For such solutions
with everywhere small curvatures, one would anticipate that standard
treatments that associate entropy with horizons would be valid.
Since each of these solutions is meant to correspond to a pure
state, the corresponding geometry should therefore be horizonless.
Much of the current fuzzball literature has focused on finding such
supergravity solutions.

On the other hand there would be many cases/states where none
of the operators dual to supergravity fields acquire a vev, or the vev is of string
scale: the corresponding solutions will then agree with the original solution up to
the string scale. One would not expect to
find fuzzball solutions representing these states in supergravity, and
indeed we will see this behavior exemplified in the 2-charge system in section
4.

There will also be cases where a large fraction of the microstates
of the original black hole have large vevs of operators dual to
supergravity fields (chiral primaries) but these vevs differ from
each other very little. Such states should not be distinguishable in
supergravity. One might find supergravity fuzzball solutions
corresponding to these states, which cannot be reliably
distinguished, as the differences between them are of the same order
as the corrections due to leading higher derivative terms. Then the
relevant fuzzball solutions have an extrapolation to the
supergravity regime, but one cannot really trust the distinctions
between similar solutions. Again we will see this behavior occurring
in the 2-charge system, and on general grounds this must persist to
other black holes with macroscopic horizons. There is also the
possibility that states sharing the same vevs of chiral primaries
and differing in the vevs of string states are best described in the
low energy regime by solutions with ``hair'' capturing the vevs of the
chiral primaries, which have either singularities or a horizon, with
area smaller than the horizon area of the original black hole. The latter
indicate that these states can not be distinguished in supergravity.
For instance, one could relate the black ring solutions to corresponding
black holes with the same charges in this way.

Clearly the low energy approximation will not suffice to describe such
cases, and to make progress one will need to work with backgrounds of the full string
theory. In particular, to even define the properties of a generic fuzzball,
one will need to address the question of the
definition of entropy outside the geometric regime: when does a string
theory background have entropy and how is this defined given the
worldsheet theory?

To date, one has relied on the
geometric definition of entropy, with the entropy being associated
with horizons. In
the supergravity limit, one uses the Bekenstein-Hawking entropy of the
horizon, with the generalization by Wald \cite{Wald2} being used when
working perturbatively with higher derivative corrections. This
geometric definition breaks down when dealing with string scale
solutions, be they fuzzballs or black holes, and needs to be
replaced.

Note that this issue can also not be circumvented in the case of
so-called small black holes which do not have horizons in
supergravity. In recent work (reviewed in \cite{Sen:2007qy}) such black holes have been treated
by evaluating the leading corrections to supergravity on (singular)
supergravity solutions, assuming that the corrected solution has a
horizon, and then computing the Wald entropy. This approach is a
priori unjustified, as the neglected higher corrections are
not small on a string scale solution, and it works unreasonably
well at reproducing the entropy of the dual CFT microstates.
Such small black holes should properly be described by a background of
the full string theory, in which the entropy would need to be defined
from the worldsheet theory as above.

Given the state of current technology, most work on the fuzzball
proposal has so far been in the context of supergravity solutions, as
will be reviewed in sections 4 and 5. The limitations of
the supergravity approximation will however be a recurrent theme
throughout, and we will return to discuss the need to
go beyond supergravity in the final sections. Next however we will
review the evidence for the fuzzball proposal obtained from finding
and analyzing fuzzball supergravity solutions.

One should note here that the fuzzball proposal has been developed
most in the context of asymptotically flat black holes in four and five
dimensions, for which the near horizon region contains $AdS_2$ or
$AdS_3$ factors. It is fuzzball solutions for these black holes which
will be discussed in the following sections. Clearly the proposal is
more generally applicable, and one could hope to make comparable
progress at finding fuzzball solutions
with other supersymmetric black hole systems. In particular,
asymptotically $AdS_5$ black holes are a natural system to explore, as
these fall into the best understood AdS/CFT duality, that with ${\cal
  N} = 4$ SYM in four dimensions.

Whilst one could carry out a detailed parallel fuzzball discussion for such black
holes, we will not explore this case here for several reasons. First
of all, it is natural to understand first black holes which are
asymptotically flat, and thus closer to astrophysical black
holes. Secondly, it turns out that the $AdS_5$ case is technically
harder than the cases we discuss. The LLM bubbling solutions
\cite{Lin:2004nb} describe all $1/2$ BPS states of the
system (visible in supergravity) but the corresponding ``black hole''
 does not have a macroscopic horizon in supergravity \cite{Myers:2001aq}, so is not a good
 test case for the fuzzball proposal.

In the asymptotically $AdS_5$ case it seems that one needs to break
the supersymmetry to $1/16$ to
obtain a black hole with macroscopic horizon area in
supergravity, see for example \cite{Gutowski:2004yv}.
With so little supersymmetry, even the counting of
the black hole microstates is rather more subtle, as the degeneracy
depends on the coupling, and the black hole entropy has not yet been
reproduced from field theory. Moreover, finding explicit fuzzball
solutions with
so little supersymmetry is likely to be hard. Thus, whilst the
fuzzball proposal should be applicable for this system, and indeed many
other interesting black hole systems, we will focus on
the better understood case of asymptotically flat black holes.

One should note that in using AdS/CFT arguments to support the
fuzzball proposal there may be certain additional subtleties that we
have not yet mentioned. Firstly, one should be slightly more careful
in relating geometries to states: the fuzzball solutions for the D1-D5 system
should be in correspondence with states in
the Higgs branch of a $(1+1)$-dimensional theory. Due to the strong infrared fluctuations
in 1+1 dimensions one should consider wavefunctions
that spread over the whole of Higgs branch
rather than continuous moduli spaces of the quantum states.
So more properly one should view the fuzzball solutions as dual to
wavefunctions on the Higgs branch. These wavefunctions, however, may be localized
around specific regions in the large $N$ limit and indeed this issue
does not seem to play a key role in any of the subsequent discussions.

Secondly, in AdS/CFT one can have multiple saddle points of the bulk
action, with the same boundary conditions, the most well known example
being thermal AdS and the Schwarzschild black hole. In this case the
onshell (renormalized) action determines the thermodynamically
preferred solution, and there is a Hawking-Page transition between the
two phases at a critical temperature. One
might wonder whether such multiple saddle points could complicate the
relationship between a given CFT microstate and a corresponding
asymptotically AdS string background. In the current context however where we
specify the state, this information determines the vevs of chiral primaries, so the boundary conditions
include both the source and vev part of the solution. As we will review
in the next section this data determines a point in the phase space of the
gravitational theory and thus there should be a unique regular
solution (if such a solution exists), see also \cite{math} and
references therein, for the corresponding discussion in
the mathematics literature on hyperbolic manifolds.

Nonetheless one should bear in mind both caveats as an
issue to be addressed in future when sharpening the definition of the fuzzball
proposal in the string regime.

\section{Holographic methods}

In this section we will summarize the status of holographic methods,
with the emphasis being on summarizing results rather on derivations.
The aim is to provide a handbook of holographic formulae and
associated prescriptions that can be readily used without having
to delve into their derivation.

The basic principles of holography were laid out in the original
papers \cite{Maldacena:1997re,Gubser:1998bc,Witten:1998qj}. In particular, the
duality maps the spectrum of string theory on asymptotically $AdS
\times X$, where $X$ is a compact space, to the spectrum of gauge
invariant operators of the dual QFT, and the string theory partition
function, which is a function of boundary conditions posed on the
conformal boundary of the spacetime, to the generating functional of
correlation functions of the dual QFT, with the fields parameterizing
the boundary conditions mapped to sources of the dual operators.

The duality relation in full generality is still very difficult to probe,
so different approximations have been developed over the years, e.g.
the low energy limit, the limit of long operators, the plane wave limit etc.
In this report we focus on the low energy limit, where string theory is
well approximated by supergravity. This limit typically corresponds
to a strong coupling limit of the boundary theory. We will further
consider the leading saddle point approximation of the bulk
path integral, where the (logarithm of the) bulk partition function
becomes equal to the on-shell supergravity action. In other words,
we suppress supergravity loops. This typically corresponds to the
large N limit in the dual theory. Within these approximations
the gravity/gauge theory duality equates the supergravity on-shell
action to the generating functional of connected QFT correlators
at strong coupling and large $N$.

Let us discuss first the case where the bulk solution is exactly
$AdS_{d+1} \times X_q$, e.g. $AdS_5 \times S^5$ or $AdS_3 \times S^3
\times X_4$. In such cases the dual theory is a $d$-dimensional
conformal field theory ($CFT_d$) and there is a one to one
correspondence between the supergravity KK spectrum and primary
operators of the dual CFT.  One can use the duality to compute
correlation functions of primary operators at strong coupling and
large $N$. Conformal field theories have vanishing 1-point
functions, so the first non-trivial computation is that of a 2-point
function and the latter can be computed holographically by solving
the linearized fluctuation equation around $AdS_p \times X_q$ with
prescribed boundary conditions at the conformal boundary of $AdS_p$.
Higher $n$-point functions can be obtained by solving the
$(n{-}1)$-th order fluctuation equations.

Solutions that are asymptotically $AdS_{d+1} \times X_q$ describe
either deformations of the $CFT_d$ or the CFT in a non-trivial
state. In such cases the most elementary question is what is
the deformation parameter and/or the state. The state of the CFT
is uniquely specified if one knows the expectation values of
all gauge invariant operators in that state. Within the supergravity
approximation we only have access to primary operators: one can
reliably describe only deformations of the original CFT
by primary operators and one can only compute the vevs of primary operators.
The latter gives partial information about the state, but this information
should be enough to specify the state within the approximations used
(strong coupling, large $N$).

The parameters of deformation and the vevs of primary operators can be
extracted by means of {\it algebraic manipulations only} from the asymptotic
expansion of the supergravity solution. The solutions we discuss
in this report describe states rather than deformations, so we will not
discuss any further the case of deformations. In the next two subsections
we will discuss the issues involved in obtaining the vevs (1-point functions)
of primary operators and describe how these are resolved leading to
general formulae for the 1-point functions. Afterwards we
focus on the case of interest, namely asymptotically $AdS_3 \times S^3$
solutions and present explicit formulae for the vevs of operators
up to dimension 2. Higher point functions can be obtained
by solving the fluctuation equations around the original
solution \cite{Bianchi:2001de,Bianchi:2001kw,Skenderis:2002wp,
Papadimitriou:2004rz} but this will not be reviewed here.

\subsection{Holographic renormalization}

The first issue one needs to address when attempting to carry out
holographic computations is that the on-shell action diverges,
essentially due to the infinite volume of spacetime. This issue is
dealt with by the formalism of holographic renormalization
\cite{Henningson:1998gx,Balasubramanian:1999re,deHaro:2000xn,
Skenderis:2000in,Bianchi:2001de,Bianchi:2001kw,Papadimitriou:2004ap,
Papadimitriou:2004rz}; for a review see \cite{Skenderis:2002wp}, and
amounts to adding local boundary covariant counterterms to cancel
the infinities. Actually the local boundary counterterms are
required, irrespectively of the issue of finiteness, by the more
fundamental requirement of the appropriate
variational problem being well posed \cite{Papadimitriou:2005ii}. As is well known
the conformal boundary of asymptotically $AdS$ spacetimes have a
well-defined conformal class of metrics rather than an induced
metric. This means that the appropriate variational problem involves
keeping fixed a conformal class and not an induced metric as in the
usual Dirichlet problem for gravity in a spacetime with a boundary.
The new variational problem requires the addition of further
boundary terms, on top of the Gibbons-Hawking term, which turn out
to be precisely the boundary counterterms, see
\cite{Papadimitriou:2005ii} for the details and a discussion of the
subtleties related to conformal anomalies.

The subject of holographic renormalization is extensively discussed
in the literature, so we will only highlight a few points and
introduce the notation to be used in later sections. The points that
should be emphasized are:
\begin{itemize}
\item To obtain renormalized correlators, the main object of interest
is the radial canonical momentum, rather than the on-shell action.
\item The source and renormalized 1-point function are a conjugate
pair and can be considered as coordinates in the phase space of the
gravitational theory.
\end{itemize}
Let us briefly discuss these points. Firstly, the asymptotic form of
bulk fields, specialized to the $D=3$ case of interest, is
\bea
ds_{3}^{2} &=& \frac{dz^2}{z^2} + \frac{1}{z^2} \left(g_{(0)uv} + z^2
\left(g_{(2)uv} + {\rm{log}}(z^2) h_{(2) uv} + ({\rm{log}}(z^2))^2
\tilde{h}_{(2) uv}\right) + \cdots\right) dx^{u} dx^v \nn \\
%&\equiv& dr^2 + \g_{ij}(x,r) dx^i dx^j; \nn \\
\Psi^1 &=& z ({\rm{log}}(z^2) \Psi^1_{(0)}(x) + \td{\Psi}^1_{(0)}(x) +
\cdots ); \label{as_exp} \\
\Psi^{k} &=& z^{2-k} \Psi^{k}_{(0)}(x) + \cdots + z^{k}
\Psi^{k}_{(2k-2)}(x) + \cdots, \hsp k \neq 1. \nn
\eea
%where $z= e^{-r}$.
In these expressions $(g_{(0)uv},  \Psi^1_{(0)}(x), \Psi^{k}_{(0)}(x))$
are sources for the stress energy tensor and scalar operators of dimension
one and $k$ respectively; as usual one must treat separately the operators of
dimension $\Delta = d/2$, where $d$ is the dimension of the boundary. Note that
the 2-dimensional boundary coordinates are labeled by $(u,v)$. We will
also use the notation $[\psi]_n$ to denote the coefficient of the $z^n$
term in the asymptotic expansion of the field $\psi$.

Correlation functions can be computed using the basic holographic
dictionary that relates the on-shell gravitational action to the
generating functional of correlators. The first variation can be
done in all generality \cite{deHaro:2000xn} yielding a relation
between the 1-point function (in the presence of sources, so higher
point functions can be obtained by further functional differentiation
w.r.t. sources) and non-linear combinations of the asymptotic
coefficients in (\ref{as_exp}). The underlying structure of the
correlators is best exhibited in the radial Hamiltonian formalism,
which is a Hamiltonian formulation with the radius playing the role
of time. The Hamilton-Jacobi theory, introduced in this context in
\cite{de Boer:1999xf}, then relates the variation of the on-shell
action w.r.t. boundary conditions, thus the holographic 1-point
functions, to radial canonical momenta. It follows that that one can
bypass the on-shell action and directly compute renormalized
correlators using radial canonical momenta $\pi$
\cite{Papadimitriou:2004ap,Papadimitriou:2004rz}.

A fundamental property of asymptotically (locally) AdS spacetimes is
that dilations are part of their asymptotic symmetries. This implies
that all covariant quantities can be decomposed into a sum of terms
each of which has definite scaling. For example, the radial
canonical momentum $\pi^k$ of scalar $\Psi^k$ has the expansion
\be\label{momentum_exp}
\pi^k=\pi_{(2-k)}^k+ \pi_{(1-k)}^k + \cdots
+ \pi_{(k)}^k+\tilde{\pi}_{(k)}^k\ \log z^2 +\cdots, \ee
where the
various terms have the dilation weight indicated by the
subscript\footnote{Note that  that the coefficient $\pi_{(k)}^k$ has
an anomalous scaling transformation, $\d \pi_{(k)}^k = -k
\pi_{(k)}^k -2 \tilde{\pi}_{(k)}^k$, due to conformal anomalies,
see \cite{Papadimitriou:2004ap}.}, i.e. $\pi_{{n}}^k$ has scaling
weight $n$. These coefficients are in one to one correspondence with the
asymptotic coefficients in (\ref{as_exp}) with the exact relation
being in general non-linear. The advantage of working with dilation
eigenvalues rather than with asymptotic coefficients is that the
former are manifestly covariant while the latter in general are not:
the asymptotic expansion (\ref{as_exp}) singles out one coordinate
so it is not covariant. Holographic 1-point functions can be
expressed most compactly in terms of eigenfunctions of the dilation
operator, and this explains the non-linearities found in explicit
computations of 1-point functions. In particular, if one considers
the case of a single scalar $\Psi^k$ in an asymptotically (locally)
AdS spacetime, the 1-point function is simply given by
\be
\< O^k \> = \pi_{(k)}^k
\ee
Thus, we indeed see that source and the
renormalized 1-point are conjugate variables and one may consider
them as coordinates in the classical phase space of the theory.

\subsection{Kaluza-Klein holography}

The discussion in the previous section involved
a $(d{+}1)$ dimensional supergravity theory which admits an $AdS_{d+1}$ solution.
However the string theory backgrounds of interest
which contain an $AdS$ factor typically
also involve a compact space, for example, $AdS_5 \times S^5$,
$AdS_3 \times S^3 \times X_4$ etc. On general grounds, one expects
that there is an effective $(d+1)$ dimensional description. Thus provided one can
obtain such a description one can use holographic renormalization
to obtain the 1-point functions of gauge invariant operators. The
method of Kaluza-Klein holography provides an explicit algorithm
for constructing the corresponding $(d+1)$ dimensional action
and extracting the vevs.

Note that generically the spheres appearing in these solutions have
a radius which is of the same order as the $AdS$ radius, so
the higher KK modes are not suppressed relative to the zero modes
and one cannot ignore them. In some cases it is
nevertheless possible to only keep a subset of modes
because the equations of motion admit a solution with all modes,
except the ones kept, set equal to zero, i.e.
there exists a ``consistent truncation''.
The existence of such a truncation signifies the existence of a
subset of operators of the dual theory that are closed under OPEs.
The resulting theory is a $(d+1)$-dimensional gauged supergravity and such
gauged supergravity theories have been the starting point for
many investigations in AdS/CFT.

However, starting from
a lower dimensional gauged supergravity is unsatisfactory on many grounds.
Firstly, gauged supergravity captures the physics of only a very small
subset of operators, typically that of the stress tensor supermultiplet.
An infinite number of other operators, namely the ones dual to KK
modes, are in principle accessible within the low energy limit,
but one excludes them a priori.
Secondly, the higher dimensional solutions are more fundamental
as reduction of regular solutions may result in singular solutions.
Thirdly, even in the case where a consistent truncation to a lower
dimensional supergravity is possible, it is often very difficult or
indeed unknown how to express known interesting higher dimensional
solutions in a non-linear reduction ansatz that produces
a corresponding solution of the gauged supergravity. Finally, a key conceptual
question for holography is how the compact part of the
geometry is encoded in QFT data and answering this question requires
keeping all modes.

For these reasons we will keep all KK modes in the reduction
(so there is also never an issue of consistency). Naively, this
results in an intractable problem of an infinite number of fields
all coupled together. Recall however that the 1-point functions are
extracted from the asymptotic expansion near the AdS boundary,
and the fall off of the fields near the AdS boundary is fixed by their
mass. This implies that to compute the
holographic 1-point function of any given operator only a finite number
of fields and a finite number of interactions are relevant. The method of
Kaluza-Klein holography systematically constructs the lower dimensional
action in a way that only the fields and interactions needed are kept
at each step.

The steps involved in this construction are the following. The starting
point is a $D$ dimensional action that admits an $AdS_{d+1} \times X^q$
solution. We further assume that the harmonic analysis on the compact
$X^q$ is known, i.e. the set of spherical harmonics is known. This clearly
is the case for $X^q= S^q$. In the first step we consider fluctuations
around this solution and expand the fluctuations in the harmonics of the
compact space. Let $\phi$ denote collectively all fields,
$\phi_o$ be the $AdS_{d+1} \times X^q$ solution and
$\delta \phi$ the fluctuation, then schematically,
\bea
\phi(x,y)&=& \phi_o(x,y) + \d \phi(x,y) \nonumber \\
\d \phi(x,y) &=& \sum_I \psi^I(x) Y^I(y) \label{har_exp}
\eea
where $x$ is a coordinate in the $(d+1)$ non-compact directions, $y$
is a coordinate in the compact directions and $Y^I$ denotes collectively
all spherical harmonics (scalar, vector, tensor and their covariant
derivatives). Precise formulae for the case of interest will be
presented in the next subsection.

The decomposition (\ref{har_exp}) is not unique because there are
coordinate transformations,
\be \label{trns}
X^M{}' = X^M - \xi^M(x,y)
\ee
where $X^M=\{x,y\}$ that transform the fluctuations $\psi^I$ to each over
or the background solution $\phi_o$. In the supergravity literature on
KK reduction one often imposes a gauge condition, most notably the
de Donder gauge condition, to eliminate this issue. We instead
construct gauge invariant combination that have the property
that in the de Donder gauge they coincide with gauge fixed variables.
This is done by working out
perturbatively in the number of fluctuations how each $\psi^I$
transformations under (\ref{trns}) and then one constructs combinations
$\hat{\psi}^I$ that transform as tensors, i.e. scalar
combinations are invariant under (\ref{trns}), $(d{+}1)$-dimensional
vector combinations $A_\mu(x)$, transform
as vectors, the metric  $g_{\m \n}$ transforms as metric etc.
Details of this procedure can be found in \cite{Skenderis:2006uy}.

The aim is now to derive the equations that the gauge invariant modes
$\hat{\psi}^I$ satisfy by substituting (\ref{har_exp}) in the $D$ dimensional
equations and working perturbatively in the number of fields. To linear
order one obtains the spectrum, to quadratic order the cubic interactions
etc. Note however that not all cubic (or higher) interactions are relevant
for the computation of the 1-point function of any given operator. Only the
ones that could modify the asymptotic coefficients that determine the
vev need to be retained.

Expanding perturbatively in fluctuations one finds
\be
{\cal L}_{\cal I} \hat{\psi}^{\cal I} = {\cal L}_{ {\cal I  J K} }
\hat{\psi}^{\cal J} \hat{\psi}^{\cal K}
+ {\cal L}_{ {\cal I  J K L}}
\hat{\psi}^{\cal J} \hat{\psi}^{\cal K} \hat{\psi}^{\cal L} + \cdots,
\ee
where $ {\cal L}_{{\cal I}_1 \cdots {\cal I}_n}$ is an appropriate differential
operator. $ {\cal L}_{{\cal I}_1 \cdots {\cal I}_n}$ involves higher
derivative terms and the set of field equations cannot generically be
integrated into an action. However, one can always define $(d+1)$-dimensional
fields $\Psi^{\cal I}$ by a non-linear Kaluza-Klein reduction map of
the fields $\psi^{\cal I}$:
\be \label{KKmap}
\Psi^{\cal I} = \hat{\psi}^{\cal I} + {\cal K}^{I}_{ {\cal J K } }
\hat{\psi}^{\cal J} \hat{\psi}^{\cal K} + \cdots,
\ee
where ${\cal K}^I_{\cal J K}$ contains appropriate derivatives. The
reduction map is such that the fields $\Psi^{\cal I}$
do satisfy field equations which can be integrated into an action.
Given this $(d+1)$-dimensional action, it is then straightforward to obtain the
one point functions of operators in terms of the asymptotics of the
fields $\Psi^{\cal I}$, using the well-developed techniques of
holographic renormalization.

This results in the following general formula for an operator
of dimension $k$
\be \label{Ovev}
\< O_k^I \> = \pi_{(k)}^I
+ \sum_{JK} a^I_{JK} \pi^J_{(k_1)} \pi^K_{(k-k_1)} + \cdots
\ee
where $a^I_{JK}$ are numerical constants and the ellipses indicate
higher powers of canonical momenta.
The non-linear terms are related to extremal correlators, i.e.
these terms are possible when the theory contains operators
with dimension $k_i$, such that $\sum k_i =k$, and the
numerical constants $a^I_{JK}$ are related to the extremal
3-point functions at the conformal point. Note that these
formulae for the vevs apply to any solution of the same
action. Given any solution one can evaluate them to find the
QFT data encoded by this solution.

To summarize, the radial canonical momenta are related
(in general non-linearly) to the coefficients in the asymptotic expansion
of the $(d+1)$-dimensional fields $\Psi$. These fields in turn are related
to the $D$-dimensional coefficients $\psi$ via the non-linear Kaluza-Klein
map (\ref{KKmap}) which relates $\Psi$ to the gauge invariant version
$\hat{\psi}$ of $\psi$, which itself is a non-linearly related to $\psi$'s.
One can now combine all these maps to produce a final formula
for the vevs which is of the schematic form
\be \label{Ovevf}
\< O_k^I (\vec{x})\> = [\psi^I(\vec{x}) ]_k
+ \sum_{JK} b^I_{JK} [\psi^J(\vec{x})]_{k_1} [\psi^K(\vec{x})]_{(k-k_1)}
+ \cdots
\ee
where  $b^I_{JK}$ are numerical coefficients,
$\vec{x}$ are $d$-dimensional (boundary) coordinates, $z$ is the
Fefferman-Graham radial coordinates and the coefficients $[\psi^I]_k$
are asymptotic coefficients in
\be  \label{dphi}
\d \phi(z,\vec{x},y) = \sum_{I,m} [\psi^I(\vec{x})]_m  z^m Y^I(y)
\ee
Thus, if we are interested in extracting the vevs of gauge invariant
operators from a given solution that is asymptotically
$AdS_{d+1} \times X^q$, the procedure is to write it as the deviation
from $AdS_{d+1} \times X^q$ in the form (\ref{dphi}), extract the
coefficients $ [\psi^I(\vec{x})]_m$ and insert those in (\ref{Ovevf}).
This procedure was carried out for  asymptotically $AdS_5 \times S^5$
solutions describing the Coulomb branch of $\cn=4$ SYM in
\cite{Skenderis:2006uy,Skenderis:2006di} and for the LLM solutions
in \cite{Skenderis:2007yb}, resulting in strong tests of gravity/gauge
theory duality away from the conformal point
and for asymptotically $AdS_3 \times S^3$ solutions relevant for the
fuzzball program in \cite{Kanitscheider:2006zf,Kanitscheider:2007wq}.
In the next section we present the detailed results for the case
of interest.

\subsection{Asymptotically $AdS_3 \times S^3$ solutions}

In what follows we will be interested in black hole and fuzzball
solutions whose decoupling regions are asymptotic to
$AdS_3 \times S^3 \times X_4$, where $X_4$ is $T^4$ or $K3$. Therefore
one can use AdS/CFT methods to extract holographic data from these
geometries and, in particular, the asymptotics of the six-dimensional
solutions near the $AdS_3 \times S^3$ boundary encode the vevs of
chiral primary operators in the dual field theory.

In the solutions of interest only zero modes of the compact space
$X_4$ are excited, so it
is convenient to first compactify the solution over $X_4$. Solutions of
type IIB supergravity compactified on $K3$
give rise to solutions of
$d=6$, $N=4b$ supergravity coupled to 21 tensor multiplets,
constructed by Romans in \cite{Romans:1986er}. Corresponding fuzzball solutions
of type IIB on $T^4$ can also be expressed as solutions of $d=6$,
$N=4b$ coupled to 5 tensor multiplets. These theories admit an
$AdS_3 \times S^3$ solution, so the program of Kaluza-Klein holography
can be applied to obtain an effective three dimensional description
of all relevant KK modes.

The bosonic field content of the $d=6$ $N =4b$ theory with $n_t$
tensor multiplets is the
graviton $g_{MN}$, 5 self-dual and $n_t$ anti-self dual tensor fields
and an $O(5,n_t)$ matrix of scalars $\cM$ which can be written in terms
of a vielbein $\cM^{-1} = V^T V$.
Following the notation of \cite{Sez98}
the bosonic field equations may be written as
\bea \label{sugraIIBK3}
R_{MN} &=& 2 P^{nr}_M P^{nr}_N + H^n_{MPQ} {H^n_N}^{PQ}
+  H^r_{MPQ} {H^r_N}^{PQ}, \nono \\
   \na^M P_M^{nr} &=& Q^{M nm} P_M^{mr} + Q^{M rs} P_M^{ns}
   + \frac{\sqrt{2}}{3} H^{n MNP} H^r_{MNP},
\eea
along with Hodge duality conditions on the 3-forms
\be
\label{sugraIIBK3_hd}
   \ast_6 H^n_3 = H^n_3, \qquad \ast_6 H^r_3 = -H^r_3,
\ee
In these equations $(m,n)$ are $SO(5)$ vector indices running from
1 to 5 whilst $(r,s)$ are
$SO(n_t)$ vector indices running from 6 to $(5+n_t)$.
The 3-form field strengths are given by
\be
\label{G3forms}
H^{n} = G^{A} V_{A}^n; \hsp
H^{r} = G^{A} V_{A}^r,
\ee
where $A \equiv \{n,r\} = 1, \cdots, (5+n_t)$; $G^{A} = db^A$ are closed and the vielbein on the
coset space $SO(5,n_t)/(SO(5) \times SO(n_t))$ satisfies
\be
 V^T \eta V = \eta, \qquad V = \left(\begin{array}{c} {V^n}_A \\
   {V^r}_A \end{array}\right), \qquad \eta = \left(\begin{array}{cc}
   I_5 & 0 \\ 0 & -I_{21} \end{array}\right).
\ee
The associated connection is
\be
\label{IIBK3conn}
d V V^{-1} =  \left ( \begin{array} {c c} Q^{mn} & \sqrt{2} P^{ms} \\
\sqrt{2} P^{rn} & Q^{rs} \end{array} \right ),
\ee
where $Q^{mn}$ and $Q^{rs}$ are antisymmetric and the off-diagonal
block matrices $P^{ms}$ and $P^{rn}$ are transposed to each other.
$Q^{mn}$ and $Q^{rs}$ are composite $SO(5)$ and $SO(n_t)$ connections
which are solvable in terms of $5 n_t$ physical scalars $\phi^{m r}$
via the Cartan-Maurer equation. 

The six-dimensional field equations (\ref{sugraIIBK3})
admit an $AdS_3 \times S^3$ solution, such that
\bea
ds_6^2 &=& \sqrt{Q_1 Q_5} \left ( \frac{1}{z^2} (-dt^2 + dy^2 + dz^2)
+ d\Omega_3^2 \right );  \label{background} \\
G^{5} &=& H^{5} \equiv G^{o5} =  \sqrt{Q_1 Q_5} (\frac{dz}{z^3} \wedge dt \wedge
dy + d\Omega_3), \nn
\eea
with the vielbein being diagonal and all other three forms (both
self-dual and anti-self dual) vanishing. In what follows it is
convenient to absorb the curvature radius $\sqrt{Q_1 Q_5}$ into an
overall prefactor in the action, and work with the unit radius $AdS_3
\times S^3$.  Perturbations of six-dimensional supergravity fields
relative to an $AdS_3 \times S^3$ background may be expressed as
\bea
g_{MN} &=& g^{o}_{MN} + h_{MN}; \hsp
G^{A} = G^{oA} + g^A; \\
V^{n}_{A} &=& \d^{n}_{A} + \f^{nr} \d_{A}^r + \half \f^{nr} \f^{mr}
\d_{A}^m; \nn \\
V^{r}_A &=& \d^{r}_{A} + \f^{nr} \d_{A}^n + \half \f^{nr} \f^{ns}
\d_{A}^s. \nn
\eea
These fluctuations can then be expanded in a basic of spherical harmonics as
follows:
\bea
h_{\m \n} &=& \sum h_{\m\n}^I (x) Y^{I} (y), \label{flc1} \\
h_{\m a} &=& \sum (h_{\m}^{I_v} (x) Y_a^{I_v} (y) +
h_{(s)\m}^{I} (x) D_a Y^I (y) ), \nn \\
h_{(a b)} &=& \sum (\rho^{I_t} (x) Y_{(ab)}^{I_t} (y) +
\rho_{(v)}^{I_v} (x) D_a Y_b^{I_v} (y) +
\rho_{(s)}^{I} (x) D_{(a} D_{b)} Y^{I} (y) ), \nn \\
h^{a}_{a} &=& \sum \pi^{I} (x) Y^{I} (y), \nn \\
g^{A}_{\m\n \r} &=& \sum 3 D_{[\m} b_{\n \r]}^{(A)I} (x) Y^{I} (y), \nn
\\
g^{A}_{\m \n a} &=& \sum ( b_{\m \n}^{(A)I} (x) D_{a} Y^{I} (y) + 2
D_{[\m} Z_{\n]}^{(A) I_{v}} (x) Y_{a}^{I_v} (y)); \nn \\
g^{A}_{\m a b} &=& \sum (D_{\m} U^{(A)I}(x) \ep_{abc} D^{c} Y^I(y) +
2 Z_{\m}^{(A) I_v} D_{[b} Y^{I_v}_{a]}); \nn \\
g^{A}_{a b c} &=& \sum (- \ep_{abc} \L^{I} U^{(A)I}(x) Y^{I} (y)) ; \nn \\
%b^{A}_{\m a} &=& \sum (Z_{\m}^{A I_v } (x) Y_{a}^{I_v }(y) +
%Z_{(s)\m}^{A I} (x) D_{a} Y^I (y) ); \nn \\
%b^{A}_{ab} &=& \sum \ep_{abc} (U^{A I} (x) D^{c} Y^I (y) + U_{(v)}^{A
%  I_{v} } (x)  Y^{c I_v } (y) ); \nn \\
\phi^{mr}  &=& \sum \phi^{(mr) I} (x) Y^{I} (y), \nn
\eea
Here $(\mu, \nu)$ are AdS indices and $(a,b)$ are $S^3$ indices,
with $x$ denoting AdS coordinates and $y$ denoting sphere coordinates.
 $\L^I$ is defined in (\ref{sph_Y}). The
subscript $(ab)$ denotes symmetrization of indices $a$ and $b$ with
the trace removed. Relevant properties of the spherical harmonics are
reviewed in appendix \ref{sphere}. We will often use a notation
where we replace the index $I$ by the degree of the harmonic $k$
or by a pair of indices
$(k,I)$ where $k$ is the degree of the harmonic and $I$ now parameterizes
their degeneracy, and similarly for $I_v, I_t$.

Imposing the de Donder gauge condition $D^{A} h_{aM} = 0$
on the metric fluctuations removes the fields with subscripts $(s,v)$.
In deriving the spectrum and computing correlation functions, this is
therefore a convenient choice. The de Donder gauge choice is however not
always a convenient choice for the asymptotic expansion of solutions;
indeed the natural coordinate choice in our applications takes us
outside de Donder gauge. As discussed in \cite{Skenderis:2006uy}
this issue is straightforwardly dealt with by
working with gauge invariant combinations of the fluctuations.

The linearized spectrum of the fluctuations was derived in
\cite{Sez98}, with the cubic interactions obtained in
\cite{Arutyunov:2000by}.
Let us briefly review the linearized spectrum
derived in \cite{Sez98}, focusing on fields dual to chiral
primaries. Consider first the scalars. It is
useful to introduce the following combinations
which diagonalize the linearized equations of motion:
\bea
s^{(r) k}_{I} &=& \frac{1}{4(k+1)} ({\phi}^{(5r) k}_{I} +2 (k+2)
{U}^{(r)k}_{I}), \label{diageqm} \\
%t^{(r)k}_{I} &=& \frac{1}{4} ({\phi}^{(5r)k}_{I} - 2 k {U}^{(r)k}_{I}),
%\nn \\
\s^k_{I} &=& \frac{1}{12 (k+1)} (6 (k+2) \hat{U}^{(5)k}_{I} -
\hat{\pi}_I^{k}), \nn
%\t^k_{I} &=& \frac{1}{12 (k+1)} ( \hat{\pi}^k_I + 6k \hat{U}^{(5) k}_{I}). \nn
\eea
The fields $s^{(r)k}$ and $\s^k$ correspond to scalar chiral
primaries, with the masses of the scalar fields being
\be \label{masses}
m_{s^{(r)k}}^2 = m_{\s^k}^2 = k (k-2), \hsp
\ee
The index $r$ spans $6 \cdots 5 + n_t$ with $n_t = 5, 21$ respectively
for $T^4$ and $K3$.
Note also that $k \ge 1$ for $s^{(r)k}$; $k \ge 2$ for $\s^k$. The hats
$(\hat{U}^{(5)k}_{I}, \hat{\pi}_I^{k})$ denote the following. As
discussed in \cite{Skenderis:2006uy}, the equations of motion for
the gauge invariant fields are precisely the same as those in de
Donder gauge, provided one replaces all fields with the corresponding
gauge invariant field. The hat thus denotes the appropriate gauge
invariant field, which reduces to the de Donder gauge field
when one sets to zero all fields with subscripts $(s,v)$. For our
purposes we will need these gauge invariant quantities only to
leading order in the fluctuations, with the appropriate
combinations being
\bea
\hat{\pi_2}^{I} &=& \pi_2^{I} + \Lambda^{2} \rho_{2(s)}^{I}; \label{gaug}
\\
\hat{U}^{(5)I}_2 &=& U^{(5)I}_2 - \half \rho^{I}_{2(s)}; \nn \\
\hat{h}^{0}_{\m\n} &=& h^{0}_{\m\n} - \sum_{\a,\pm} h_{\m}^{1 \pm \a}
h_{\n}^{1 \pm \a}. \nn
\eea
Next consider the vector fields. It is useful to introduce
the following combinations which diagonalize the equations of motion:
\be
h^{\pm}_{\m I_v} = \half (C_{\m I_v}^{\pm} - A^{\pm}_{\m
  I_v}), \hsp
Z_{\m I_v}^{(5) \pm} = \pm \qu (C_{\m I_v}^{\pm} + A_{\m I_v}^{\pm}).
\ee
For general $k$ the equations of motion are Proca-Chern-Simons
equations which couple $(A_{\m}^{\pm}, C^{\pm}_{\m})$ via a first
order constraint \cite{Sez98}. The three dynamical fields at each degree $k$
have masses $(k-1, k+1, k+3)$, corresponding to dual operators of
dimensions $(k,k+2,k+4)$ respectively; the operators of dimension $k$
are vector chiral primaries. The lowest dimension operators
are the R symmetry currents, which couple to the $k=1$ $A^{\pm \a}_{\m }$
bulk fields. The latter satisfy the Chern-Simons equation
\be \label{cs11}
F_{\m \n}(A^{\pm \a}) = 0,
\ee
where $F_{\m\n}(A^{\pm \a})$ is the curvature of the connection and
the index $\a = 1,2,3$ is an $SU(2)$ adjoint index. We will here only
discuss the vevs of these vector chiral primaries.

Finally there is a tower of KK gravitons with $m^2 = k (k+2)$ but
only the massless graviton, dual to the stress energy tensor, will
play a role here. Note that it is the
combination $\hat{H}_{\m \n} = \hat{h}^{0}_{\m \n} + \pi^{0} g^{o}_{\m \n}$ which
satisfies the Einstein equation; moreover one needs the appropriate
gauge covariant combination $\hat{h}^0_{\m\n}$ given in (\ref{gaug}).

\subsection{Vevs of gauge invariant operators}

Let us now give the expressions for the vevs of gauge invariant operators
up to dimension two. These are expressed in terms of coefficients in
the asymptotic expansions of the fields near the conformal boundary at $z=0$.

Let us denote by $({\cal O}_{S^{(r) k}_I}, {\cal O}_{\Sigma^{k}_I})$
the chiral primary operators dual to the fields $(s^{(r) k}_I,
  \s^{k}_I)$ respectively. The vevs of the scalar operators with dimension two or less
can then be expressed in terms of the coefficients in the
  asymptotic expansion as
\bea
\left < {\cal O}_{S^{(r)1}_i} \right > &=& \frac{2 N}{\pi} \sqrt{2}
[s^{(r)1}_i]_1; \hsp
\left < {\cal O}_{S^{(r)2}_I} \right > = \frac{2 N}{\pi} \sqrt{6}
[s^{(r)2}_I]_2; \label{sc-1} \\
\left < {\cal O}_{\Sigma^{2}_I} \right > &=& \frac{N}{\pi} \left (2
  \sqrt{2} [\s^2_I]_2 - \frac{1}{3} \sqrt{2} a_{Iij}
  \sum_{r} [s^{(r) 1}_{i}]_1 [s^{(r) 1}_{j}]_1 \right ). \nn
\eea
Here $[\psi]_n$ denotes the coefficient of the $z^n$ term in the
  asymptotic expansion of the field $\psi$. The coefficient $a_{Iij}$
refers to the triple overlap between spherical harmonics, defined in
  (\ref{ap-ov0}). Note that dimension one scalar spherical harmonics
have degeneracy four, and are thus labeled by $i = 1, \cdots 4$.
In deriving these vevs one needs to follow the steps outlined in the
previous section, in particular making use of the cubic coupling and
Kaluza-Klein reduction formulae obtained in \cite{Arutyunov:2000by}.

Now consider the stress energy tensor and the R symmetry currents. The
three dimensional metric, which is given by $g^{o}_{\m\n} + \hat{H}_{\m\n}$,
and the Chern-Simons gauge fields admit the
following asymptotic expansions:
\bea
ds_{3}^{2} &=& \frac{dz^2}{z^2} + \frac{1}{z^2} \left(g_{(0)\bar{\mu}\bar{\nu}} + z^2
\left(g_{(2) \bar{\mu}\bar{\nu}} + {\rm{log}}(z^2) h_{(2)
  \bar{\mu}\bar{\nu}} + ({\rm{log}}(z^2))^2
\tilde{h}_{(2) \bar{\mu}\bar{\nu} }\right) + \cdots\right) dx^{\bar{\mu}} dx^{\bar{\nu}}; \nn \\
A^{\pm \a} &=& {\cal A}^{\pm \a} + z^2 A_{(2)}^{\pm \a} + \cdots
\eea
The vevs of the R symmetry currents $J^{\pm \a}_u$ are then given in terms
  of terms in the asymptotic expansion of $ A^{\pm \a}_{\m}$ as
\be
\left < J^{\pm
  \a}_{\bar{\mu}} \right >  = \frac{N}{4 \pi} \left (g_{(0) \bar{\mu} \bar{\nu}}  \pm \ep_{\bar{\mu}
  \bar{\nu}} \right ) {\cal A}^{\pm \a \bar{\nu}},  \label{j1}
\ee
where $\ep_{\bar{\mu} \bar{\nu}}$ is the constant antisymmetric tensor
such that $\ep_{01} = 1$.
The vev of the stress energy tensor $T_{\bar{\mu}\bar{\nu}}$ is given by
\bea
\< T_{\bar{\mu}\bar{\nu}} \> &=& \frac{N}{2 \pi} \left (
g_{(2)\bar{\mu}\bar{\nu}} + \half R g_{(0) \bar{\mu}\bar{\nu}}
+ 8 \sum_r [\td{s}^{(r)1}_{i}]_1^2 g_{(0) \bar{\mu}\bar{\nu}} + \qu ({\cal A}^{+ \a}_{(\bar{\mu}}
{\cal A}^{+ \a}_{\bar{\nu})} + {\cal A}^{- \a}_{(\bar{\mu}}
{\cal A}^{- \a}_{\bar{\nu})}) \right) \label{t1}
\eea
where parentheses denote the symmetrized traceless combination of indices.

\bigskip

This summarizes the expressions for the vevs of chiral primaries with
dimension two or less which were derived in
\cite{Kanitscheider:2006zf}. Note that these operators correspond to
supergravity fields which are at the bottom of each Kaluza-Klein
tower. The supergravity solution of course
also captures the vevs of operators dual to the other fields in each
tower.  Expressions for these vevs involving all contributing non-linear
terms were not explicitly derived in
\cite{Kanitscheider:2006zf}: in general the vev of a dimension $p$ operator will include
contributions from terms involving up to $p$ supergravity
fields. Computing these in turn requires the field equations (along
with gauge invariant combinations, KK reduction maps etc) up to
$p$th order in the fluctuations.

Note that using only the linear term in a vev generically gives
qualitatively the wrong answer. For example, when supersymmetry is
unbroken only operators which are the bottom components of
supermultiplets can acquire vevs. Evaluating only the linearized
expressions for the vevs of operators which are not bottom
components of a supermultiplet in a supersymmetric background
however usually gives a non-zero answer, which cannot be correct. In
\cite{Alday:2005xj} linearized expressions for holographic vevs of
operators were evaluated in the decoupling $AdS_3 \times S^3 \times
X_4$ limit of the 2-charge supersymmetric black ring background.
Non-zero vevs for found for the operators ${\cal O}_{\rho^{I_t}}$
dual to the scalars $\rho^{I_t}$ given in (\ref{flc1}). However,
these operators (for dimension $\Delta \ge 3$) sit in the middle
component of the spin 2 supermultiplet whose lowest component is a
vector chiral primary (as can inferred from Table 3 and Figure 1 of
\cite{Sez98}) and cannot acquire a vev in a supersymmetric state.
Thus the linearized expressions indeed give qualitatively wrong
answers in this case.

Given an asymptotically $AdS$ supergravity solution, one can extract
the vevs, and the deformation parameters, by purely algebraic
manipulations. This information characterizes the state of the dual
theory and is the natural information to extract first. Higher point
functions can be extracted by extending the holographic one-point functions to include sources,
and then solving fluctuation equations in the geometry to sufficient order in the sources.

\section{Two charge system}

\subsection{Naive geometry and D-branes}

The 2-charge supersymmetric D1-D5 supergravity solution is
\bea
ds^2 &=& \frac{1}{\sqrt{h_1 h_5}} (- dt^2 + dz^2)
+ \sqrt{h_1 h_5} dx^m dx^m + \sqrt{\frac{h_1}{h_5}} ds^2(X_4); \\
C &=& (h_1^{-1} - 1) dt \wedge dz + \ast_{4} d h_5, \qquad
e^{-2 \Phi} = \frac{h_5}{h_1}. \nn
\eea
Here $ds^2(X_4)$ denotes the metric on $T^4$ or $K3$ respectively
and the functions $(h_1,h_5)$ are harmonic on $R^4$, which has
coordinates $x^m$. The notation $\ast_{4}$ denotes the Poincar\'{e}
dual in the flat $R^4$ metric. Our conventions for the supergravity field
equations are given in appendix \ref{conv}. The solution describes
D5-branes wrapping $X_4$ and intersecting D1-branes over the string
directions $(t,z)$, with $z$ being a periodic coordinate\footnote{Note
we called $z$ the Fefferman-Graham radial coordinate in the previous
section. It should be clear from the context whether $z$ is a Fefferman-Graham
coordinate or the compact $S^1$ coordinate that the D1 and D5 brane wraps.
I hope this wil not cause any confusion.}
 with radius $R_z$.
Consider now single-centered solutions such that $h_i = (1+ Q_i/r^2)$;
the charges $Q_i$ are related to the integral charges
$N_i$ as
\be
Q_1 = \frac{(\a')^3 N_1}{V}; \qquad
Q_5 = \a' N_5,
\ee
where we have set the string coupling $g =1$ and
the volume $v$ of $X_4$ is $(2 \pi)^4 V$.
Reducing over $X_4$ and $z$ leads to the five dimensional metric
\be \label{5dmetric}
ds^2 = \l^{-2/3} dt^2 +\l^{1/3} (dr^2 + r^2 d \Omega_3^2)
\ee
where
\be
\l= h_1 h_5 = (1 + \frac{Q_1}{r^2})(1 + \frac{Q_5}{r^2})
\ee
This is an asymptotically flat spacetime which has a naked singularity
at $r=0$: the Ricci scalar is given by
\be \label{curv1}
R = - \frac{1}{6} \left ( \frac{2 \pa_r^2 \l}{\l^{4/3}} + \frac{6
  \pa_r \l}{r \l^{4/3}} - \frac{(\pa_r \l)^2}{\l^{7/3}} \right ),
\ee
and thus for $\l \sim r^{-2p}$ near $r=0$
\be \label{curv2}
R = \frac{4}{3} p (p+1) r^{2p/3 - 2}.
\ee
In the 2-charge case, $p = 2$ and the curvature diverges at $r=0$.
As we shall see when we discuss the 3-charge case the
corresponding $5d$ metric is of exactly the same form but in that
case $\l$ contains the product of 3-harmonic functions. Thus for this
case $p=3$ and the curvature is finite at $r=0$: there is an
extremal horizon rather than a naked singularity at $r=0$.

Consider now the decoupling limit:
\be
\alpha' \to 0, \qquad \r = \frac{r}{\a'}\ {\rm fixed}, \qquad R_z,
\frac{v}{\a'^2}\ {\rm fixed}
\ee
In this limit the constant in the harmonic functions drops out and one is focusing
on the region near $r=0$. After rescaling the coordinates as
\be
\r \to \r \frac{R_z}{\sqrt{Q_1 Q_5}}, \qquad
z \to \frac{z}{R_z}, \qquad t \to \frac{t}{R_z}
\ee
the metric becomes
\be \label{D1D5nearhor}
ds^2 = \sqrt{Q_1 Q_5} \left (  \rho^2 (-dt^2 + dz^2) +
\frac{d\rho^2}{\rho^2} + d \Omega_3^2 \right )
+ \sqrt{\frac{Q_1}{Q_5}} ds^2(X_4),
\ee
where $d\Omega_3^2$ is the metric on a unit 3-sphere and $z$ has
period $2 \pi$ (and as usual we have set $\a'=1$ after the limit
was taken). The
six-dimensional metric is thus the product of a locally $AdS_3$
spacetime and a 3-sphere,  with $AdS$ and sphere radii both
equal to $(Q_1 Q_5)^{1/4}$. Had the $z$ coordinate been
non-compact the non-compact part would have been $AdS_3$ in Poincar\'{e}
coordinates and the apparent horizon at $\rho = 0$ would have
merely been a coordinate horizon, which can be removed by introducing global
coordinates for $AdS_3$. Since $z$ is compact, however, the three
dimensional non-compact spacetime is instead the zero mass BTZ black
hole. The spacetime (\ref{D1D5nearhor}) does not have a curvature
singularity but there are geodesics that terminate at $r=0$.
Extremal BTZ black holes have a horizon radius proportional to their
mass and a singularity at $r=0$. In the massless limit the position
of the horizon coincides with the singularity, so we end up
with a naked singularity. In the 3-charge case one instead obtains
a massive extremal BTZ black hole with horizon radius proportional to
the new charge. The area of the horizon is proportional to the horizon radius
and thus vanishes for the 2-charge system. Note that reducing the decoupled
region over $X^4$ and $z$ leads to a spacetime with a curvature
singularity,  explaining the singularity structure
of (\ref{5dmetric}), but this is due to the fact that we reduce over a
circle of vanishing proper length.

Thus the 2-charge black hole does not have a macroscopic horizon
within supergravity. As we will momentarily review, however, the entropy
of the 2-charge system is $S \sim \sqrt{N_1 N_5}$
and one would expect that the D1-D5 solution with higher derivative
corrections taken into account should
have a horizon which accounts for this entropy. In this section we
will see that supergravity fuzzball solutions for this case
can be explicitly constructed, and matched with CFT
microstates. However, the average fuzzball in this system is necessarily
string scale, and thus the system can only provide an indicative toy
example for the proposal as a whole.

\subsection{CFT microstates}

The dual CFT describing the low
energy physics of the D1-D5 system is believed to be a deformation of
the ${\cal N}=(4,4)$ supersymmetric sigma model with target space
$S^N(X_4)$, where $N= N_1 N_5$ and the compact space $X_4$
is either $T^4$ or $K3$, see \cite{David:2002wn} for a review.
 Most of the discussion below will
be for the case of $T^4$, although the results extend simply to $K3$.
The orbifold point is roughly the analogue
of the free field limit of ${\cal N}=4$ SYM in the context of $AdS_5/CFT_4$
duality.

The Hilbert space of the orbifold theory decomposes into twisted
sectors, which are labeled by the conjugacy classes of the symmetric
group $S(N)$, the latter consisting of cyclic groups of various
lengths. The various conjugacy classes that occur and their multiplicity are
subject to the constraint
\be
\sum_i n_i m_i = N,
\ee
where $n_i$ is the length of the cycle (or the twist) and $m_i$ is the
multiplicity of the cycle. As emphasized in \cite{Moore} there is a
direct correspondence between the combinatorial description of the
conjugacy classes and a second quantized string theory, the long and
short string picture advocated in \cite{Das:1996ug,Mald}.

The point is that one can view the supersymmetric sigma model on
$S^N(X_4)$ as the transversal fluctuations of a string with target
space $R \times S^1 \times S^N(X_4)$, in lightcone gauge, with the
string winding once around $S^1$. However, the partition function of
this single string decomposes into several distinct topological
sectors, corresponding to the number of ways the string can be
disentangled into separate strings that wind one or more times around
$X_4 \times S^1$. The factorization of the conjugacy classes into a
product of irreducible cyclic permutations $n_i$ determines the
decomposition into several strings of winding number $n_i$. Whilst the
string picture is intuitive and useful for qualitative features, one
will need the explicit orbifold CFT
description for computations and it is therefore the latter we will use.

The chiral primaries and R ground states can be precisely described
at the orbifold point. In particular the R ground states are
associated with the cohomology of the internal manifold, $T^4$ or
$K3$, and therefore the corresponding chiral primaries in the NS
sector obtained from spectral flow are also labeled by the cohomology.
For our discussions only the states associated with the even
cohomology will be relevant; let the NS sector chiral primaries
be labeled as ${\cal{O}}_{n}^{(p,q)}$ where $n$ is the twist and
$(p,q)$ labels the associated cohomology class. The degeneracy of the
operators associated with the $(1,1)$ cohomology is $h^{1,1}$ of the
compact manifold.

The complete set of chiral primaries associated with the even cohomology
is then built from products of the form
\be
\prod_{l}({\cal{O}}_{n_l}^{p_l,q_l})^{m_l}, \hsp
\sum_{l} n_l m_l = N, \label{oper-1}
\ee
where symmetrization over the N copies of the CFT is implicit.
Spectral flow maps these chiral primaries in the NS sector to R ground states, where
\bea
h^R &=& h^{NS} - j_3^{NS} + \frac{c}{24}; \nonumber \\
j_3^R &=& j_3^{NS} - \frac{c}{12}, \label{spe_fl}
\eea
where $c$ is the central charge of the CFT. Each of the operators in (\ref{oper-1})
is mapped  by spectral flow to a (ground state) operator of definite R-charge
\bea \label{Rop}
& & \prod_{l=1} ({\cal {O}}^{(p_l,q_l)}_{n_l})^{m_l} \quad
  \rightarrow \qquad
\prod_{l=1} ({\cal {O}}^{R (p_l,q_l)}_{n_l})^{m_l}, \\
j_3^R &=& \half \sum_l (p_l - 1) m_l, \qquad
\bar{j}^R_3 =\half \sum_l (q_l - 1) m_l. \nn
\eea
Note that Ramond operators which are obtained from spectral flow of primaries
associated with the $(1,1)$ cohomology have zero R charge. Explicit
free field representations of these operators are reviewed in
\cite{David:2002wn} but these will not be needed in what follows.

The degeneracy of the Ramond ground states can be computed by counting
the partitions of $N$ into pairs of integers; for large $N$ this
results in a total degeneracy $d$ such that
\be
\log (d) \approx 2 \pi \sqrt{\frac{C}{6} N},
\ee
where $C = 12$ for $T^4$ and $C = 24$ for K3. (In counting the former
one needs to include the odd dimensional cohomology.) Note that the
degeneracies are clearly those of type IIB strings on $T^4$ and
heterotic strings on $T^4$ respectively, with left moving excitation level $N$. It
is also useful to note here that the distribution of ground states is
highly localized around those with zero R charge: the degeneracy of ground states with zero R charge,
$d_0$, is given by \cite{Kanitscheider:2007wq}, see also \cite{Russo:1994ev},
\be
d_0 \approx d/N
\ee
for large $N$. Thus as already mentioned the 2-charge black hole
has an entropy of order $\sqrt{N}$. By adding momentum one obtains a
black hole with a regular horizon in supergravity; the corresponding
microstates are left moving excited and right moving ground states,
whose degeneracy for high excitation level can be computed via the
Cardy formula. We will postpone detailed discussions of these states
until section \ref{3cft}.

\bigskip

The AdS/CFT dictionary relates the supergravity spectrum about the
$AdS_3 \times S^3 \times X_4$ background to these chiral primary
operators. In particular in what follows we will make use of the
relationship between (scalar) supergravity fields and
chiral primaries. This map was established recently
\cite{Taylor:2007hs} by matching three
point functions computed from supergravity with those computed in the
orbifold CFT \cite{Jevicki:1998bm,Lunin:2001pw}, assuming non-renormalization of the correlators. Such
non-renormalization is supported by the matching of the correlators computed from the
worldsheet theory with those computed in the orbifold CFT
\cite{Gaberdiel:2007vu,Dabholkar:2007ey,Pakman:2007hn}. Labeling
the orbifold operators by their twist and $(j_3,\bar{j_3})$ charges
$(m, \bar{m})$ respectively, the correspondence
is
\be
\left ( \begin{array} {c} {\cal O}^{S}_{(n-1) m   \bar{m}} \\
{\cal O}^{\Sigma}_{(n-1) m   \bar{m}} \end{array} \right )
\leftrightarrow {\cal M}
\left ( \begin{array} {c} {\cal O}^{0,0}_{n m \bar{m}} \\
{\cal O}^{2,2}_{(n-2) m \bar{m}} \end{array} \right ),
\ee
and
\be
{\cal O}^{(r)1,1}_{n m \bar{m}} \leftrightarrow  {\cal O}^{S^{(r)}}_{n m
  \bar{m}};
\ee
with the matrix ${\cal M}$ being
\be
{\cal M} = \frac{1}{\sqrt{2\D}} \left ( \begin{array} {c c }
(\D + 1)^{1/2} & - (\D - 1)^{1/2} \\
(\D - 1)^{1/2} & (\D + 1)^{1/2}
\end{array} \right )
\ee
for $\D \ge 2$. Here ${\cal O}^{\Phi}_{\Delta m \bar{m}}$ denotes the
operator of dimension $\Delta$ and R charges $(m,\bar{m})$ coupling to
the bulk supergravity field $\Phi$ of corresponding mass and $SO(4)$ charges.

The formulae for the holographic vevs allow us to extract the
expectation values of these operators, along with those of the
stress energy tensor and R currents, from a given asymptotically
$AdS_3 \times S^3 \times X_4$ supergravity solution. Given a proposed
correspondence between the supergravity solution and a field theory
state, one can use this data to test the correspondence. More
generally $n$-point functions extracted from the bulk can also
be compared $n$-point functions in a given state, but there is no reason to
anticipate that such correlation functions are non-renormalized.

\subsection{Fuzzball solutions}

The key observation in constructing fuzzball solutions for the 2-charge system
is that D1-D5 is related by dualities to a fundamental string carrying
momentum, and supergravity solutions for the latter have been known
for more than a decade: The relevant solutions are general chiral null models. For the heterotic
string, the general such model takes the form
\bea \label{het1}
   ds^2 &=& H^{-1}(-dudv + (K-2\alpha^\prime
   H^{-1}N^{(c)}N^{(c)})dv^2+ 2A_I dx^I dv) + dx_I dx^I \nn \\
   {B}^{(2)}_{uv} &=& \hp(H^{-1}-1), \qquad {B}^{(2)}_{vI} = H^{-1} A_{I}, \\
\qquad {\Phi} &=& -\hp \ln H,  \qquad
   {V}_v^{(c)} = H^{-1} N^{(c)}, \nn
\eea
where $I=1,\cdots,8$ labels the transverse directions and ${V}_m^{(c)}$ are
Abelian gauge fields, with
$((c)=1,\cdots,16)$ labeling the elements of the Cartan of the gauge group.
The equations of motion for the heterotic string are given in appendix
\ref{conv}; here the defining functions satisfy
\be
   \Box H (x,v) = \Box K (x,v) = \Box A_I (x,v) = (\pa_I A^I (x,v) - \pa_v
   H (x,v)) = \Box N^{(c)} = 0.
\ee
For the solution to correspond to a charged heterotic
string with generic wave profile, one takes the following solutions
\bea
\label{K3harmonics}
   H &=& 1+\frac{Q}{|x-F(v)|^6},
\qquad A_I = - \frac{Q\dot{F}_I(v)}{|x-F(v)|^6}, \qquad N^{(c)} =
\frac{q^{(c)}(v)}{|x-F(v)|^6}, \nono \\
   K &=& \frac{Q^2 \dot{F}(v)^2+2\alpha^\prime q^{(c)}q^{(c)}(v)}{Q|x-F(v)|^6},
\eea
where $F^I(v)$ is an arbitrary null curve in $R^8$; $q^{(c)}(v)$ is an
arbitrary charge wave and $\dot{F}_I(v)$ denotes $\pa_v F_I(v)$. Such
solutions were first discussed in
\cite{Callan:1995hn,Dabholkar:1995nc}, although the above has
a more generic charge wave, lying in $U(1)^{16}$ rather than $U(1)$.
Chiral null model solutions of type IIB are obtained by setting
$N^{(c)} = 0$. More general solutions in which the fundamental strings
carry condensates of fermion bilinears were considered in
\cite{Taylor:2005db}; such solutions are necessary to account for all D1-D5
microstates, but for brevity we will not discuss them in
detail here.

\bigskip

The F1-P solutions described by such chiral null models can be
dualized to give corresponding solutions for the D1-D5 system as
follows \cite{Lunin:2001jy}. Compactify four of the transverse directions on a torus,
such that $x^i$ with $i=1,\cdots,4$ are coordinates
on $R^4$ and $x^{\r}$ with $\r = 5,\cdots,8$ are coordinates on $T^4$.
Then let $v = (t - z)$ and $u = (t+ z)$ with the coordinate $z$ being
periodic with length $L_z \equiv 2 \pi R_z$,
and smear all harmonic functions over both this circle and
over the $T^4$, so that they satisfy
\be \label{equ_harmonics}
 \Box_{R^4} H (x) = \Box_{R^4} K (x) = \Box_{R^4} A_I (x) = \Box_{R^4}
 N^{(c)} = 0, \hsp
 \pa_{i} A^i = 0.
\ee
Thus the harmonic functions appropriate for describing strings with
only bosonic condensates are \cite{Lunin:2001fv}
\bea
H &=& 1 + \frac{Q}{L_z} \int_0^{L_z} \frac{dv}{|x-F(v)|^2};
\qquad A_I = -\frac{Q}{L_z}\int_0^{L_z} \frac{dv \dot{F}_I(v)}{|x-F(v)|^2}; \\
N_{(c)} &=& -\frac{Q}{L_z} \int_0^{L_z} \frac{dv q_{c}(v)} {|x-F(v)|^2}; \qquad
   K = \frac{Q }{L_z} \int_0^{L_z} \frac{dv (\dot{F}_i(v)^2 +
\dot{ F}_{\rho} (v)^2)} {|x-F(v)|^2}. \nn
\eea
Here $|x-F(v)|^2$ denotes $\sum_{i} (x_i - F_i(v))^2$.

Next one follows an appropriate chain of dualities to obtain a D1-D5
solution. For D1-D5 on $T^4$ one starts with type IIB F1-P solutions on $T^4$
whilst for D1-D5 on $K3$ one starts with heterotic F1-P solutions and
uses heterotic/IIA duality. The computational details of this duality
chain are given in \cite{Kanitscheider:2007wq}.
The final result in both cases can be written as
\bea
\label{equ_D1D5K3pot}
   ds^2 &=& \frac{f_1^{1/2}}{\td{f_1} f_5^{1/2} }[-(dt-A_i
     dx^i)^2+(dz-B_i dx^i)^2] + f_1^{1/2} f_5^{1/2} dx_i dx^i
+  f_1^{1/2} f_{5}^{-1/2} ds^2(X_4),  \nn \\
   e^{2\Phi} &=& \frac{f_1^2}{f_5 \td{f_1} }, \qquad B^{(2)}_{tz} =
   \frac{{\cal A}}{f_5 \td{f}_1},
\qquad B^{(2)}_{\bar{\mu} i} = \frac{{\cal A} {\cal B}^{\bar{\mu}}_i}{f_5 \td{f_1} }, \\
B^{(2)}_{ij} &=& \l_{ij} +
\frac{2 {\cal A} A_{[i} B_{j]}} {f_5 \td{f}_1 }, \qquad
B^{(2)}_{\rho\sigma} = f_{5}^{-1} k^{\gamma}
\w_{\rho \sigma}^{\gamma}, \qquad  C^{(0)} = - f_{1}^{-1} \cal A, \nono \\
   C^{(2)}_{tz} &=& 1-\td{f}_1^{-1}, \qquad C^{(2)}_{\bar{\mu}i} = -
   \td{f}_1^{-1} {\cal B}^{\bar{\mu}}_i, \qquad C^{(2)}_{ij} = c_{ij} - 2 \td{f}_1^{-1}
A_{[i}B_{j]}, \nono \\
   C^{(4)}_{tzij} &=& \l_{ij} + \frac{{\cal A}}{f_5 \td{f_1}}(c_{ij} + 2 A_{[i}B_{j]}),
\qquad C^{(4)}_{\bar{\mu}ijk} = \frac{3 {\cal A}}{f_5 \td{f}_1}
{\cal B}^{\bar{\mu}}_{[i}c_{jk]}, \nono \\
   C^{(4)}_{tz \rho \sigma}
&=& f_{5}^{-1} k^{\gamma} \w^{\gamma}_{\rho \sigma},
\qquad C^{(4)}_{ij \rho \sigma} = (\l^{\gamma}_{ij} +
f_{5}^{-1} k^{\gamma}c_{ij})\w^{\gamma}_{\rho \sigma}, \qquad
   C^{(4)}_{\rho \sigma \tau \pi} = f_{5}^{-1} \cal A \e_{\rho
     \sigma \tau \pi}, \nn
\eea
where we introduce a basis of self-dual and anti-self-dual 2-forms
$\w^{\gamma} \equiv (\w^{\a_+}, \w^{\a_-})$ with $\gamma =
1,\cdots,b^2$ on the compact manifold $X_4$. For both $T^4$ and $K3$
the self-dual forms are labeled by
$\a_{+} = 1,2,3$ whilst the anti-self-dual forms are labeled by
$\a_{-} =1,2,3$ for $T^4$ and $\a_{-} = 1,\cdots 19$ for $K3$. The
intersections of these forms are defined by
\be
\label{K3dABdef}
   d_{\g\d} = \frac{1}{(2 \pi)^4 V} \int_{X_4} \w_2^{\g} \wedge \w_2^{\d}.
\ee
%(\ref{tt1}), (\ref{tt2}) and (\ref{K3dABdef}).
The solutions are expressed
in terms of the following combinations of harmonic functions
$(H,K,A_{i},{\cal A}, {\cal A}^{\a_-})$
\bea
\label{D1D5K3aux}
f_{5} &=& H; \qquad \td{f}_1 = 1 + K - H^{-1} (\cA^2 + \cA^{\a_-} \cA^{\a_-}); \qquad
f_1 = \td{f}_1 + H^{-1} \cA^2; \nn \\
k^{\gamma} &=& (0_3, \sqrt{2} \cA^{\a_-});
\qquad dB = -\ast_4 dA; \qquad
   dc = - \ast_4 df_5; \\
d\l^{\gamma} &=& \ast_4 dk^{\gamma}; \qquad d\l = \ast_4 d {\cal A}; \qquad
{\cal B}^{\bar{\mu}}_i = (-B_i,A_i), \nn
\eea
where $\bar{\mu} = (t,z)$ and the Hodge dual $\ast_{4}$ is defined over (flat)
$R^4$, with the Hodge dual in the Ricci flat metric on
the compact manifold being denoted by $\ep_{\r \s\t \pi}$.
The constant term in $C_{tz}^{(2)}$ is chosen so that the
potential vanishes at asymptotically flat infinity.

We are interested in solutions for which the defining harmonic functions are given by
\bea
H &=& 1 + \frac{Q_5}{L} \int_0^L \frac{dv}{|x-F(v)|^2};
\qquad A_i = -\frac{Q_5}{L}\int_0^L \frac{dv \dot{F}_i(v)}{|x-F(v)|^2},
\label{harm} \\
   {\cal A} &=& -\frac{Q_5}{L}\int_0^L \frac{dv
     \dot{\cal F} (v)}{|x-F(v)|^2}; \qquad
\cA^{\a_-} = -\frac{Q_5}{L} \int_0^L \frac{dv \dot{\cal{F}}^{\a_-}(v)}{|x-F(v)|^2}, \nono \\
   K &=& \frac{Q_5}{L} \int_0^L \frac{dv (\dot{F}(v)^2 +
\dot{\cal F}(v)^2 + \dot{\cal{F}}^{\a -}(v)^2 )}{|x-F(v)|^2}. \nn
\eea
In these expressions $Q_5$ is the 5-brane charge and
$L$ is the length of the defining curve in the D1-D5 system, given
by
\be
L = 2 \pi Q_5/R_z,
\ee
where $R_z$ is the radius of the $z$ circle. Note that $Q_5$ has
dimensions of length squared and is related to the integral charge via
\be \label{q5}
Q_5 = \a' N_5
\ee
(where $g_s$ has been set to one).
%The curves
%$(\dot{\cal F} (v),\dot{\cal{F}}^{\a_-}(v))$ do not have zero modes,
The D1-brane charge $Q_1$ is given by
\be \label{d1-charg}
Q_1 = \frac{Q_5}{L} \int_0^L dv (\dot{F}(v)^2 +
\dot{\cal F}(v)^2 + \dot{\cal{F}}^{\a_-}(v)^2 ),
\ee
and the corresponding integral charge is given by
\be \label{q1}
Q_1 = \frac{N_1 (\a')^3}{V},
\ee
where $(2 \pi)^4 V$ is the volume of the compact manifold.

\bigskip

Solutions with no internal
excitations, that is $({\cal F}(v), {\cal F}^{\a_-}(v)) = 0$, were
obtained in \cite{Lunin:2001jy}. These Lunin-Mathur solutions involve
only the graviton, dilaton and RR 2-form:
\bea
   ds^2 &=& \frac{1 }{ (f_1 f_5)^{1/2} }[-(dt-A_i
     dx^i)^2+(dz-B_i dx^i)^2] + (f_1 f_5)^{1/2} dx_i dx^i
+  f_1^{1/2} f_{5}^{-1/2} ds^2(X_4),  \nn \\
   e^{2\Phi} &=& \frac{f_1}{f_5}, \label{lun-mat} \\
   C^{(2)}_{tz} &=& 1- {f}_1^{-1}, \qquad C^{(2)}_{\bar{\mu}i} = -
   {f}_1^{-1} {\cal B}^{\bar{\mu}}_i, \qquad C^{(2)}_{ij} = c_{ij} - 2 {f}_1^{-1}
A_{[i}B_{j]}, \nono
\eea
and are defined in terms of harmonic functions sourced on the curve
$F^{i}(v)$ in $R^4$:
\be
f_5 = 1 + \frac{1}{L} \int^{L}_{0} \frac{ Q_5 dv}{ |x -
  F|^2} ; \qquad
f_1 = 1 + \frac{1}{L} \int^{L}_{0} \frac{ Q_5 (\pa_v F)^2  dv}{ |x -
  F|^2} ; \label{lm-func}
\qquad A = \frac{1}{L} \int_{0}^{L} \frac{ Q_5 \pa_v F}{| x- F |^2},
\ee
with $dB = - \ast_4 dA$. Note that corresponding 2-charge
solutions carrying KK monopole charge and momentum which are related
by dualities to these solutions were found in \cite{Srivastava:2006xn}.

The mapping of the parameters from the original F1-P systems to
the D1-D5 systems was discussed in \cite{Lunin:2001jy}.
The fact that the solutions take exactly the same form, regardless of
whether the compact manifold is $T^4$ or $K3$, is unsurprising given
that only zero modes of the compact manifold are excited.
The solutions defined in terms of the harmonic functions (\ref{harm})
describe the complete set of two-charge fuzzballs for the D1-D5 system on
$K3$. In the case of $T^4$, these describe fuzzballs with only
bosonic excitations; the most general solution would include fermionic
excitations and thus more general harmonic functions of the type
discussed in \cite{Taylor:2005db}.

Given a generic fuzzball solution, one would first like to check whether
the geometry
is indeed smooth and horizon-free. There are no horizons, since the
defining functions are positive definite. For the fuzzballs with no internal
excitations the question of singularity was discussed in
\cite{Lunin:2002iz}, the
conclusion being that the solutions are non-singular unless the
defining curve $F^i(v)$ is non-generic and self-intersects. A generic fuzzball
solution with internal excitations is also non-singular provided that the
defining curve $F^i(v)$ does not self-intersect and $\dot{F}_i(v)$
only has isolated zeroes \cite{Kanitscheider:2007wq}.

One can illustrate that the fuzzballs are non-singular as
follows \cite{Lunin:2002iz}. The fuzzballs are defined in terms of harmonic functions, and
potential singularities of the solutions are at the sources of these
harmonic functions. Consider
for simplicity the case of fuzzballs with no internal excitations,
for which the defining harmonic functions are sourced on a closed
curve $F^{i}(v)$ in $R^4$. In the neighborhood of a specific point on the
curve, the harmonic function picks up contributions from a line of
sources, and thus effectively behaves like a three-dimensional
harmonic function. That is, parameterizing the $R^4$ as
\be
ds_4^2 = dw^2 + dr^2 + r^2 d \Omega_2^2,
\ee
the direction $w$ is aligned with the curve. The defining
harmonic functions behave as
\be
\int_{-\infty}^{\infty} \frac{dw}{w^2 + r^2}  = \frac{\pi}{r},
\ee
so that $f_a = Q_a/r$, whilst the defining one form behaves as
\be
w = a \frac{dz}{r}.
\ee
Then the metric behaves as
\bea
ds^2 &=& - \frac{2a}{\sqrt{Q_1 Q_5}} dw dt + \sqrt{Q_1 Q_5}
\frac{dw^2}{r} (1 - \frac{a^2}{Q_1 Q_5}) \\
&& + \frac{r}{\sqrt{Q_1 Q_5}} (dz + a \cos \q d \phi)^2 + \frac{\sqrt{Q_1
  Q_5}}{r} (dr^2 + r^2 d \Omega_2^2) + \frac{\sqrt{Q_1}}{\sqrt{Q_5}} ds^2(X_4). \nn
\eea
When $a = \sqrt{Q_1 Q_5}$, which is exactly the condition that follows
from the explicit forms of the full harmonic functions, this metric is
the product of $R^{1,1} \times X_4 \times TN$ with $TN$ Taub-Nut and
thus the metric is non-singular at the curve location. We should
contrast this behavior with that of the naive solution (\ref{D1D5nearhor}).
In the naive geometric the circle $S^1$ parameterized by $z$ is
non-contractible but it has vanishing proper length in the interior
leading to singular behavior. Instead, in the fuzzball geometries
the $z$ circle is contractible and part of a Taub-NUT geometry, so
the vanishing of its radius in the interior
is akin to the behavior of polar coordinates
near the origin. A generalization of this
analysis shows that generic fuzzballs defined by non-intersecting
curves are also non-singular \cite{Lunin:2002iz,Kanitscheider:2007wq}.

Note that if there are no transverse excitations, $F^i(v) = 0$, the solution
collapses to the naive singular solution. When there are only internal
excitations, the corresponding microstates in the CFT are such that no
supergravity operators acquire vevs. Thus fuzzballs corresponding to
such states are not visible in the supergravity approximation at all.

\bigskip

Fuzzballs for which the defining curve is a circle in $R^4$ have been
extensively used in the literature; these solutions are a special case
of the Lunin-Mathur solutions (\ref{lun-mat}).
One chooses $F^i(v)$ to be a (multiwound) circle
\cite{Balasubramanian:2000rt,Maldacena:2000dr,Lunin:2001fv},
\be \label{fcirc}
F^1 = \mu_n \cos \frac{2 \pi n v}{L}, \qquad F^2 =
\mu_n \sin \frac{2 \pi n v}{L},
\qquad F^3=F^4=0,
\ee
with the internal excitations being zero.
In this case one can integrate the
expressions for the harmonic functions to obtain the solution in the
form
\bea
ds^2 &=& f_{1}^{- 1/2} f_{5}^{- 1/2} \left ( - (d{t} - \frac{\mu_n
  \sqrt{Q_1 Q_5}}{{r}^2 + \mu_n^2 \cos^2 {\theta}} \sin^2 {\q}
d \phi)^2 + (d{z} -
\frac{\mu_n \sqrt{Q_1 Q_5}}{{r}^2 + \mu_n^2 \cos^2 {\theta}}
\cos^2 {\q} d \psi)^2
\right ) \nn \\
&& + f_{1}^{1/2} f_{5}^{1/2} \left ( ({r}^2 + \mu_n^2 \cos^2 {\q}) (
\frac{d{r}^2}{{r}^2 + \mu_n^2} + d {\q}^2) + {r}^2 \cos^2
{\q} d \psi^2 + ({r}^2 + \mu_n^2) \sin^2 {\q} d \phi^2 \right )
\nn \\
&& +
f_{1}^{1/2} f_{5}^{- 1/2} ds^2(X_4); \label{ez2} \\
e^{2 \Phi} &=& f_1 f_5^{-1}, \nn \\
f_{1,5} &=& 1 + \frac{Q_{1,5}}{{r}^2 + \mu_n^2 \cos^2 {\q}}, \nn
\eea
whilst the RR 2-form field is as in (\ref{equ_D1D5K3pot}) with
\be
A = \mu_n \frac{\sqrt{Q_1 Q_5}}{({r}^2 + \mu_n^2 \cos^2 {\q})} \sin^2
{\q} d\phi; \hsp
B = - \mu_n \frac{\sqrt{Q_1 Q_5}}{({r}^2 + \mu_n^2 \cos^2 {\q})} \cos^2
{\q} d\psi.
\ee
The parameters $(n,\mu_n)$ labeling the curve are
related to the charges via
\be \label{ch-ab}
n \mu_n = \frac{L}{2 \pi} \sqrt{\frac{Q_1}{Q_5}} = \frac{\sqrt{Q_1
    Q_5}}{R_z} \equiv \mu.
\ee
These solutions have been used for illustrative purposes, since
the defining functions are in analytic form and the geometries
preserve a $U(1)^2$ symmetry of the transverse $R^4$.
Furthermore, the fluctuation equations are separable, due to the
symmetry, and thus two point functions can be obtained analytically.

One needs to bear in mind, however, that these solutions are not
representative of generic fuzzballs. In this case the defining
curves are multi-wound, and therefore self-intersect, violating
the condition for non singularity given above. Indeed except for the
case of $n=1$ the solutions have orbifold singularities at the curve
location.

\bigskip

{}From the asymptotics one
can see that the fuzzball solutions have the same mass and D1-brane,
D5-brane charges as the naive solution; the latter are given in
(\ref{q5}) and (\ref{q1}) whilst the ADM mass is
\be
M = \frac{R_z V}{(\a')^4} (Q_1 + Q_5).
\ee
The two charge fuzzball geometries
have the correct charges to describe Ramond ground states in the
CFT. At the same time, to make quantitative progress with the fuzzball
program, one would like to understand the detailed correspondence
between these geometries and microstates. Holographic methods provide
a powerful tool to address this question by extracting field theory
data from a given geometry.

The first step is thus to extract the decoupling asymptotically $AdS$
regions of the fuzzball geometries; this decoupling region is obtained
by dropping the constant from $(\td{f}_1,f_5)$. Note that the very existence of
such a region is a test that a given geometry corresponds to a CFT
ground state. If by contrast a geometry with the correct D1 and D5
charges does not have such a throat region, then it should not be
interpreted as a 2-charge black hole microstate; it fails the first
test.

It may be useful to comment here that {\it all} CFT quantities
should be interpreted in the asymptotically $AdS$ region of the
geometry, be they conserved charges or scattering amplitudes. CFT data
reconstructs the inner $AdS$ region of the asymptotically flat
geometry, but there is no known holographic description which
reconstructs the full geometry. Indeed, if there were, an immediate
corollary would be a holographic description of flat spacetime!

Given the asymptotically $AdS_3 \times S^3 \times X_4$ type IIB
solutions, one can compactify on $X_4$ to obtain asymptotically $AdS_3
\times S^3$ solutions of six dimensional ${\cal N} = 4b$
supergravity. (Note that whilst a general $T^4$ compactification would
give a solution of ${\cal N} =8$ supergravity in six dimensions the solutions
of interest here solve the equations of motion of the truncated
theory.) These six-dimensional solutions can be analyzed in detail using
holographic technology.

\subsection{Map between geometries and microstates}

The key question one would like to address is the correspondence
between a given fuzzball solution and a Ramond ground state in the
CFT. Motivated by the duality relation of the fuzzball solutions
to the fundamental string system, a precise proposal was made in
\cite{Skenderis:2006ah,Kanitscheider:2006zf,Kanitscheider:2007wq} for the
correspondence between
geometries and ground states; see also \cite{Alday:2006nd}.

The fuzzball geometries are determined by giving a curve
$F^I \equiv (F^i(v),{\cal F}(v), {\cal F}^{\a_-}(v))$ in an
$N$ dimensional space, where $N=8$ when $X_4 = T^4$ and $N=24$
when $X_4 = K3$. The corresponding state is now determined
via the following steps.
\begin{enumerate}
\item Fourier expand the curve
\be
F^I(v) = \sum_{n > 0} \frac{1}{\sqrt{n}} (\a^I_n e^{-i n \s^+} +
  (\a^I_n)^{\ast} e^{in \s^+}),
\ee
and consider the coherent state
\be
\left | F^I \right ) = \prod_{n,I} \left | \a^I_n \right ),
\ee
where $\left | \a^I_n \right )$ is a standard coherent state, i.e.
it satisfies
\be
\hat{a}^I_n \left | \a^I_n \right ) =
\a^I_n \left | \a^I_n \right )
\ee
At this stage $\hat{a}^I_n$ are just auxiliary harmonic oscillators.
\item Expressing the coherent states in terms of Fock states,
$\left | \a \right ) = e^{-|\a|^2/2} \sum_k \frac{\a^k }{k!}
(\hat{a}^\dagger)^k
|0 \rangle$,  we identify in $\left | F^I \right )$ the Fock
states that satisfy the constraint
\be
\prod (\hat{a}^I_{-n_I})^{m_I} \left | 0 \right >, \hsp
\sum n_I m_I = N_1 N_5, \quad (n_I>0).   \label{constr}
\ee
\item We retain only these terms from  $\left | F^I \right )$ and map the
harmonic oscillators to CFT R operators via the dictionary
\bea
\frac{1}{\sqrt{2}} (\hat{a}^1_n \pm i \hat{a}^2_n)  & \leftrightarrow &
\hat{\cal O}^{R (\pm 1 + 1),(\pm 1 + 1)}_{n}; \label{fp-d} \\
\frac{1}{\sqrt{2}} (\hat{a}^3_n \pm i \hat{a}^4_n)  & \leftrightarrow &
\hat{\cal O}^{R (\pm 1 + 1),(\mp 1 + 1)}_{n}; \nn \\
\hat{a}^{\tilde{\a}}_{n} & \leftrightarrow &
\hat{\cal O}^{R (1,1)}_{\tilde{\a} n}. \qquad \nn
\eea
where $F^{\tilde{\a}} =({\cal F}(v), {\cal F}^{\a_-}(v))$.
This state is proposed to correspond via AdS/CFT to the fuzzball geometry,
i.e. the vevs of gauge invariant operators in this state should
agree with the one extract from the asymptotics of the solution.
\end{enumerate}

The motivation for this map comes from the duality between the fuzzball
solutions and the supergravity solutions describing a
fundamental string carrying momentum. The FP solutions (\ref{het1})
solve the supergravity equations coupled to a macroscopic classical
string source with the profile given by the curve $F^I$. On general grounds,
one expects that
the classical string source is produced by a coherent state of
string oscillators. For the case at hand, these are BPS solutions
and the auxiliary oscillator $\hat{a}_n^I$ are
identified with the left moving string oscillators. We now want to
apply the duality to map the system to the D1-D5 frame. The FP states
satisfying (\ref{constr}) are precisely the states that map across to
D1-D5 states and this is the reason we retained only this part
of the coherent state. The one to one correspondence between string oscillators
and R operators of the symmetric orbifold CFT in (\ref{fp-d}) goes back to
\cite{Vafa:1994tf}. Note however that while the stringy oscillators
satisfy a Heisenberg algebra, the operator algebra of R ground states
is more complicated.

\subsection{Tests of correspondence}

In this section we will describe the methods used to probe the proposed
correspondence between geometries and black hole microstates.

\subsubsection{Holographic one point functions}

Holographic methodology can be used to systematically test the correspondence.
In particular, one can extract the holographic one point
functions for chiral primaries from the geometry, and compare
them to the vevs in the proposed dual state. Such computations were
carried out in \cite{Skenderis:2006ah,Kanitscheider:2006zf,Kanitscheider:2007wq}.
One first needs to reduce the asymptotically $AdS_3 \times S^3 \times X_4$ type IIB
solutions on $X_4$ to obtain asymptotically $AdS_3
\times S^3$ solutions of six dimensional ${\cal N} = 4b$
supergravity (\ref{sugraIIBK3}), and then one can apply the holographic formulae given in
(\ref{sc-1}), (\ref{j1}), (\ref{t1}) to extract the vevs from the spacetime asymptotics.

Let us briefly sketch the steps involved. The compactification is
straightforward although the explicit reduction
formulae are rather complicated, particularly for $X_4 = K3$; details may be found in
\cite{Kanitscheider:2006zf,Kanitscheider:2007wq}. The resulting
six-dimensional metric and three form fields are given by
\bea
\label{D1D56d}
   ds^2 &=& \frac{1}{\sqrt{f_5 \td{f}_1}} [ -(dt-A_idx^i)^2 + (dy-B_i
     dx^i)^2)] + \sqrt{f_5 \td{f}_1}
dx_idx^i, \\
   G_{tyi}^A &=& \pa_i \left ( \frac{m^A}{f_5 \td{f}_1 } \right ) , \qquad
   G_{\bar{\mu}ij}^A = - 2 \pa_{[i} \left( \frac{m^A}{f_5 \td{f}_1 }
{\cal B}^{\bar{\mu}}_{j]}\right), \nono \\
%   G_{yij}^A &=& -2 \pa_{[i} \left(\frac{m^A}{f_5 \td{f}_1 } A_{j]}\right), \nono \\
   G_{ijk}^A &=& \e_{ijkl} \pa^l m^A + 6\pa_{[i} \left(\frac{m^A}{f_5
       \td{f}_1 } A_j B_{k]}\right), \nn
\eea
with
\bea
   m^{n} &=& \left(0_4,\qu (f_5 + F_1) \right),\\
 m^{r} &=& \frac{1}{4}\left((f_5 -  F_1),
 - 2 A_{\a} ,
- \sqrt{2} N^{(c)}, 2 A_5 \right) \nono \\
&\equiv & \frac{1}{4}\left((f_5 - F_1),  - 2 {\cA}^{\a_-}, 2 \cA
\right). \nn
\eea
Here $n=1 \cdots 5$ and $r = 6 \cdots (n_t+5)$ with
the index $\a=6,7,8$ and $F_1 = (1 + K)$. The corresponding $SO(5,n_t)$ matrix of scalar fields may be found in
\cite{Kanitscheider:2006zf,Kanitscheider:2007wq}.

Before giving the vevs let us remark that in solutions with only
transverse excitations there are only two non-zero tensors
$(G^5,G^6)$. Such solutions involve only an $SO(1,1)$ truncation of
the  ${\cal N} = 4b$ fields. Moreover, if one sets $f_1 = f_5$ then
$G^6 = 0$, and the resulting solution solves the equations of minimal
supergravity in six dimensions (i.e. with no additional tensors coupled).
Looking ahead to the next section, we will see that
almost all known fuzzball solutions for three charge black holes
involve only the $SO(1,1)$ truncation of
the  ${\cal N} = 4b$ fields; there are no analogues of the fuzzballs
with internal excitations. It would be interesting to explore whether
one can generate more general three charge solutions using the
$SO(5,n_t)$ symmetry of the ${\cal N} = 4b$ theory.

The decoupling asymptotically $AdS_3 \times S^3$ region of the geometry is
obtained by dropping the constant terms in the functions $(f_1,
\td{f}_1, f_5,F_1)$. Next one can expand the defining functions for
large radius and extract the appropriate fields (\ref{flc1})
characterizing the perturbation with respect to $AdS_3 \times
S^3$. Substituting these perturbations in the formulae (\ref{sc-1},
\ref{j1},\ref{t1}) gives the holographic vevs.

The vevs of the stress energy tensor and of the R symmetry currents are
\be
\left < T_{{\mu}{\nu}} \right > = 0; \qquad
\left < J^{\pm \a} \right > = \pm \frac{N}{2 \pi} a^{\a \pm} (dz \pm
dt), \label{j-char}
\ee
where $a^{\a \pm}$ is a constant defined below in (\ref{a_def}).
The vanishing of the stress energy tensor is as anticipated, since
these solutions should be dual to R vacua; however, the cancellation is very
non-trivial. The vevs of the low lying chiral primaries were also
computed, and expressed in terms of the harmonics of the curves
defining the solutions:
\bea
   f_{kI}^5 &=& \frac{1}{L (k+1)}  \int_{0}^L dv (C^I_{i_1\cdots i_k}
   F^{i_1} \cdots F^{i_k}) , \label{coeff} \\
   f_{kI}^1 &=& \frac{Q_5}{L (k+1)Q_1} \int_{0}^L dv \left (\dot{F}^2 + \dot{\cal{F}}^2 +
     (\dot{\cal{F}}^{\a_-})^2 \right ) C^I_{i_1\cdots i_k} F^{i_1} \cdots
   F^{i_k}  , \nono \\
   (A_{kI})_i &=& -\frac{1}{L (k+1)} \int_{0}^L  dv
\dot{F}_i C^I_{i_1\cdots i_k} F^{i_1} \cdots F^{i_k}, \nono \\
   (\cA_{kI}) &=& -\frac{\sqrt{Q_5}}{\sqrt{Q_1} L(k+1)} \int_{0}^L
dv \dot{\cal{F}} C^I_{i_1\cdots i_k} F^{i_1} \cdots F^{i_k} , \nono \\
   \cA^{\a_-}_{kI} &=& -\frac{\sqrt{Q_5}}{\sqrt{Q_1} L(k+1)} \int_0^L dv
\dot{\cal{F}}^{\a_-} C^I_{i_1\cdots i_k} F^{i_1} \cdots F^{i_k} .\nn
\eea
Here the $C^I_{i_1\cdots i_k}$ are orthogonal symmetric traceless rank $k$
tensors on $\mathbb{R}^4$ which are in one-to-one correspondence with
the (normalized) spherical harmonics $Y_k^I(\theta_3)$ of degree $k$
on $S^3$. Fixing the center of mass of the whole system implies that
\be
   (f_{1i}^1 + f_{1i}^5) = 0.
\ee
The leading term in the asymptotic expansion of the transverse gauge
field $A_i$ can be written in terms of degree one vector harmonics as
\be
  A = \frac{Q_5}{r^2}(A_{1j})_iY_1^j dY_1^i \equiv
\frac{\sqrt{Q_1Q_5}}{r^2}(a^{\a-}Y_1^{\a-} + a^{\a+} Y_1^{\a+}),
\ee
where $(Y_1^{\a-},Y_1^{\a+})$ with $\a=1,2,3$ form a basis for
the $k=1$ vector harmonics and we have defined
\be \label{a_def}
a^{\a\pm}
= \frac{\sqrt{Q_5}}{\sqrt{Q_1}} \sum_{i > j} e^\pm_{\a ij} (A_{1j})_i,
\ee
where the spherical harmonic triple overlap $e^\pm_{\a ij}$ is
defined in (\ref{ap-ov3}).

Note that the vevs of the R symmetry currents increase with the
radius of the curve in $R^4$. For example, the vevs for geometries
based on the circular curves (\ref{fcirc}) are
\be
\left < J^{\pm 3} \right > = \pm \frac{N}{2 \pi n}  (dz \pm
dt), \label{j-char2}
\ee
where the symmetry implies that only the $J^{\pm 3}$ components
acquire vevs. In this case the R symmetry charges are proportional to
the radius of the curve. Typical Ramond states have small R charges,
and necessarily correspond to geometries sourced by curves of small
radii.

The vevs of the scalar operators dual to the fields $(s^{(6)k}_I,\s^k_I)$
are given in terms of the transverse fluctuations:
\bea
\left < {\cal O}_{S^{(6)1}_i} \right > &=& \frac{N}{4 \pi}
(- 4 \sqrt{2} f^{5}_{1i}); \label{vv2} \\
\left < {\cal O}_{S^{(6)2}_I} \right > &=& \frac{N}{4 \pi} ( \sqrt{6} (f^1_{2I} -
f^5_{2I}) ); \nn \\
\left < {\cal O}_{\Sigma^2_I} \right > &=& \frac{N}{4 \pi} \sqrt{2}
( -  (f^1_{2I} +f^5_{2I}) + 8 a^{\a -} a^{\b +} f_{I\a\b} ). \nn
\eea
The internal excitations of the fuzzball solutions are
captured by the vevs of operators dual to the fields
$s^{(r)k}_I$ with $r > 6$:
\bea
\left < {\cal O}_{S^{(5 + n_t)1}_i} \right > &=& - \frac{N}{\pi}
\sqrt{2} (\cA_{1i}); \hsp
\left < {\cal O}_{S^{(6+ \a_-)1}_i} \right > =
\frac{N}{\pi} \sqrt{2} \cA^{\a_-}_{1i}; \label{vv-new} \\
\left < {\cal O}_{S^{(5+n_t)2}_I} \right > &=&  - \frac{N}{2 \pi}
\sqrt{6} (\cA_{2I}); \hsp
\left < {\cal O}_{S^{(6+ \a_-)2}_I} \right > =
\frac{N}{2 \pi} \sqrt{6} \cA^{\a_-}_{2I}. \nn
\eea
Here $n_{t} = 5,21$ for $T^4$ and $K3$ respectively, with $\a_{-} =
1,\cdots,b^{2-}$ with $b^{2-} = 3,19$ respectively.
Thus each curve $({\cal F}(v), {\cal F}^{\a_-}(v))$ induces
corresponding vevs of operators associated with the middle cohomology
of $M^4$. Note the sign difference for the vevs of operators which are
related to the distinguished harmonic function ${\cal F}(v)$.

\bigskip

One would like to reproduce these vevs from the field theory, using
the proposed correspondence between geometries and Ramond ground
states. Substantial progress in this direction was obtained in
\cite{Skenderis:2006ah,Kanitscheider:2006zf,Kanitscheider:2007wq}
where the detailed structure of these
vevs was reproduced from the corresponding field theory
calculations. The point is that one point functions in a given state
may be related to three point functions at the conformal
point. Consider a general state such that
$|\Psi\rangle = O_\Psi(0)|0\rangle$. Then the vev of an operator
$O_k$ of dimension $k$ in the this state is given by
\be
\langle \Psi| O_k(\l^{-1}) |\Psi \rangle
= \langle 0| (O_\Psi(\infty))^\dagger O_k(\l^{-1}) O_\Psi(0)|0\rangle,
\ee
where $\l$ is a mass scale.
For scalar operators the 3-point function is uniquely determined by
conformal invariance and the above computation yields
\be \label{fusion}
\langle \Psi| O_k(\l^{-1}) |\Psi \rangle  = \l^k C_{\Psi k \Psi}
\ee
where $C_{\Psi k \Psi}$ is the fusion coefficient. Similarly, the
expectation value of a symmetry current measures the charge
of the state
\be
\langle \Psi| j(\l^{-1}) |\Psi \rangle
= \langle 0| (O_\Psi(\infty))^\dagger j(\l^{-1}) O_\Psi(0)|0\rangle
= q \l  \langle \Psi|\Psi \rangle
\ee
where $q$ is the charge of the operator $ O_\Psi$ under $j$.

To reproduce the vevs one needs to consider states $|\Psi \rangle$
which are ground states in the Ramond sector, or equivalently chiral
primaries in the NS sector. The stress energy tensor and the R current
expectation values are immediately reproduced from the proposed
correspondence between curves and superpositions of Ramond ground states.
Although the relevant
three point functions of scalar chiral primaries have not been computed in
full generality, the structure of the vevs given above is reproduced
using selection rules for the three point functions, along with the large
$N$ expansion.

Selection rules are responsible for determining
which scalar operators acquire a vev in a given superposition of Ramond
ground states. For example, one can see from (\ref{vv-new}), (\ref{coeff})
that primaries associated with the $(1,1)$ cohomology of $X_4$ acquire a
vev only if both internal and transverse fluctuations of the fuzzball
are non-zero. Moreover, the internal and transverse curves must share
common Fourier modes for the dimension one operators to acquire vevs;
further selection rules on the Fourier modes are needed for dimension
two operators to acquire vevs. This structure can be reproduced in the field
theory, using selection rules for the three point functions. A basic property of such
three point functions is that they are only
non-zero when the total number of operators ${\cal O}^{(1,1)}_{\td{\a}}$ with
a given index $\td{\a}$ in the correlation function is even. This
implies that the $(1,1)$ chiral primaries only acquire vevs when the fuzzball has internal
excitations. When there are no internal internal fluctuations, there
are no ${\cal O}^{(1,1)}_{\td{\a}}$ operators in the superposition defining
the state, so  total number of operators ${\cal O}^{(1,1)}_{\td{\a}}$ with
a given index $\td{\a}$ in the three point function is never even. By
contrast when there are internal excitations there are
${\cal O}^{(1,1)}_{\td{\a}}$ operators in the superposition defining
the state, and thus $(1,1)$ chiral primaries generically acquire vevs.

One can also easily see why the operator only acquires a vev if there
are transverse excitations as well. All Ramond ground states
associated with the middle cohomology have
zero R charge, with the corresponding chiral primaries in the NS
sector having the same charge  $j_{3}^{NS} = \half
N$. Thus a superposition involving only ${\cal O}^{(1,1)}$ operators has a
definite R charge, and a charged operator cannot acquire a
vev. Including transverse excitations means that the superposition
of Ramond ground states contains charged operators, associated
with the other cohomology, does not have definite R
charge and therefore a charged operator can acquire a vev. Note that
this point also explains why superpositions
involving only ${\cal O}^{(1,1)}$ operators
do not have dual fuzzball descriptions within supergravity: no
operator which is dual to a supergravity mode can acquire a vev.

In fact one can go even further in matching field theory one point
functions to the holographic vevs, using the fact that the most
disconnected component of any correlator dominates at large $N$. In
practice this implies that for many calculations the operators involved in the
superposition behave like free harmonic oscillators - the deviation of
the algebra from the free algebra is subleading in $N$ -  and thus
even the numerical values of the vevs can be reproduced by harmonic oscillator
algebra! This point was illustrated in detail in \cite{Kanitscheider:2006zf,Kanitscheider:2007wq}
using specific examples, such as elliptic curves.
Given the recently refined dictionary relating supergravity fields to chiral
primaries \cite{Taylor:2007hs}, and the proposed non-renormalization of the chiral ring,
one could hope to push these calculations even further.

Another issue which is nicely demonstrated by the calculations in
\cite{Kanitscheider:2006zf,Kanitscheider:2007wq} is the
indistinguishability of many solutions with supergravity: typically
the vevs extracted from a solution defined by any given curve are
small, comparable with the next order corrections to
supergravity. Thus even fuzzballs which are non-singular in
supergravity generally suffer from indistinguishability.

To summarize: the detailed holographic data extracted from these two charge
geometries supports their interpretation as superpositions of Ramond
ground states. Moreover, the data reinforces the general expectation
that supergravity does not suffice to describe all the fuzzballs. Some
cannot be seen at all in the supergravity approximation, whilst others
cannot be reliably distinguished.

As we shall see later there is as yet no complete understanding of the
space of fuzzball solutions for the 3-charge system: candidate
solutions within supergravity seem to be characterized by a (constrained) set of curves along
with additional parameters and the correspondence with states in the
CFT is unclear. Extracting vevs from such geometries may
thus provide a useful indicator as to the correspondence with black hole
microstates.

\subsubsection{Flight time and mass gap}

Apart from matching the one-point functions, one could also test the
correspondence by matching higher point functions. To make precision
tests is rather hard in general. Firstly to compute even a two-point
function one needs to solve a partial differential equation on the
asymptotically $AdS_3 \times S^3$ background, which for a general
fuzzball solution is not even separable due to lack of
symmetry. Secondly, $n$-point functions in a given geometry are
related to $(n+2)$-point functions at the conformal point, and there
is no reason to expect that these are non-renormalized even for chiral
primaries. Thus the
correlation functions computed in the orbifold CFT will not necessarily match
those in the geometry.

Bearing in mind these caveats, estimates of the two point functions in
symmetric geometries can give indicative information about the
correspondence with CFT states, see
\cite{Lunin:2001jy,Lunin:2002qf,Giusto:2004ip,Mathur:2005zp,Mathur:2005ai}. For example, in the geometric
approximation for the two point function of a massless scalar one
considers null geodesics propagating inwards and then returning back
to the boundary. The (finite) affine length of such a geodesic defines
a characteristic scale or inverse mass gap.
To give an example: one can consider the geometries obtained from a
circular curve in $R^4$ given in (\ref{ez2}). The decoupled region in the
geometry has a metric:
\bea
ds^2 &=& \sqrt{Q_1 Q_5} \left ( - (r^2 + \mu_n^2) d\hat{t}^2 + r^2 d\hat{z}^2 +
\frac{dr^2}{(r^2 + \mu_n^2)} \right ) \label{nh} \\
&& + \sqrt{Q_1 Q_5}
\left ( d{\q}^2 + \sin^2 {\q} (d \phi + \mu_n d \hat{t})^2 +
\cos^2 {\q} (d \psi - \mu_n d \hat{z} )^2 \right ) + \sqrt{Q_1/Q_5}
ds^2 (X_4), \nn
\eea
where $\mu_n = \mu/n = \sqrt{Q_1 Q_5}/(nR_z)$ and $\hat{z}$ is periodic
with periodicity $2\pi \mu^{-1}$. Null radial geodesics such that $\q
= 0$ and $(\hat{z},\psi)$ are constant satisfy
\be
\left ( \frac{dr}{d \hat{t}} \right )^2 = (r^2 + \mu_n^2)^2,
\ee
and thus the total affine length of a geodesic propagating inwards from radial
infinity and back out again is
\be
\hat{T} = \int_{0}^{\infty} \frac{dr}{(r^2 + \mu_n^2)} = \frac{\pi}{2
  \mu_n}.
\ee
The corresponding energy scale $\mu_n$ characterizes the
mass gap of the spectrum in this sector
of the CFT and since $\mu^{-1}$ is the scale of $\hat{z}$,
the dimensionless energy scale is $1/n$. (This is
the same energy scale that is obtained from the holographic
computation of the R charges, see (\ref{j-char2}).)
This fits with the conjectured
correspondence between these geometries and dual R charge eigenstates,
$(\hat{\cal O}^{R(p,q)}_n)^{N/n}$: excitations in the twist $n$ sector
are of the scale $1/n$. In the effective string picture,
the excitation energy  is also
related to the time taken by excitations to
propagate around the circle wound by the string.

In such a simple example, one does not acquire new information from
this computation. In more complicated fuzzball geometries, however,
such as 3-charge bubbling solutions to be discussed later, this
estimation of the mass gap may be a useful indicator of the dual black
hole microstate. At the same time, in geometries which break
rotational symmetry and in which there are many parameters, the flight
time is sensitive to the specific geodesic under consideration and has
to be interpreted with care.

\subsubsection{Geometric quantization} \label{geo-quant}

The fuzzballs are characterized by a curve in an $8$ ($T^4$) or $24$
($K3$) dimensional space. One would like to enumerate them and compare
them with the number of dual microstates, but naively one has an
infinite number of solutions since any classical curve determines a solution.
The map between fuzzball solutions and microstates effectively discretizes the space
of fuzzball solutions. From our earlier discussion it is clear that
the classical curves are due to the states being approximately coherent; the
coefficients in the Fourier expansion of the curve are eigenvalues of quantum
harmonic oscillators.

Alternatively, note that the space of curves is the classical phase space of these
gravitational solutions.
The number of the corresponding quantum states should be obtained by appropriate
quantization. These states should be in correspondence with the states in the
dual CFT. One can thus consider
quantizing this moduli space, and comparing the latter to the
Ramond ground states in the CFT. For fuzzballs with only
transverse excitations, namely (\ref{lun-mat}), such that the
solutions are determined by
a smooth closed non-intersecting curve in four dimensions, the
geometric quantization was carried out in \cite{Rychkov:2005ji}.
It was found that the Fourier coefficients of the curve satisfy chiral boson
commutation relations, with the value of the effective Planck constant
agreeing with that obtained from dualising the fundamental string.

The basic idea of this quantization procedure is as follows. Consider
any classical dynamical system with phase space coordinates satisfying
the standard Poisson brackets
\be
\{ q^I, p^J \} = \d^{IJ}.
\ee
Now restrict to a subspace $M$ of the full phase space, parameterized by
some coordinates $x^A$. The induced Poisson brackets can be extracted
from pullback of the symplectic structure onto this subspace. The
symplectic structure on the phase space is given by
\be
\Omega = d p^{I} \wedge dq^{I},
\ee
and the pullback $\Omega_M$ of this symplectic structure onto a subspace is given
by
\bea
\Omega_M &=& \omega_{AB}(x) d x^{A} \wedge dx^B, \\
\omega_{AB} &=& \left ( \frac{\pa p^I}{\pa x^{A}} \frac{\pa q^I}{\pa
  x^{B}} -  \frac{\pa p^I}{\pa x^{B}} \frac{\pa q^I}{\pa
  x^{A}} \right ) . \nn
\eea
Then the induced Poisson brackets are given by the inverse of $\w$,
\be
\{ x^A, x^B \} = \half \w^{AB}.
\ee
Once one has extracted the Poisson brackets from the symplectic form, one
can quantize in the standard way to find the commutation relations.

In the case of interest, one considers the symplectic form of
supergravity, restricted to the fuzzball solutions of interest, and
extracts the appropriate Poisson brackets. Starting from the
relevant part of the IIB supergravity action (in Einstein frame)
\be
S_{IIB} = \frac{1}{2 \k_{10}^2} \int d^{10}x \sqrt{-g} (R - \half (\pa
\Phi)^2 - \half e^{\Phi} F_3^3),
\ee
the symplectic form is
\be
\Omega = \frac{1}{2 \k_{10}^2} \int d\Sigma_{m} J^m
\ee
where the integration is over a Cauchy surface $\Sigma$ and the
symplectic current $J^m$ is given by \cite{Crnkovic,Wald,Rychkov:2005ji}
\bea
J^{m} &=& - \d \G^{m}_{ab} \wedge \d (\sqrt{-g} g^{ab}) + \d \G^{b}_{ab}
\wedge \d (\sqrt{-g} g^{am}) - \d (\sqrt{-g} e^{-\Phi} F^{mab} )
\wedge \d C_{ab}  \nn \\
&& - \delta (\sqrt{-g} \pa^m \Phi) \wedge \d \Phi.
\eea
Restricting to the fuzzball solutions (\ref{lun-mat}) characterized by a curve
$F_i(v)$ in $R^4$, the pullback $\Omega_M$ of the symplectic form can be
explicitly evaluated to give
\be \label{op-s}
\Omega_M = \frac{1}{2 \a} \int \d (\pa_v F_i(v)) \wedge \d F_{i}(v) dv
\ee
which implies the Poisson bracket
\be
\{ F_i(v), \pa_v F_j(v') \} = \a \d_{ij} \d (v - v').
\ee
The parameter $\a$ is found by explicit computation of the symplectic
form to be given by
\be
\a = \pi \frac{g_s^2}{R_z^2 V},
\ee
in agreement with the value obtained by duality from the fundamental
string. Thus as previously stated the defining curve of the fuzzball
solution satisfies the standard Poisson bracket relations of chiral
bosons. One can next promote the Poisson bracket to a quantum
commutation relation, and introduce harmonic oscillators
$(\hat{a}^i_n, (\hat{a}^i_n)^{\dagger})$ for the
curve. To count the number of black hole microstates obtained from
these geometries, one must then restrict to curves satisfying
the relation constraining the total D1-brane charge:
\be
Q_1 = \frac{Q_5}{L} \int_{0}^L | \pa_v F^i |^2 dv.
\ee
As expected this reduces to counting the degeneracy of states at level $N_1
N_5$ in a system of four chiral bosons, namely counting the degeneracy
of states satisfying
\be
N_1 N_5 = \sum_{n =1}^{\infty} n \langle (\hat{a}^i_{n})^{\dagger} \hat{a}^i_n \rangle,
\ee
which gives an entropy
\be
S \approx 2 \pi \sqrt{\frac{2}{3} N_1 N_5}.
\ee
It would be interesting to see if the quantization of  general fuzzball
geometries with internal excitations gives an analogous
answer, corresponding to eight or twenty-four bosons for $T^4$ and
$K3$ respectively. Given that fuzzballs with only internal excitations
are not visible in supergravity, it seems possible that the symplectic
form in the general case would have a more complicated structure than
in (\ref{op-s}), but the calculation has not been carried out.

Whilst the geometric quantization reproduces the expected entropy of
the 2-charge microstates, one must clearly view this calculation with
some caution. The 2-charge system does not have a horizon in
supergravity, so the average fuzzball geometry must have a scale
which is sub-string scale and thus higher derivative
corrections to supergravity are non-negligible for these
geometries. It is not clear that quantizing the extrapolation of
fuzzball geometries to the supergravity regime will in general give the
correct counting, even though it does in this case.

Note that in this case the number of microstates accounted for by the
fuzzball solutions was also estimated using (probe) supertubes
\cite{Mateos:2001qs} in the dual D0-F1 system. In \cite{Palmer:2004gu} and \cite{Bak:2004kz},
the quantum states of the supertube were counted
by directly quantizing the linearized Born-Infeld action near a
round tube. This gives an entropy $S = 2\pi \sqrt{2
  N_{D0}N_{F1}}$, where $N_{D0}$ is the D0-brane charge and $N_{F1}$
is the fundamental string charge. Backreacting a supertube with given
profile in $R^4$ onto
the supergravity fields, and then dualizing, should give the D1-D5 fuzzball
solutions characterized by the same profile in $R^4$. Thus the entropy
obtained from quantizing the Born-Infeld action indicates the number
of 2-charge fuzzball solutions, with the entropy matching the expected
value. This approach is less direct than that discussed above, and is
hard to generalize to the 3-charge system where there is no such
duality to a supertube description.

\subsection{Open questions}

Even given the detailed analysis already made in this system, there
remain many interesting open questions and lessons for black holes
with macroscopic horizons. Firstly, the recent demonstration that the chiral ring is
non-renormalized, along with the matching between orbifold CFT
operators and supergravity fields \cite{Taylor:2007hs}, would allow the vevs to be computed
systematically at large $N$, and for the proposed correspondence to be
tested to even higher precision using the holographic methods.

Furthermore, there is a detailed map from smooth geometries to black
hole microstates, but not vice versa. Put differently, it is
not understood directly in the D1-D5 system how a curve emerges.
It is clearly important to understand this better,
not least because this might lead to clues as to the defining data of
3-charge D1-D5-P fuzzball solutions. Note that a given R charge
eigenstate does not in general appear to have a smooth supergravity
description, whereas the smooth geometries correspond to large
superpositions of R charge eigenstates with the specific superposition
determined by the classical curve. Thus the natural geometric basis
does not coincide with the natural field theory basis, and the
geometric basis does not sample well states with very small R charge.

Given the matching between geometries and microstates, one could
consider coarse-graining over the fuzzball geometries, to understand
how black hole properties emerge. Of course in this specific system
the resulting black hole does not have a regular horizon in
supergravity, so the coarse-graining can at best provide an indication
of what happens for macroscopic black holes, but nonetheless it may be
informative to explore this issue.

In general the fact that many of the fuzzball geometries have very
small parameters and are not well described using only supergravity
could be interpreted as an indication that they will develop a horizon
when higher derivative corrections are taken into account. For fuzzballs
characterized by a generic curve this however seems unlikely: there is
no singularity in the supersymmetric supergravity solution, so the
higher derivative corrections are bounded.

Results in this section indicate that a more complete understanding of this
system rests on understanding fuzzballs in the stringy regime. This is
not unexpected, as the black hole does not have a macroscopic horizon,
and thus one would clearly like to develop the fuzzball proposal for
black holes with macroscopic horizons. As we shall see, however, even
for large black holes, one is unlikely to evade the need to go beyond
supergravity. Thus this 2-charge system may be useful as the simplest test case
for both sharpening the definition of fuzzballs outside the
supergravity regime and for developing calculational techniques to
handle stringy fuzzballs.

\section{Three and four charge systems}

In the previous section we have discussed fuzzball solutions for the two
charge system, and shown how AdS/CFT methodology can be used to make
detailed tests of the correspondence between such geometries and CFT
states. At the same time, the two charge black hole does not have a
macroscopic horizon, so one would like to make progress with a system
which does have a horizon in supergravity.

As a first step one would like to develop the fuzzball picture for
asymptotically flat, (near) supersymmetric,
black holes in four and five dimensions which have
finite area horizons in supergravity. This class
includes static five-dimensional black holes with three charges, the
first black hole whose entropy was explained microscopically in string
theory by Strominger and Vafa \cite{Strominger:1996sh}, along with the rotating BMPV
generalization \cite{Breckenridge:1996is}.
In four dimensions black holes with a macroscopic horizon have four
charges; for reviews of black hole solutions see \cite{Youm:1997hw}.

There are a variety of reasons for developing first the picture for
supersymmetric black holes. Perhaps most important is that
in constructing candidate fuzzball
solutions for supersymmetric black holes, one is not forced to solve
the (highly non-linear) supergravity equations. One instead first solves the
first order supersymmetry equations and then enforces where
necessary additional restrictions arising from the equations of
motion. Moreover, one can make use of the systematic classifications
of supersymmetric solutions of supergravity that have been derived in
recent years; in fact, almost all candidate three charge solutions
derive from the classification of minimal supergravity in five
dimensions \cite{Gauntlett:2002nw}. Once supersymmetry is broken, one loses these powerful
tools, and, as we will later review, very few fuzzball solutions are
known.

Detailed microscopic explanations of the entropy are also only
currently possible for supersymmetric black holes. Following the
steps of Strominger and Vafa, one relates states in a D-brane system at weak
coupling to a black hole at strong coupling, but the degeneracy of
states in the weakly coupling theory can only be compared to the
entropy of the black hole at strong coupling if supersymmetry implies
non-renormalization. This is the case for (the leading order term in)
the entropy of supersymmetric black holes with large charges.

Without detailed knowledge of the properties of black hole
microstates, it is hard to probe whether candidate fuzzball
geometries do indeed correspond to such microstates. Even when
one has a description in terms of D-branes at weak coupling, one
cannot generically make a precise map between all data in this
theory and that extracted from the fuzzball geometry.

One can however develop a precision map in cases where one can use AdS/CFT
technology. Thus to push the fuzzball proposal further it is natural
to consider black holes which admit an $AdS_3$ factor
in their near horizon regions, and relate the fuzzball geometries to
states in the dual two dimensional conformal field theories. AdS/CFT
technology is rather less developed for five and four dimensional
black holes with $AdS_2$ near horizon regions, and thus we will not
focus on such black holes here, although candidate fuzzball geometries
have been constructed in these cases.

Before moving on to reviewing relevant properties of the black holes,
one should note that whilst supersymmetric solutions can have finite
horizon area they do not have non-zero temperature. Thus one does not
expect to address how temperature and Hawking radiation emerges in
such a system. We will return to this issue in section \ref{open}.

\subsection{Black hole geometry and D-branes}

In this section we will review relevant
properties of 3-charge black hole solutions. For what follows it is
convenient to discuss solutions both of M theory compactified on a
Calabi-Yau manifold, and of type IIB compactified on $X_4$ which is
either $T^4$ or $K3$. Our conventions for the supergravity field
equations are given in appendix \ref{conv}.

\subsubsection{D1-D5 solutions}

Let us begin with the type IIB solutions. Supersymmetric rotating (BMPV)
strings wrapping a circle and carrying D1 charge $Q_1$, D5 charge
$Q_5$ along with momentum charge
$Q_p$ along the circle have a metric of the form
\bea
ds^2 &=& \frac{1}{\sqrt{h_1 h_5}} \left ( - (dt^2  - dz^2) +
\frac{Q_p}{r^2} (dz - dt)^2 \right ) \nn  \\
&&
+ \sqrt{h_1 h_5} \left ( dr^2  +
r^2 (d \q^2 + \cos^2 \q d \psi^2 + \sin^2 \q d \phi^2) \right )
\label{bmpv1} \\
&&
+ \frac{2 \sqrt{Q_1 Q_5 Q_p}}{r^2 \sqrt{h_1 h_5}} (L_1 - L_2) ( dt -dz ) (
\cos^2 \q d \psi - \sin^2 \q d \phi)
+ \sqrt{\frac{h_1}{h_5}} ds^2(X_4), \nn
\eea
with $h_i = 1 + Q_i/r^2$ and where
$ds^2(X_4)$ denotes the metric on $T^4$ or $K3$ respectively. The
corresponding RR 2-form potential
and dilaton are given by
\bea
e^{-2 \Phi} &=& \frac{h_5}{h_1}; \\
C &=& (h_1^{-1} - 1) dt \wedge dz - Q_5 \cos^2 \q d \psi \wedge d
\phi \nn \\
& & + \frac{\sqrt{Q_1 Q_5 Q_p}}{(r^2 + Q_1)} (L_2 - L_1) (dt -dz) (
\cos^2 \q d \psi - \sin^2 \q d \phi). \nn
\eea
 The
angular momentum charge is proportional to $(L_1 - L_2)$ and will be given
below.

This solution follows from the supersymmetric limit of general non-extremal 3-charge black
hole (actually black string) solutions of type IIB with rotation constructed
in \cite{Cvetic:1996xz}. The metric for these solutions is
\bea
ds^2 &=& \frac{1}{\sqrt{H_1 H_5}} \left ( - f(dt^2  - dz^2) + m (s_p dz
- c_p dt)^2 + m (a_1 \cos^2 \q d \psi + a_2 \sin^2 \q d \phi)^2 \right
) \nn  \\
&&
+ \sqrt{H_1 H_5} \left ( \frac{r^2 dr^2}{ (r^2 + a_1^2)(r^2 + a_2^2) -
  m r^2} + d \q^2 + \cos^2 \q d \psi^2 + \sin^2 \q d \phi^2 \right )
\label{cvetic-youm} \\
&&
+ \frac{2m}{\sqrt{H_1 H_5}} \left ( \cos^2 \q ( (a_1 c_1 c_5 c_p - a_2
s_1 s_5 s_p) dt + (a_2 s_1 s_5 c_p - a_1 c_1 c_5 s_p) dz) d \psi\right
) \nn \\
&& + \frac{2m}{\sqrt{H_1 H_5}} \left (
\sin^2 \q ( (a_2 c_1 c_5 c_p - a_1
s_1 s_5 s_p) dt + (a_1 s_1 s_5 c_p - a_2 c_1 c_5 s_p) dz) d \phi
\right ) \nn \\
&& + (a_2^2 - a_1^2) \frac{H_1 + H_5 - f}{\sqrt{H_1 H_5}} (\sin^4 \q d
\phi^2 - \cos^4 \q d \psi^2) + \sqrt{\frac{H_1}{H_5}} ds^2(X_4), \nn
\eea
where the solution is parameterized by $(\d_1,\d_5,\d_p)$ and we use
the abbreviations $s_i = \sinh \d_i$, $c_i = \cosh \d_i$.
The functions $(H_1,H_5,f)$ are given by
\be
H_i = f + m \sinh^2 \d_i, \qquad
f = r^2 + a_1^2 \sin^2 \q + a_2^2 \cos^2 \q.
\ee
The RR 2-form potential and the dilaton are given by
\bea
C &=& \frac{m \cos^2 \q}{H_1} \left ( ( a_2 c_1 s_5 c_p - a_1
s_1 c_5 s_p) dt + (a_1 s_1 c_5 c_p - a_2 c_1 s_5 s_p) dz \right )
\wedge d\psi \\
&& + \frac{m \sin^2 \q}{H_1} \left ( ( a_1 c_1 s_5 c_p - a_2
s_1 c_5 s_p) dt + (a_2 s_1 c_5 c_p - a_1 c_1 s_5 s_p) dz \right )
\wedge d\phi \nn \\
&& - \frac{m s_1 c_1}{H_1} dt \wedge dz - \frac{m s_5 c_5}{H_1} (r^2 +
a_2^2 + m s_1^2) \cos^2 \q d\psi \wedge d\phi; \nn  \\
e^{-2 \Phi} &=& \frac{H_5}{H_1}. \nn
\eea
Thus the solution is labeled by six parameters $(m,\d_i,a_1,a_2)$
along with the coupling constants and the
two moduli $(R_z,V)$. The former is the radius of the $z$ circle and
the latter is related to the volume $v$ of $X_4$ by $v = (2 \pi)^4 V$. The
solution is asymptotically flat, approaching $R^{4,1} \times S^1
\times X_4$ as $r \rightarrow \infty$; here $S^1$ is the $z$ circle.
The horizons are located at
\be
(r^2 + a_1^2) (r^2 + a_2^2) = m r^2,
\ee
and the topology of the horizon is $S^1 \times S^3 \times X_4$ where
again the $S^1$ direction is associated with the $z$ circle. This
solution is hence a black string, and compactifying to five dimensions
over $S^1 \times X_4$ produces a five-dimensional three charge
rotating black hole.

The conserved charges and angular momenta are given by
\bea
%M &=& \frac{\pi m}{4 G_5} (s_1^2 + s_5^2 + s_p^2 + \frac{3}{2}) \\
Q_1 &=& m s_1 c_1 = \frac{g \a^{'3} N_1}{V}; \label{charges} \\
Q_5 &=& m s_5 c_5 = g \a' N_5; \nn \\
Q_p &=& m s_p c_p = \frac{g^2 (\a')^4}{V R_z^2} N_p, \nn \\
J_{\psi} &=&  - \frac{\pi m}{4 G_5} (a_1 c_1 c_5 c_p - a_2 s_1 s_5
s_p) \nn \\
J_{\phi} &=&  - \frac{\pi m}{4 G_5} (a_2 c_1 c_5 c_p - a_1 s_1 s_5
s_p) , \nn
\eea
where $(N_1,N_5,N_p)$ are the integral quantized charges and
\be
\frac{\pi}{4 G_5} = \frac{R_z V}{g^2 (\a')^4}.
\ee
The mass is given by
\be
M = \frac{\pi m}{4 G_5} (s_1^2 + s_5^2 + s_p^2 + \frac{3}{2}).
\ee
Regular extremal black holes which saturate a BPS bound
are obtained in the limit $m \rightarrow 0$ and $\d_i \rightarrow
\infty$ with $Q_i = m s_i^2 $, $L_1 = a_1/\sqrt{m_1}$ and $L_2 =
a_2/\sqrt{m_2}$ held fixed. The solution in this limit becomes
(\ref{bmpv1}).

The decoupling limit of these solutions is obtained by focusing on the
region $r^2 \ll Q_1, Q_5$; then $H_1$ and $H_5$ may be approximated by
$H_1 \approx Q_1$ and $H_5 \approx Q_5$. One also focuses on near
extremal solutions for which $Q_1, Q_5 \gg m,a_1^2,a_2^2$ so that
$m \sinh \d_1 \sinh \d_5 \approx m \cosh \d_1 \cosh \d_5 \approx
\sqrt{Q_1 Q_5}$. In this limit the metric can be written as
\bea \label{btzz}
ds^2 &=& - l^2 (\rho^2 - M_3 + \frac{J_3^2}{4 \rho^2}) d\t^2 +
 ({\rho^2} - M_3 + \frac{J_3^2}{4 \rho^2})^{-1} l^2 d\rho^2 +
l^2 \rho^2 (d\varphi - \frac{J_3}{2 \rho^2} d\t)^2 \nn \\
&& + l^2 d \q^2 + \sin^2 \q (d \phi + \frac{R_z}{l} (a_1 c_p - a_2
s_p) d\varphi + \frac{R_z}{l} (a_2 c_p - a_1
s_p) d\tau)^2  \\
&& + l^2 \cos^2 \q (d\psi + \frac{R_z}{l} (a_2 c_p - a_1
s_p) d\varphi + \frac{R_z}{l} (a_1 c_p - a_2
s_p) d\tau)^2  + \frac{\sqrt{Q_1}}{\sqrt{Q_5}} ds^2(X_4), \nn
\eea
with new coordinates $\varphi = z/l R_z$ and $\tau = t /l R_z$ with $l^2
= \sqrt{Q_1 Q_5}$ and
\be
\rho^2 = \frac{R_z^2}{l^2} (r^2 +(m - a_1^2 - a_2^2) s_p^2 + 2 a_1 a_2
s_p c_p).
\ee
The non-trivial six-dimensional metric is a twisted fibration of $S^3$
over the BTZ black hole, with
the mass and angular momentum parameters $(M_3,J_3)$ of the BTZ metric being
\bea
M_3 &=& \frac{R_z^2}{l^2} (  (m-a_1^2 - a_2^2) \cosh 2 \d_p + 2 a_1
a_2 \sinh 2 \d_p); \\
J_3 &=& \frac{R_z^2}{l^2} (  (m-a_1^2 - a_2^2) \sinh 2 \d_p + 2 a_1
a_2 \cosh 2 \d_p). \nn
\eea
Restricting to the extremal limit in which $M_3 = J_3$ gives
\bea
ds^2 &=& - l^2 ({\rho} - \frac{M_3}{2 \rho})^2 d\t^2 +
({\rho} - \frac{M_3}{2 \rho})^{-2}
l^2 d\rho^2 +
l^2 \rho^2 (d\varphi - \frac{M_3}{2 \rho^2} d\t)^2 \nn \\
&& + l^2 (d \q^2 + \sin^2 \q (d \phi + \frac{R_z}{l} \sqrt{Q_p} (L_1
- L_2) (d\varphi - d\tau ) )^2  \\
&& + l^2 \cos^2 \q (d\psi + \frac{R_z}{l} \sqrt{Q_p} (L_1 - L_2)
(d\tau - d \varphi))^2  + \frac{\sqrt{Q_1}}{\sqrt{Q_5}} ds^2(X_4), \nn
\eea
which is a twisted fibration of $S^3$ over the extremal BTZ black
hole.

The entropy of black hole solutions is given by
\be
S = \frac{2 \pi \rho_+}{4 G_3}
= 2 \pi \left ( \sqrt{ N N_p - \qu (J_{\phi} - J_{\psi})^2} + \sqrt{N
  \bar{N}_p - \qu (J_{\phi} + J_{\psi})^2} \right ). \label{entropy1}
\ee
where the effective three dimensional Newton constant is
\be
\frac{1}{4 G_3} = N = N_1 N_5
\ee
and $\rho_+$ is the location of the outer event horizon. The Hawking
temperature of the black hole is given by
\be \label{btz-temp}
T_{H} = \frac{(\r_{+}^2 - \rho_{-}^2)}{2 \pi \r_{+}},
\ee
where $\rho_{-}$ is the location of the inner event horizon of the black hole.

\bigskip

Given the asymptotically $AdS_3 \times S^3$ region, one can obtain the
corresponding data in the dual conformal field theory. In particular,
making use of the formulae for the holographic vevs previously derived, one
can extract the non-zero components of the stress energy tensor and R
symmetry currents. Recall that the vevs are
\bea
\langle T_{ij} \rangle &=& \frac{N}{2 \pi} (g_{(2) ij} + \qu {\cal
  A}^{+\a}_{(i} {\cal A}^{+\a}_{j)} + \qu {\cal
  A}^{-\a}_{(i} {\cal A}^{-\a}_{j)} ); \nn \\
\langle J^{i \pm \a} \rangle &=& \frac{N}{8 \pi} (g_{(0) ij} \pm
\ep_{ij}){\cal
  A}^{\pm\a j} ,
\eea
where $g_{(n) ij}$ and ${\cal A}^{\pm \a}_{ij}$ are terms in the
asymptotic expansion of the three dimensional metric and gauge fields
respectively. The relevant three dimensional metric is the BTZ metric
in the first line of (\ref{btzz}) which in Fefferman-Graham form is
\bea
ds^2 &=& \frac{l^2}{z^2} (dz^2 + dx^i dx^j (g_{(0) ij} + z^2 g_{(2)ij}
+ \cdots)); \\
&=& \frac{l^2}{z^2} (dz^2 + (-d\t^2 + d\varphi^2) + \half M_3 z^2 (d\t^2 +
d\varphi^2) + J_3 z^2 d\t d\varphi + \cdots), \nn
\eea
whilst the three dimensional gauge
fields are respectively
\bea
{A}^{+ 3} &=& {\cal A}^{+3} + \cdots = \frac{R_z}{l} e^{\d_p} (a_1
  -a_2) (d\varphi - d\t); \\
{A}^{- 3} &=& {\cal A}^{-3} + \cdots = \frac{R_z}{l} e^{-\d_p} (a_2
+ a_1) (d\varphi + d\t). \nn
\eea
Putting these values into the holographic formulae gives
\bea \label{h-vevs2}
\langle T \rangle &=& \frac{1}{2\pi} (N_p (dx^+)^2 + \bar{N}_p
(dx^-)^2); \\
\langle J^{+3} \rangle &=&  \frac{\sqrt{N N_p}}{2\pi} (L_1 - L_2) dx^+ \equiv
\frac{1}{2 \pi} j^{+} dx^+; \nn \\
\langle J^{-3} \rangle &=&  - \frac{\sqrt{N \bar{N}_p}}{2 \pi} (L_1 + L_2) dx^- \equiv
\frac{1}{2 \pi} j^{-} dx^-, \nn
\eea
where we define
\be
N_p  =  \frac{V R_z^2}{4 g^2 (\a')^4} m e^{2 \d_p}; \qquad
\bar{N}_p =  \frac{V R_z^2}{4 g^2 (\a')^4} m e^{-2 \d_p},
\ee
and introduce light cone coordinates $dx^{\pm} = l  (d\tau \pm d
\varphi)$, which have periodicity $2\pi$. Define $h = \int \langle
T_{++} \rangle$ and $\bar{h} = \int \langle T_{--} \rangle$, and similarly
$j^{+} = \int \langle J^{+3} \rangle dx^{+}$, $j^{-} = \int \langle
J^{-3} \rangle dx^-$. Then for the product of global $AdS_3$ with
$S^3$, $j^{+} = j^{-} = L_1 = L_2 = 0$ and
$h = \bar{h} = N_p = \bar{N}_p = - N/4$. For an extremal BTZ
black hole, such that $M_3 = J_3$ one obtains
$\bar{N}_p = 0$ and
\be
h = N_p; \qquad
j^{+} = \sqrt{N N_p}  (L_1 - L_2) \equiv J_{\phi}; \qquad
\bar{h} = j^- = 0,
\ee
where $J_{\phi}$ is the angular momentum with respect to
asymptotically flat infinity given in (\ref{charges}).

\bigskip

One can reduce the black string solutions over the $z$
circle and $X_4$ to produce a three charge rotating
black hole solution of five-dimensional supergravity, as was indeed
done in the original BMPV paper \cite{Breckenridge:1996is}.
For the supersymmetric case this leads to a five dimensional metric
of the form
\be \label{5d3metric}
ds^2 = \l^{-2/3} (dt + k) ^2 + \l^{1/3} (dr^2 + r^2 d \Omega_3^2)
\ee
where
\bea
\l &=& h_1 h_5 h_P = (1 + \frac{Q_1}{r^2})(1 + \frac{Q_5}{r^2}) ( 1 +
\frac{Q_p}{r^2}); \\
k &=& \frac{\sqrt{Q_1 Q_5 Q_p}}{r^2} (L_2 - L_1) (
\cos^2 \q d \psi - \sin^2 \q d \phi). \nn
\eea
Setting $k=0$ gives the 3-charge Strominger-Vafa black hole \cite{Strominger:1996sh},
with the metric being of the same form as the 2-charge metric
(\ref{5dmetric}). In this case there is no curvature singularity at $r=0$,
see (\ref{curv1}), (\ref{curv2}), but instead a regular horizon. The
same is true for the rotating BMPV 3-charge black hole. The near
horizon region of the five-dimensional solutions is $AdS_2 \times S^3$
which is less amenable to detailed holographic analysis\footnote{
Note though that in this case the $AdS_2$ factor originates from a
reduction of the extremal BTZ over the compact boundary direction,
so one would anticipate the dual CFT to be a chiral
CFT \cite{Boonstra:1998yu} and thus that there is
better control over the duality.}
than the
six-dimensional $AdS_3 \times S^3$ near horizon region.
For this reason, we will focus on fuzzball geometries
for the black string solutions.

\subsubsection{M theory solutions}

It is also useful to review briefly certain three charge solutions of M theory
compactified on a Calabi-Yau. For simplicity let us first give
the supersymmetric eleven-dimensional solution for a toroidal
compactification describing orthogonally intersecting M2-branes:
\bea
ds^2 &=& - \left ( \frac{1}{H_1 H_2 H_3} \right )^{2/3} (dt + k)^2 + (H_1 H_2
H_3)^{1/3} dx^m dx^m \\
&& + \left ( \frac{H_2 H_3}{H_1^2} \right )^{1/3} (dy_1^2 + dy_2^2) +
 \left ( \frac{H_1 H_3}{H_2^2} \right )^{1/3} (dy_3^2 + dy_4^2)
+ \left ( \frac{H_1 H_2}{H_3^2} \right )^{1/3} (dy_5^2 + dy_6^2), \nn
\eea
where $x^m$ are coordinates on $R^4$ and the four form is
\be
F = F^1 \wedge dy_1 \wedge dy_2 + F^2 \wedge dy_3 \wedge dy_4 +
F^3 \wedge dy_5 \wedge dy_6.
\ee
where the two forms can be written as
\be
F^{a} = \frac{1}{2} d (H_{a}^{-1} (dt + k)),
\ee
with $a = 1,2,3$. The solutions are defined by three harmonic
functions $(H_1,H_2,H_3)$ on $R^4$ along with the one form $k$ on $R^4$:
\bea
H_a &=& \left (1 + \frac{Q_a}{r^2} \right ); \label{data222} \\
k &=& \sqrt{Q_1 Q_2 Q_3} \frac{\w}{r^2} (\cos^2 \q d \psi - \sin^2 \q d\phi),
\nn
\eea
where the metric on $R^4$ is
\be
ds^2 = dr^2 + r^2 d \q^2 + \cos^2 \q d\psi^2 + \sin^2 \q d \phi^2
\ee
and $dk$ is self-dual in this metric, namely $dk - \ast_4 dk = 0$.

The charges $Q_a$ are related to the
integral M2-brane charges via $Q_a = l_p^2 V_a/(V_1 V_2 V_3)$ where
$(2\pi)^2 V_a$ is the volume of each $T^2$ and $l_p$ is the Planck
length, related to the eleven-dimensional Newton constant $G_{11}$  via
\be
\frac{1}{16 \pi G_{11}} = \frac{1}{(2 \pi)^8 l_p^9}.
\ee
This metric describes an extremal 3-charge
rotating black hole, with the decoupling region geometry being
$AdS_2 \times S^3 \times T^6$. The 3 charges are the $Q_a$ with the
angular momenta being
\be
J_{\psi} = - J_{\phi} = \frac{\pi}{4 G_5} \w \sqrt{Q_1 Q_2 Q_3} = \w
\sqrt{N_1 N_2 N_3},
\ee
and the entropy is given by
\be
S = 2 \pi \sqrt{N_1 N_2 N_3 (1 - \w^2)}.
\ee
Clearly this is the same formula as the extremal BPS limit of
(\ref{entropy1}). Indeed the type IIB on $T^4$ and M theory on $T^6$ solutions are
related by dualities; first reduce the latter on the M theory circle $y_6$
to obtain a $(D2_{y_1 y_2} \perp D2_{y_3 y_4} \perp F1_{y_5})$ solution of type IIA and
then T dualise on $(y_3,y_4,y_5)$ to obtain a $(D4_{y_1 y_2 y_3 y_4 y_5}
\perp D1_{y_5} \perp P_{y_5})$ solution of type IIB.  Here $Dp_{y_1
  \cdots y_p}$ denotes the spatial directions wrapped by the
$Dp$-brane.

Analogous three charge rotating five-dimensional black hole
solutions for Calabi-Yau compactifications are well-known; the entropy
of such wrapped M-brane configurations was first discussed in
\cite{Maldacena:1997de}. Moreover by
replacing $R^4$ by a Taub-NUT space and compactifying on the NUT circle one
can obtain an extremal four-charge solution in four
dimensions. However, for us the key feature of all of these solutions
is that they give a near horizon
geometry with an $AdS_2$ factor, rather than $AdS_3$.

Candidate fuzzball solutions have been constructed for
these black holes; indeed as we shall see almost all supersymmetric
fuzzball solutions
are constructed from a set of defining data, which can be used to
build fuzzballs for both type IIB and M theory black holes.
Detailed holographic analysis of type IIA and M theory systems
is however more subtle than in the IIB solutions which have $AdS_3$ decoupling
regions. One can easily build other three charge M theory solutions which
do have $AdS_3$ near horizon regions, for example, by intersecting
$(M2 \perp M5 \perp P)$ along a string. %, see \cite{Boonstra:1998yu}.
However the dual conformal
field theory is far less understood in such cases, so again it is more
natural to explore first the D1-D5-P system of type IIB.

\subsection{CFT microstates} \label{3cft}

To find the microstates of the (non-extremal) 3-charge D1-D5-P black
hole, we need to consider states in the D1-D5 CFT
with $h = N_{p}$ and $\bar{h} = \bar{N}_p$. The asymptotic number of
distinct states of this CFT can be obtained immediately by Cardy's formula \cite{Cardy}
\bea
d &=& \Omega \bar{\Omega}; \\
\Omega &=& \exp (2 \pi \sqrt{c N_p/6}); \qquad
\bar{\Omega} = \exp (2 \pi \sqrt{c \bar{N}_p/6}), \nn
\eea
where $c = 6 N_1 N_5$ is the central charge. Clearly this formula
exactly reproduces the Bekenstein-Hawking entropy of the non-rotating
black hole, which is somewhat surprising, as the states are far from
supersymmetric when $\bar{N}_p \gg 1$.

Cardy's formula gives the entropy of a 3-charge
black hole in a canonical ensemble, in which the R-charges are not
fixed. However, just as for the 2-charge black hole, the majority of
states being counted have zero R charge. Thus the difference
between the density of states with zero R charge $d_0$ and the total
density of states $d$ is subleading for large $N$. The density of
(supersymmetric) states with fixed and large $j^{+}$ and $h =
N_{p}$ was computed by \cite{Breckenridge:1996is}, in the case where
$X_4 = K3$. This gives
\be
S = 2 \pi \sqrt{N N_p - (j^{+})^2},
\ee
exactly reproducing the Bekenstein-Hawking entropy (\ref{entropy1}).

Many subsequent works have been devoted to refining the computation of
the black hole entropy; in particular, in recent times, substantial
effort has gone into reproducing subleading terms in the entropy on
both sides of the correspondence, see the review \cite{Sen:2007qy}. In the bulk this involves evaluating
higher derivative corrections on the leading order solution, whilst
in the CFT one needs subleading terms in the asymptotic expansion of
the degeneracy of states. In computing the latter, one can use an appropriate
supersymmetric index, the elliptic genus in the case of $K3$ and a
topological partition function in the case of $T^4$.

Whilst the index is clearly a useful tool in the counting of BPS states,
it is important to derive more properties of
the microstates being counted, in order to deduce the corresponding
properties of
fuzzball solutions. There is surprisingly little literature on the
relevant CFT states; for example, no correlation functions for
non-primary operators have been explicitly computed even in the
orbifold theory.

At the same time, one can already infer information about typical
black hole microstates from the long/short string picture introduced in
\cite{Mald}, which more formally corresponds to the decomposition of
the Hilbert space of the orbifold theory into twisted sectors. The
supersymmetric 3-charge microstates have left moving excitation level
$N_p$ and no right moving excitations. In the sector of twist $n_i$
the momentum is quantized in units of
$1/n_i$, although the total momentum is necessarily integral. This
immediately implies \cite{Mald} that the entropy is dominated by the highest twist
(twist $N$) sector, or equivalently by the longest strings, as the momentum
$N_p$ can be partitioned maximally in this sector. Thus to establish
properties of a typical microstate of a Strominger-Vafa black hole it
may suffice to explore states in this sector.

\subsection{Survey of fuzzball solutions}

The key step in constructing fuzzball solutions for the two charge
D1-D5 system was to dualize known supergravity solutions for a
fundamental string carrying an arbitrary profile wave. The three and
four charge systems cannot however be dualized to any analogous
system, and thus there is no systematic construction of fuzzball
solutions corresponding to black hole microstates.
Instead families and isolated examples of horizon-free, non-singular
solutions with the correct conserved charges have been constructed
using standard techniques for finding supergravity solutions. In
particular, the two principle construction techniques are:

\bigskip

\noindent {\bf 1. Supersymmetric classification techniques:}
In recent years there has been considerable progress in classifying
supersymmetric solutions of supergravity theories. In the context of
the fuzzball proposal, the classification of solutions of
minimal (ungauged) supergravity in five dimensions
\cite{Gauntlett:2002nw}, and its extension to minimal supergravity
coupled to Abelian vector multiplets, has proved particularly useful. 
As we will review below, every supersymmetric
solution is built from a set of defining data, a four-dimensional
hyper-K\"{a}hler space along with certain functions and forms on this
space. Specific choices of this defining data reproduce the previously known
supersymmetric rotating black hole and black ring solutions, but more
general choices of defining data lead to horizon-free non-singular
fuzzball solutions. Note that the classifications of six dimensional
supergravity solutions found in
\cite{Gutowski:2003rg,Cariglia:2004kk} have also proved useful.

The five-dimensional fuzzball or, as they are often called,
{\it bubbling}, solutions can be embedded into
eleven-dimensional supergravity compactified on a Calabi-Yau or torus,
in which case they should be related to microstates of M-brane black
holes. By taking the hyper-K\"{a}hler space to be asymptotically
locally flat (Taub-NUT), and reducing on the Taub-NUT circle one also
obtains candidate geometries for four-dimensional four-charge black
holes. Moreover, using a chain of dualities, the same defining data
generates a D1-D5-P solution of type IIB on $T^4$ or $K3$, and thus
gives candidate fuzzball geometries for the microstates of this system.

Below we will discuss in detail which black hole microstates are
likely to be captured by such solutions. Let us remark here, however,
that it is clear a priori that
solutions of minimal supergravity coupled to Abelian vector multiplets
will not be sufficient to
capture all fuzzball solutions for a given black hole. In the two
charge case we saw that a finite fraction of the black hole entropy
was associated with fuzzballs which also had internal excitations,
along the compact part of the geometry. Such fuzzballs with internal
excitations excite all type II supergravity fields, and when compactified along the
compact manifold, give rise to generic solutions of {\it extended} rather than
minimal supergravities in five or six dimensions.

In the three charge system we would also expect to find fuzzball
solutions with internal excitations, as solutions of extended
supergravity theories in five or six dimensions. However, there is to
date no complete classification of solutions of the relevant
extended supergravity theories, so finding fuzzballs with internal
excitations is an open problem. We should emphasize here that such
fuzzballs are likely to have qualitatively different properties to
those without internal excitations (as they did in the two charge
system) and thus one will need to know these properties before
successfully coarse-graining over geometries.

\bigskip

\noindent {\bf 2. Horizon-less limits of known black hole solutions:}
The second construction technique is to begin with known supergravity
solutions describing rotating charged black objects, and then take
careful limits of the parameters in the solutions to obtain
horizon-free non-singular solutions. Note that this technique is applicable to
non-supersymmetric solutions, representing microstates of non-extremal
black holes, and has been used to generate the few known fuzzball
solutions for non-extremal black holes.

We will discuss below which black hole microstates are likely to be
captured by such techniques, but let us already mention the main
limitation of the technique. The most general
non-extremal black hole and black ring
solutions in five dimensions are labeled by a discrete number of
parameters, and admit three Killing vectors (time, plus two rotational
symmetries). The fuzzball geometries obtained by careful limits of the
parameters are thus also highly symmetric, and are labeled by only
a few parameters in addition to their conserved charges. Thus one does
not obtain families of solutions, parameterized by arbitrary functions,
as in the two charge case; one obtains only a discrete (small) number
of fuzzball geometries. Moreover, the high degree of symmetry often
allows one to identify uniquely the dual microstate, and it is found
to be atypical.

\bigskip

There is by now a substantial literature on fuzzball solutions for the
three and four charge cases. The aim of this section will not be to
give a comprehensive discussion of every known solution, but instead
to give an overview of the known solutions, emphasizing the
connections between them. A comprehensive
review of the known three charge fuzzball geometries was
given in the review \cite{Bena:2007kg}.

Solutions for three charge black holes and black rings have been
discussed in
\cite{Mathur:2003hj,Bena:2004wt,Lunin:2004uu,Giusto:2004id,Giusto:2004ip,Bena:2004de,
Giusto:2004kj,Bena:2005ay,Giusto:2006zi,Ford:2006yb}.
In particular bubbling solutions for black holes and black rings, to be described below,
were developed in
\cite{Bena:2005va,Bena:2005zy,Berglund:2005vb,Bena:2006is,Bena:2006kb,Cheng:2006yq,Bena:2007ju,Bena:2007qc,
Gimon:2007mha,Bena:2008wt}.
Four charge solutions in four dimensions have been discussed in
\cite{Saxena:2005uk,Balasubramanian:2006gi}.
Non-supersymmetric solutions were found and their properties explored in
\cite{Jejjala:2005yu,Gimon:2007ps,Giusto:2007tt}.

\subsubsection{Supersymmetric solutions of M theory}

Let us begin with supersymmetric solutions of eleven-dimensional
supergravity compactified on $T^6$, which were obtained in \cite{Bena:2004de}
using the classification
of solutions of minimal ungauged supergravity coupled to Abelian
vector multiplets in five dimensions.
The eleven-dimensional solutions of interest have a metric of the form
\bea
ds^2 &=& - \left ( \frac{1}{Z_1 Z_2 Z_3} \right )^{2/3} (dt + k)^2 + (Z_1 Z_2
Z_3)^{1/3} h_{mn} dx^{m} dx^{n} \label{m-bub} \\
&& + \left ( \frac{Z_2 Z_3}{Z_1^2} \right )^{1/3} (dy_1^2 + dy_2^2) +
 \left ( \frac{Z_1 Z_3}{Z_2^2} \right )^{1/3} (dy_3^2 + dy_4^2)
+ \left ( \frac{Z_1 Z_2}{Z_3^2} \right )^{1/3} (dy_5^2 + dy_6^2), \nn
\eea
with the four form being
\be
F = F^1 \wedge dy_1 \wedge dy_2 + F^2 \wedge dy_3 \wedge dy_4 +
F^3 \wedge dy_5 \wedge dy_6.
\ee
Thus the solutions are defined by three functions $(Z_1,Z_2,Z_3)$, one
form $k$ and three two-forms $(F^1,F^2,F^3)$. The two forms can be
written as
\be
F^{a} = \theta^{a} - \frac{1}{2} d (Z_{a}^{-1} (dt + k)),
\ee
with $a = 1,2,3$. Then supersymmetric solutions are such that
$h_{mn}$ is hyper-K\"{a}hler, and the forms $\theta^{a}$ are
self-dual and closed on the hyper-K\"{a}hler manifold $H_4$, namely
\be
\theta^{a} = \ast_4 \theta^{a}; \hsp
d \theta^a = 0,
\ee
with the Hodge dual taken in the metric $h_{mn}$. As we will see later
in this section, smooth fuzzball 
solutions will require choosing pseudo hyper-K\"{a}hler base spaces,
in which the signature changes in certain regions of the space. 
The functions $Z_a$ and the one form $k$ then satisfy
\bea
\Box Z_a &=& 2 | \ep^{abc}| ( \theta_b \cdot \theta_c); \\
dk + \ast_4 dk &=& 2 \sum_{a} \theta^a Z_a, \nn
\eea
where $|\ep^{abc} |$ is the absolute value of the totally
antisymmetric tensor; $\Box$ is the Laplacian  on $H_4$ and for
2-forms on $H_4$, $(\alpha \cdot \beta) \equiv \frac{1}{2} \a^{mn} \b_{mn}$.
Note that setting $Z_1 = Z_2 = Z_3$ and reducing on the torus gives a
solution of minimal ungauged supergravity in five dimensions; the
effective five-dimensional solution is thus within the framework of that general
classification \cite{Gauntlett:2002nw}.

The Killing spinors of these solutions are given by
\be
\ep = (Z_1 Z_2 Z_3)^{-1/6} \ep_{0},
\ee
where the $32$-dimensional Majorana spinor satisfies the following
projection conditions
\be \label{proj}
\G^{0 56} \ep_{0} = \G^{0 78 } \ep_0 = \G^{0 9 (10)} \ep_0 = - \ep_0,
\ee
and furthermore the spinor $\ep_{0}$ is covariantly constant on the
hyper-K\"{a}hler space. Note that since $\G^{0 1234 56789 (10)} = 1$
the projection conditions also imply that $\G^{1234} \ep_0 =
\ep_0$. The proof is reviewed in appendix \ref{spinors}.

\bigskip

Reducing on the torus to five dimensions for general $(Z_1,Z_2,Z_3)$
gives solutions of ${\cal N} = 8$ ungauged supergravity in five
dimensions, which involve only three scalars, three gauge fields and the
metric. Thus the solutions involve only a subset of the
${\cal N} = 8$ fields; more general fuzzball solutions involving
excitations along the $T^6$ would need to involve additional ${\cal N}
= 8$ fields and have not yet been constructed in the three charge
case.

Replacing $|\ep_{abc}|$ by the intersection form $C_{abc}$ of a Calabi-Yau gives
a solution of ${\cal N} = 2$ ungauged supergravity, corresponding to
eleven-dimensional supergravity compactified on a Calabi-Yau. Further
reducing the five-dimensional solutions along a compact isometry in
the hyper-K\"{a}hler manifold results in a four-dimensional solution,
corresponding to reduction of eleven-dimensional supergravity solutions on
the product of a Calabi-Yau and a circle.

Solutions of this type, built from the defining data of a
4-dimensional hyper-K\"{a}hler space, along with harmonic forms and
functions on this space, account for almost all known bubbling
solutions for M theory and type IIA two, three and four charge black
holes. The main class of exceptions are the two charge
fuzzballs with internal excitations \cite{Kanitscheider:2007wq},
given in (\ref{equ_D1D5K3pot}), which switch on many additional fields
in the compactified theory. More recently in \cite{Bena:2008wt}
coordinate transformations have also been used
to generate from known bubbling solutions new smooth three charge
solutions which involve internal excitations.

In what follows it is useful to work directly
with the eleven-dimensional solution, rather than its dimensional
reduction to five or four dimensions. In
the next section we will see how these solutions are dualized to the type IIB
frame. Moreover, when analyzing regularity
conditions, it is regularity in the uplifted solution which is
relevant, not that in the dimensionally reduced solution. There are
many known examples where the dimensionally reduced solution is
singular, where the uplifted solution is not. Conditions for
the absence of closed timelike curves can differ, and where they do
again it is the conditions in the uplifted solution which are relevant.

\subsubsection{Bubbling solutions of type IIB}

The 11d solutions can be related to solutions of type IIB on $T^4$ by
a simple chain of dualities. First reduce on the $y_6$ circle to
obtain an F1-D2-D2 solution of type IIA, then T-dualize on
$(y_3,y_4,y_5)$ to obtain a D1-D5-P type IIB solution on $T^4$.

The solutions in the type IIB frame can then be written in the form:
\bea
ds^2 &=& - \frac{1}{ \sqrt{Z_1 Z_2} Z_{3}} (dt + k)^2 + \frac{Z_3}{
\sqrt{Z_1 Z_2}} (dz + {\cal A}_3)^2 + \sqrt{ Z_1 Z_2 } dx_4^2 + \sqrt{ \frac{Z_2}{Z_1}}
dz_4^2, \nn \\
e^{2 \Phi} &=& \frac{Z_2}{Z_{1}}; \qquad
F^{(3)} = (Z_1^{-2} Z_2 Z_3)^{-2/3} \ast_{5} F_1 + F_{2} \wedge (dz +
{\cal A}_3), \label{gen1}
\eea
where $dz_4^2$ is the metric on $T^4$ or $K3$
and the dual $\ast_5$ is defined in the five-dimensional metric
\be
ds_5^2 = - (Z_1 Z_2 Z_3)^{-2/3} (dt + k)^2 + (Z_1 Z_2 Z_3)^{1/3}
dx_4^2.
\ee
They are therefore defined in terms of three functions $Z_a$ with $a =
1,2,3$ and four one forms $(\w,{\cal A}_a)$ with ${\cal F}_a = d{\cal
  A}_a$, along with the metric on the four-dimensional
hyper-K\"{a}hler space. Introducing
three harmonic self-dual one forms $\eta_a$ on the base space
the connections ${\cal A}_a$ are
\be
{\cal A}_a = - Z_a^{-1} (dt + k) + \eta_a.
\ee
The form $k$ satisfies
\be
dk + \ast dk = \sum_a Z_a d \eta_a, \label{p2}
\ee
whilst the functions $Z_a$ satisfy
\be
\Box Z_a =  \half | \ep_{abc} | (d\eta_b)_{ij}
(d\eta_c)^{ij} , \label{p3}
\ee
where $\Box$ is the Laplacian on the hyper-K\"{a}hler space. Clearly
the defining data is the same as for the M theory and type IIA solutions.

\bigskip

The Killing spinors in this case can be written as
\be
\ep = (Z_1 Z_2)^{-1/8} (Z_3)^{-1/4} \ep_0, \label{spiniib}
\ee
where the spinor $\ep_0$ is covariantly constant on the
hyper-K\"{a}hler space and satisfies the projection conditions
\be
\ep_0 = \G^{01} \ep_0 = \G^{016789} \ep_0 = - i \ep_0^{\ast}.
\ee
These projection conditions along with the Majorana-Weyl condition
imply that the covariantly constant (complex) spinor $\ep_0 = \ep^1_0
+ i \ep^2_0$  is such that $\ep^1_0 = - \ep^2_0 \equiv \eta_0$ and
\be
\G^{01} \eta_0 = \G^{016789} \eta_{0} = \G^{2345} \eta_0 = \eta_0.
\ee
Again we include for convenience the derivation of these spinors in appendix
\ref{spinors}.

\subsubsection{Types of supersymmetric solutions}

We have seen that the supersymmetric solutions of both type IIB and M theory
are specified by a set of data, the metric $h_{mn}$ on the
hyper-K\"{a}hler space, along with three functions $Z_a$ and one-forms
$(k,\eta_a)$, satisfying (\ref{p2}) and (\ref{p3}). We will now review
various black hole and fuzzball solutions characterized by different choices of the forms and
functions. In view of what will follow we will focus on solutions in
the type IIB frame.

\bigskip

\noindent {\bf a. Explicit integration for Gibbons-Hawking base space}

\noindent The most general defining data can be given
in terms of three harmonic functions $h_a$ and three harmonic one
forms $\eta_a$ on the hyper-K\"{a}hler base space. An
additional one form $k_{-}$ on the hyper-K\"ahler space, whose field
strength is anti-self-dual, appears as an integration constant.

That is, the scalar functions $Z_a$ can be formally expressed as
\cite{Bena:2004de,Bena:2005ay}
\be
Z_a = h_a + \half | \ep_{abc} | \Box^{-1} \left ( (d\eta_b)_{ij}
(d\eta_c)^{ij} \right ),
\ee
whilst the three forms ${\cal A}_a$ are given by
\be
{\cal A}_a = Z_a^{-1} (dt + k) + \eta_a.
\ee
For the asymptotically flat limit of the metric to be manifest one can
use a (constant) gauge transformation on
${\cal A}_3$ and write it as
\be
{\cal A}_3 = (Z_{3}^{-1} - 1) (dt + k) + (k + \eta_3).
\ee
Since the form $k$ satisfies
\be \label{wq1}
dk + \ast d k = - \sum_a Z_a d \eta_a.
\ee
the solution implicitly includes an
integration constant. There is always
the freedom to add to $k$ any form $k_{-}$ which satisfies
\be \label{int-const}
d k_{-} + \ast_4 d k_{-} = 0.
\ee
Explicit integrated solutions are not known in general. In the case of
hyper-K\"ahler base spaces with a $U(1)$ isometry which is preserved,
the equations can however be integrated \cite{Gauntlett:2004wh}.
Writing the base space in the
Gibbons-Hawking form \cite{Gibbons:1979zt,Gibbons:1987sp}, the metric is
\be \label{gh2}
ds_4^2 = V^{-1} (d \psi + A)^2 + V dx^{i} dx^{i}
\ee
where $x^i$ with $i=2,3,4$ are coordinates on $R^3$ and the connection $A$ satisfies
$\ast_3 dA = dV$. Then the solution may be written in terms
of seven harmonic functions $(K_a,L_a,M)$ on $R^3$ as
\bea
Z_a &=& V^{-1} K_b K_c + L_a; \label{u1-sol} \\
\eta_a &=& V^{-1} K_a (d \psi + A) - \ast_3 d K_a, \nn \\
k &=& k_{\psi} (d \psi + A) + \hat{k}_i dx^i; \nn \\
k_{\psi} &=& V^{-2} K_a K_b K_c + \half V^{-1} \sum_a K_a L_a + M; \nn
\\
\ast_3 d \hat{k} &=& V d M - M dV + \half \sum_a (K_a d L_a - L_a d
K_a). \nn
\eea
Here $\ast_3$ denotes the Hodge dual on $R^3$. Note that the harmonic
functions $(K_a,M)$ define $(\eta_a,k_-)$ respectively. This explicit integration
is central to most of the fuzzball solutions which have been
constructed, with the strategy being to choose harmonic data on $R^3$
such that regularity conditions are satisfied.

Finding explicit solutions for the case in which the
hyper-K\"ahler base space does not have a $U(1)$ isometry is an open
problem. In the 2-charge system fuzzballs in which the
symmetry of the transverse $R^4$ is completely broken form a less
singular and more representative basis. One might expect the same to
be true in the 3-charge system, and thus one would like to solve
explicitly in the non-symmetric case.

\bigskip

\noindent {\bf b. Black holes and black rings}

\noindent Supersymmetric three charge rotating black hole and black ring
geometries can be realized as specific solutions within this
framework. In fact the defining data for the supersymmetric rotating
black holes was already given in (\ref{data222}): three harmonic
functions sourced at the origin of an $R^4$ base space, along with one anti-self
dual form (the integration constant $k_{-}$).

Let us now consider the supersymmetric black rings found and analyzed in
\cite{Elvang:2004rt,Gauntlett:2004wh,Bena:2004de, Elvang:2004ds,Gauntlett:2004qy}.
The defining functions can be written as
\bea
Z_a &=& 1 + \frac{Q_a}{\Sigma} - \half | \ep_{abc} | \frac{q_b q_c R^2 \cos 2 \q}{\Sigma^2}; \\
\Sigma &=& (r^2 + R^2 \cos^2 \q); \nn \\
{\cal A}_{a} &=& Z_a^{-1} (dt + k) + \frac{q_a R^2}{\Sigma} (\sin^2 \q d \varphi -
\cos^2 \q d \phi); \nn \\
k_{\phi} &=& - \frac{r^2 \cos^2 \q}{2 \Sigma^2} \left (\sum_a q_a Q_a - q_1 q_2
q_3 ( 1 + \frac{2 R^2 \cos^2 2 \q}{\Sigma}) \right ); \nn \\
k_{\varphi} &=& - \frac{\sum_a q_a R^2 \sin^2 \q}{\Sigma} + (1 + \frac{R^2}{r^2})
\tan^2 \q k_{\phi}, \nn
\eea
with $a = 1,2,3$ and the base space being $R^4$ with the metric
\be \label{met-r4}
dx_4^2 = \Sigma \left (\frac{dr^2}{(r^2 + R^2)} + d \q^2 \right ) + (r^2 + R^2) \sin^2
\q d\varphi^2 + r^2 \cos^2 \q d\phi^2.
\ee
The defining harmonic functions and harmonic forms in this case are
simply
\bea
h_a &=& \left (1 + \frac{Q_a}{\Sigma} \right ); \\
\eta_a &=& \frac{q_a R^2}{\Sigma}  (\sin^2 \q d \varphi - \cos^2 \q d \phi). \nn
\eea
In the form $k$ is included an integration constant (\ref{int-const})
\be
k_{-} = - \frac{\sum_a q_a R^2}{2 \Sigma}  (\sin^2 \q d \varphi + \cos^2 \q d \phi).
\ee
In contrast to the black hole solution, the harmonic functions in the
black ring are sourced on a circle in $R^4$, located at $r = 0$, $\q = \pi/2$. Since the solution
preserves $U(1)^2$ isometries of the hyper-K\"{a}hler base ($R^4$), it
can also be rewritten as a solution of type (a) on a Gibbons-Hawking
base space. The metric on $R^4$ given in (\ref{met-r4}) can be rewritten in
Gibbons-Hawking form via the coordinate transformations
\be
r \cos \theta = 2 \sqrt{\rho} \cos (\half \theta_3); \qquad
r^2 \sin^2 \theta + R^2 = 4 \rho; \qquad
\phi = \half (\psi + \phi_3); \qquad
\varphi = \half (\psi - \phi_3),
\ee
to give the Gibbons-Hawking metric
\be
dx_4^2 = V^{-1} (d \psi + \cos \q_3  d \phi_3)^2 + V dx^i dx_i,
\ee
where the $R^3$ metric is written as
\be
dx_i dx^i = d \rho^2 + \rho^2 (d \theta_3^2 + \sin^2 \theta_3 d \phi_3^2)
\ee
and the Gibbons-Hawking potential is $V = 1/\rho$. Thus the black ring solutions
written in this coordinate system are such that the harmonic functions
are sourced at $\rho = \qu R^2$ and $\q_3 = \pi$. The source circle thus wraps
the fibre, and is located at a point on $R^3$. Letting the Cartesian
coordinates be $x^1 = \rho \cos \theta_3$, $x^2 = \rho \sin \theta_3
\cos \phi_3$ and $x^3 = \rho \sin \theta_3 \sin \phi_3$, then the
source is located at $\vec{x}^i_0 \equiv (- \qu R^2,0,0)$. Noting that
\be
(r^2 + R^2 \cos^2 \q) = 4 (\rho^2 + \half R^2 \rho \cos \theta_3 +
\frac{R^4}{16})^{1/2} = 4 |x - x_0|,
\ee
the defining harmonic functions in $R^3$ are given by
\bea
h &=& \frac{1}{|x - x_0|}; \qquad
L_a = 1 + \qu (Q_a  - q_b q_c) h; \\
K_a &=& -\half q_a h; \qquad M = \qu \sum_a q_a (1 - \qu R^2 h). \nn
\eea
Multi-centered supersymmetric black rings, and black saturns, may
then be immediately constructed by choosing instead defining harmonic
functions in $R^3$ which are multi-centered
\cite{Gauntlett:2004wh,Bena:2004de,Gauntlett:2004qy}. Note however that each
point source in $R^3$ corresponds to a circle of sources in $R^4$,
except when the source is located at the origin of $R^3$, in which
case the radius of the wrapped fiber is zero, and the harmonic
function is sourced at the origin of $R^4$. The latter choice of
harmonic function gives as above a supersymmetric black hole.

Let us consider the solutions in the type IIB frame, where they are
understood in the context of the D1-D5-P system.
The single supersymmetric black ring is characterized by seven
parameters, $(Q_a,q_a,R)$ along with the moduli $(R_z,V)$, where
$R_z$ is the radius of the $z$ circle and $v = (2 \pi)^4 V$ is the volume of
$X_4$. The integral charges $N_a$ of the black ring are given by
\cite{Elvang:2004rt,Bena:2004de, Elvang:2004ds}
\be
Q_1 = \frac{g (\a')^3}{V} N_1; \qquad
Q_2 = g \a' N_5; \qquad
Q_3 = \frac{g^2 (\a')^4}{V R_z^2} N_p,
\ee
where $(N_1,N_5,N_p)$ are the D1-brane, D5-brane and momentum charges respectively.
The dimensionless (non-conserved) dipole charges $n_a$ are given by
\be
q_1 = \frac{g\a'}{R_z} n_1; \qquad
q_2 = \frac{g (\a')^3}{V R_z} n_2; \qquad
q_3 = R_z n_3,
\ee
and the angular momenta at asymptotically flat infinity are
\be
J_{\phi} = \half \sum_a n_a N_a - \half n_1 n_2 n_3; \qquad
J_{\varphi} = J_{\phi} + \frac{R_z V}{(\a')^4 g^2} (q_1 + q_2 + q_3) R^2.
\ee
The entropy of the black ring can be written as
\be
S = 2 \pi \sqrt{n_1 n_2 n_3 \d - \g^2}
\ee
where
\bea
\g &=& \half (n_3 N_3 - n_1 N_1 - n_2 N_2 + n_1 n_2 n_3); \\
\d &=& \frac{N_1 N_2}{n_3} - n_1 N_1 - n_2 N_2 + n_1 n_2 n_3 -
\frac{q_3 R^2}{C}, \nn \\
C &=& \frac{(\a')^4}{R_z V}. \nn
\eea
Given that the supersymmetric black ring has the same charges as the
three charge black hole, it should be interpreted as a specific
mixed state in the D1-D5 CFT, with momentum $N_p$, characterized by
the parameters $(n_a,R)$. Explicitly identifying this mixed state is
an open problem. Presumably candidate 3-charge fuzzball geometries
with the same angular momentum as the black ring should correspond to
microstates of both the 3-charge black hole and the 3-charge black
ring. However, identifying which geometries would be relevant for the black ring
is likely to be rather difficult, unless the corresponding CFT mixed
state can be identified.

In the context of the fuzzball proposal, one might wonder whether
there exist horizon-less non-singular limits of the black ring
solutions. These would have D1-D5-P charges and could correspond to
microstates of both the black hole and of the black ring.
The case $\d = 0 = \g$ is known to give a solution without a horizon,
but with an orbifold singularity \cite{Bena:2004tk}. Clearly other restrictions on the
parameters also give solutions with zero entropy, such as for example
$\g = \d = n_1 n_2 n_3$, and it is possible that specific restrictions
could give rise to regular horizon-free geometries. This issue has not
yet been systematically investigated.

\bigskip

\noindent {\bf c. Lunin-Mathur solutions}

\noindent It is useful to show explicit how
the two-charge Lunin-Mathur fuzzball solutions (\ref{lun-mat}) are contained within this
solution set. Let us first restrict the defining data to three harmonic functions
$h_a$ and a single one form $\eta^3  \equiv \eta$ on the
hyper-K\"{a}hler space, taken now to be $R^4$.
In these solutions the remaining three one-forms
${\cal A}_a$ are defined in terms of $(h_a,\eta)$ as
\bea
{\cal A}_1 &=& h_1^{-1} (dt  + k); \qquad {\cal A}_2 = h_{2}^{-1} (dt + k);
\label{res1} \\
{\cal A}_3 &=& (h_3^{-1} - 1) (dt + k) + (k - \eta) \equiv
(h_3^{-1} - 1) (dt + k) + b, \nn
\eea
Substituting these expressions into (\ref{gen1}) the three form can be
written as
\be \label{above}
F^{(3)} = d (h_{2}^{-1} (dt + k) \wedge (dz + b)) + \ast_4 dh_1.
\ee
This is the same form as in the Lunin-Mathur solutions,
as one would expect, since the gravitational wave does not couple to
the RR field strengths. However, the form $k$ satisfies the relation
\be \label{eq2}
d k + \ast_4 d k = h_3 d \eta,
\ee
and therefore $dk$ {\it cannot} be harmonic except when $h_3$ is
constant. On setting $\eta = 0$ and choosing the harmonic
functions to be of the standard (single-centered) form
\be
h_a = \left (1 + \frac{Q_a}{r^2} \right )
\ee
one clearly recovers the three charge static extremal black
string. Choosing $\eta$ to be non-zero, for the same choice of
harmonic functions, reproduces the three charge rotating extremal
BMPV black string.

On setting $h_3 =1$ one recovers the Lunin-Mathur geometries as
follows. Let $d \eta = d A + \ast_4 d A \equiv dA + dB$, where the
definition of $B$ is that $dB = \ast_4 dA$, so that $\eta = A + B$.
Then (\ref{eq2}) is solved by letting $k = A$ which implies that in (\ref{above})
$b = B$. Choosing the explicit forms of the harmonic
functions to be those given in (\ref{lm-func}) one indeed obtains the
solutions given in (\ref{lun-mat}).

It is important to note however that the two
charge fuzzball solutions with internal excitations cannot be obtained within
this framework, as they involve many additional fields in type IIB, and
hence in the compactified theory.

\bigskip

\noindent{\bf d. Bubbling geometries}

\noindent We now turn our attention to the main class of fuzzball
geometries which have been constructed, the so-called bubbling
solutions
\cite{Bena:2005va,Bena:2005zy,Berglund:2005vb,Bena:2006is,Bena:2006kb,Cheng:2006yq,Bena:2007ju,Bena:2007qc,
Gimon:2007mha,Bena:2008wt,Balasubramanian:2006gi}.
Given the detailed discussion of these backgrounds in other works,
for example the review \cite{Bena:2007kg}, we will simply
highlight their main features here.

The basic idea of the bubbling geometries is to use a non-trivial
``ambipolar'' $U(1)$ invariant
hyper-K\"{a}hler base space, whose signature changes in
different regions. That is, the potential for the Gibbons-Hawking
space is
\be
V = \sum_{i} \frac{q_i}{|x - x_i|}, \qquad \sum_i {q_i} = 1.
\ee
The constraint on the total charge is necessary for the space to be
asymptotically flat. In regions where $V$ changes sign the signature
of the base space changes from $(+,+,+,+)$ to $(-,-,-,-)$. Denote the
2-dimensional surfaces where $V$ changes sign by $\Sigma_{\a}$; then
$V(\Sigma_{\a}) = 0$. For the
complete metric signature to remain unchanged this means that
\be \label{po2}
Z_a V \ge 0
\ee
everywhere, which in turn requires that the $Z_a$ change sign at
$\Sigma_{\a}$ also. Taking the remaining seven harmonic functions
$(L_a,K_a,M)$ defined in (\ref{u1-sol})
to be sourced at the same locations $x_i$, so that
\be
L_a = 1 + \sum_{i} \frac{l_a^i}{|x - x_i|}; \qquad
K_a = \sum_{i} \frac{k_a^i}{|x - x_i|}; \qquad
M =  m + \sum_{i} \frac{m^i}{|x - x_i|},
\ee
one then chooses the parameters $(l_a^i,k_a^i,m^i)$ so that
$(Z_a,\eta_a,k)$ are finite at the locations of the harmonic function
sources, $x_i$. In particular this implies that
\be \label{po3}
l_a^i = - \frac{k_b^i k_c^i}{q_i}; \qquad m^i = \frac{k_1^i k_2^i
    k_3^i}{2 (q_i)^2}; \qquad m = - \half \sum_{i,a} k_a^i.
\ee
These values are chosen so as to cancel the poles in the
defining functions.

Note that the functions $Z_a$ for large $|x|$ are expanded as
\be
Z_a = 1 + \frac{1}{|x|} \left (\sum_i l_a^i + \sum_{i,j} k_b^i k_c^i
\right ) +
\cdots \equiv 1 - \frac{1}{|x|} \sum_{i} \frac{\td{k}_b^i
  \td{k}_c^i}{q_i} + \cdots,
\ee
where
\be
\td{k}_a^i = k_a^i - q_i \sum_{j} k_a^j; \qquad \sum_{i} \td{k}_a^i =
0.
\ee
One can show that the supergravity solutions are invariant under
\be
K_a \rightarrow K_a + c_a V,
\ee
for any constant $c_a$, see \cite{Bena:2005va,Berglund:2005vb}, and thus physical quantities such as mass
which are expressed in terms of $Z_a$ must depend on the invariant quantities
$\td{k}_a^i$ rather than $k_a^i$.

For the supergravity solutions to be regular requires:
\begin{enumerate}
{\item{Absence of singularities, in particular (a) at the locations of the sources
    in the harmonic functions and (b) where the base space signature
    changes.}}
{\item{Absence of closed timelike curves and Dirac-Misner strings.}}
\end{enumerate}
Conditions (\ref{po2}) and (\ref{po3}) are
sufficient to ensure that the solution is non-singular, see the
detailed analysis of \cite{Berglund:2005vb}. Note that although the
defining functions in the solution are built
from harmonic functions which have sources at the Gibbons-Hawking
centers there are no delta function sources in the defining
functions \cite{Bena:2005va,Berglund:2005vb}. The solution is therefore regular at
the locations of the sources. One can also show that the solution is
regular where the base space signature changes, namely where $V
\rightarrow 0$. Consider the five-dimensional part of the metric
\be
ds_5^2 = - (Z_1 Z_2 Z_3)^{-2/3} (dt + k)^2 + (Z_1 Z_2 Z_3)^{1/3} \left
( V^{-1} (d \psi  + A)^2 + V dx^i dx^i \right )
\ee
in a neighborhood of a hypersurface
$\Sigma_a$ where $V(\Sigma_a) = 0$. Using the
explicit forms of the functions one can show that in the neighborhood of a
point on $\Sigma_a$
\be \label{sigma-beh}
ds^2_5 \approx - 2 dt d \varphi + dx^i dx^i.
\ee
Here $d{\varphi} = d \psi + A_{\Sigma_a}$ with
$A_{\Sigma_a}$ the Gibbons-Hawking gauge field and
$(t,x^i)$ have been appropriately rescaled.
The metric is regular at $\Sigma_a$, with the
Killing vector $\pa_t$ becoming null on this surface. A detailed
analysis of the regularity of the solutions on these hypersurfaces can
be found in \cite{Berglund:2005vb}.

Removing closed timelike curves restricts the parameters further; for
example, in the eleven-dimensional geometries one has to ensure that
\be
Z_1 Z_2 Z_3 V^2 - k_{\psi}^2 V \ge 0
\ee
globally. Recall that the $k_{\psi}$ was given in (\ref{u1-sol}).
Solving the constraints this equation
makes on the parameters $(k_a^i,q_i,\vec{x}_i)$ in general is rather difficult.
However, one can derive constraints which remove Dirac-Misner
strings from $\hat{k}$, defined in (\ref{u1-sol}); these so-called {\it bubble equations} are
\bea
\sum_{j \neq i} \Pi_{ij}^1 \Pi_{ij}^2 \Pi_{ij}^3 \frac{q_i q_j}{r_{ij}}
&=& - 2 (m q_i + \half \sum_{a} k_a^i); \\
\Pi_{ij}^a &=& \left ( \frac{k_a^j}{q_j} - \frac{k_a^i}{q_i} \right );
\qquad r_{ij} = |x_i - x_j|. \nn
\eea
In specific examples one finds that satisfying these equations is sufficient to
guarantee the global absence of CTCs, but this is not always the case
and identifying
systematically the conditions required to remove closed timelike
curves remains an open problem.

The three conserved charges with respect to asymptotically flat infinity can
be expressed in terms of the defining data
$(\td{k}_a^i,q_i,\vec{x}_i)$ as
\be
Q_a = - 2 {\cal N} | \ep_{abc} | \sum_{i=1}^N \frac{\td{k}^b_i
  \td{k}^c_i}{q_i},
\ee
where the prefactor ${\cal N}_a$ depends on whether one is considering
the type IIB or M theory case. In the former case, the $Q_a$ are the conserved
D1, D5 and momentum charges, and the normalization is
\be
{\cal N}_{IIB} = \frac{R_z V}{(\a')^4 g^2}.
\ee
In the M theory case, the $Q_a$ are the conserved membrane charges,
and the appropriate normalization is
\be
{\cal N}_{M} = \frac{\pi}{4 G_5} = \frac{\pi V_{6}}{4 G_{11}},
\ee
with $G_{d}$ the $d$-dimensional Newton constant and $V_6$ the volume
of the six torus (or of the Calabi-Yau). The solutions also have conserved
angular momenta with respect to asymptotically flat infinity. The
generic solution breaks the $SO(4)$ rotational invariance of the
transverse $R^4$ at infinity to $U(1)$, and therefore there are
non-zero angular momenta $(J_{\psi},\vec{J})$, where $\vec{J}$ defines
a three dimensional vector in the $R^3$ of the Gibbons-Hawking metric
(\ref{gh2}). Then one finds that
\bea
J_{\psi} &=& \frac{4}{3} {\cal N} | \ep_{abc} | \sum_{i=1}^N
\frac{\td{k}^a_i \td{k}^b_i \td{k}^c_i}{q_i^2}; \nn \\
\vec{J} &=& 8 {\cal N} \vec{D} , \qquad
\vec{D} \equiv \sum_{i=1}^{N} \vec{D}_i, \qquad
\vec{D}_i \equiv \sum_{a} \td{k}^a_i \vec{x}^i. \nn
\eea
$J_{\psi}$ and $J_{\phi_3}$ correspond to the left and right moving components
$J_{+}$ and $J_{-}$ of the CFT R symmetry currents,
respectively. More generally $\vec{J}$ should correspond to the right
moving $SU(2)$ CFT R current $J^a_{-}$.
Thus all the bubbling solutions should correspond to
R charge eigenstates in the left moving (excited) sector, but need not
be R charge eigenstates in the right moving (ground state) sector. Note
however that the angular momenta given above are with respect to
asymptotically flat infinity and do not automatically coincide precisely
with the vevs of the R symmetry currents, extracted from the decoupling
region in the geometry; one needs to extract the vevs of the R
symmetry currents using the holographic formulae.

The main goal of the literature has been to find data sets
$(\td{k}_a^i,q_i,\vec{x}_i)$ such that the supergravity solution is
regular and has the same charges as a black hole or black ring with macroscopic
horizon area. For example, recall that a supersymmetric
BMPV black hole with integral D1-D5-P
charges $(N_1,N_5,N_p)$ and angular momentum $J_{+} = j_{+}$ has entropy
\be
S = 2 \pi \sqrt{N_1 N_5 N_p - j_{+}^2},
\ee
so for a black hole with macroscopic horizon area one needs $j_{+}^2 \ll
N_1 N_5 N_p$. The regular solutions first constructed had $j_{+}^2 \sim
N_1 N_5 N_p$, and thus could not correspond to microstates of a
macroscopic BMPV black hole. Later, however, it was observed that
$j_{+}^2 = k N_1 N_5 N_p$ (with the fraction $k$ being smaller than one)
could be achieved in scaling solutions \cite{Bena:2006kb,Bena:2007ju,Bena:2007qc}, namely
solutions in which all the centers are at
\be
\vec{x}_i = \mu \vec{y}_i,
\ee
with $\mu \rightarrow 0$ and $\vec{y_i}$ finite. These so-called deep
microstate solutions were explored numerically and analytically in
\cite{Bena:2006kb,Bena:2007ju,Bena:2007qc}.

Recalling the
behavior in the 2-charge system it is perhaps unsurprising that the bubbling
points need to be clustered at the origin. In the 2-charge geometries
the radius of the curve in $R^4$ determining the solution was related
to the angular momenta, with solutions of small angular momenta
deriving from curves of small radii, see the discussion around
(\ref{j-char2}). Most of the Ramond ground states
have small R charge, and thus the typical fuzzball should be
characterized by a curve of small radius. Here we see analogous
behavior in the 3-charge system: the defining harmonic functions
are sourced on curves in the four-dimensional base space, which must
have small radii for the fuzzball to have typical R charges.

It also seems natural that one needs to cluster the centers. In
the 2-charge system it seems unlikely that solutions characterized by
disconnected curves describe bound states of D1 branes and D5
branes. They can have the same charges and angular momenta as typical
black hole microstates, but are most likely related to Coulomb branch
physics. Here in the 3-charge system one sees that a scaling solution
in which the centers cluster is needed to even obtain a solution with
typical charges.

Note that candidate fuzzball geometries for black
rings can be found by taking a single center at the origin with $q_1 =
1$ and clustering the remaining points at some distance $\r$ from the
origin. The scale $\r$ determines the radius of the corresponding
black ring. In section \ref{corresp} we will consider the
correspondence between such geometries and D1-D5-P microstates.

\subsubsection{Limits of black hole solutions and spectral flow}

The second technique used to find candidate fuzzball geometries
involves starting with known non-extremal charged rotating black hole
solutions, and then taking careful limits of the parameters to obtain
solutions with no horizons or singularities. More generally one could
start with any rotating solution of Einstein equations, apply
boosts and dualities to obtain solutions carrying the required charges
and then restrict the parameters to obtain horizon-free non-singular
solutions. Thus as previously mentioned one could also look for fuzzball limits
of black ring solutions. A systematic exploration of the possibilities
has not yet been carried out.

The principal advantage
of this technique is that it does not rely on supersymmetry, and thus
allows one to find fuzzball solutions for non-extremal black
holes. The main drawback, however, is that the known black hole and
black ring solutions are highly symmetric and characterized by only a
small number of
parameters, which are further restricted by demanding regularity and
absence of horizons. Thus the resulting fuzzball solutions typically
have only a few parameters in
addition to the required conserved charges, and correspond to rather
atypical black hole microstates.

Another related solution generating technique is ``spectral flow'': one
begins with a given near horizon geometry and makes a coordinate
transformation, which preserves the $AdS$ asymptotics.
In particular, one can generate certain 3-charge D1-D5-P geometries
from 2-charge D1-D5 geometries in this way
\cite{Lunin:2004uu,Giusto:2004id,Giusto:2004ip,Giusto:2004kj}.
Gluing back the asymptotically flat region then gives a fuzzball
geometry of the 3-charge D1-D5-P black hole. One should note that these
solution generating
techniques are often called spectral flow transformations, although it
is unclear whether all such named transformations indeed correspond to
spectral flow in the CFT.

Before the development of the bubbling solutions using the
classification of supergravity solutions, most candidate fuzzball
solutions for the 3-charge and 4-charge systems were found by
techniques of this kind. Supersymmetric fuzzballs for the D1-D5-P system were found in
\cite{Lunin:2004uu,Giusto:2004id,Giusto:2004ip,Giusto:2004kj}, whilst
supersymmetric fuzzballs for the D1-D5-KK system were found in
\cite{Bena:2005ay} and for the 3-charge supersymmetric black ring in
\cite{Giusto:2006zi}. In \cite{Ford:2006yb} time dependent solutions
carrying 3 charges were found and in \cite{Giusto:2007tt}
non-supersymmetric fuzzball solutions of the D1-D5-KK system were
found and analyzed.

To illustrate these techniques, let us focus on the example of
the non-extremal D1-D5-P solutions found in \cite{Jejjala:2005yu} and
analysed further in \cite{Gimon:2007ps}. General non-extremal 3-charge black
hole solutions with rotation were given in equation (\ref{cvetic-youm}).
Smooth geometries with no horizons can be obtained by demanding that
the singularity where $g_{rr}^{-1} = 0$ be nothing but a
coordinate singularity, analogous to that of polar coordinates at the
origin of $R^2$. There are four conditions on the parameters needed to
ensure regularity (assuming the momentum charge is non-zero):
\bea \label{con-1}
m &=& a_1^2 + a_2^2 - a_1 a_2 \frac{c_1^2 c_5^2 c_p^2 + s_1^2 s_5^2
  s_p^2}{s_1 c_1 s_5 c_5 s_p c_p}; \\
\frac{j + j^{-1}}{s + s^{-1}} &=& (l - n), \qquad
\frac{j - j^{-1}}{s - s^{-1}} = (l + n), \qquad (l,n \in Z);  \nn \\
R_z &=& \frac{m s_1 c_1 s_5 c_5 \sqrt{s_1 c_1 s_5 c_5 s_p
  c_p}}{\sqrt{a_1 a_2} (c_1^2 c_5^2 c_p^2 - s_1^2 s_5^2 s_p^2)},
\nn
\eea
where $R_z$ is the radius of the $z$ direction and
the parameters $(j,s)$ are given by
\be
j \equiv (\frac{a_2}{a_1})^{1/2}, \qquad
s \equiv \left (\frac{s_1 s_5 s_p}{c_1 c_5 c_p} \right )^{1/2} \le 1.
\ee
The conserved charges and angular momenta are given by
\bea
%M &=& \frac{\pi m}{4 G_5} (s_1^2 + s_5^2 + s_p^2 + \frac{3}{2}) \\
Q_1 &=& m s_1 c_1 = \frac{g \a^{'3} N_1}{V}; \qquad
Q_5 = m s_5 c_5 = g \a' N_5; \\
Q_p &=& m s_p c_p = \frac{g^2 (\a')^4}{V R_z^2} N_p, \qquad N_p = N_1
N_5 l n; \nn \\
J_{\psi} &=&  - \frac{\pi m}{4 G_5} (a_1 c_1 c_5 c_p - a_2 s_1 s_5
s_p) =  - N_1 N_5 l \nn \\
J_{\phi} &=&  - \frac{\pi m}{4 G_5} (a_2 c_1 c_5 c_p - a_1 s_1 s_5
s_p) = N_1 N_5 n, \nn
\eea
where $(N_1,N_5,N_p)$ are the integral charges and
\be
\frac{\pi}{4 G_5} = \frac{R_z V}{g^2 (\a')^4}.
\ee
The first expressions for $(Q_p,J_{\psi},J_{\phi})$ hold generally,
with the restrictions to the solutions of interest being given in
terms of the integers $(l,n)$. The mass is given by
\be
M = \frac{\pi m}{4 G_5} (s_1^2 + s_5^2 + s_p^2 + \frac{3}{2}).
\ee
Restricting the parameters to ensure regularity, and focussing on near
extremal configurations, this can be rewritten as
\be
M =  \frac{\pi}{4 G_5} (Q_1 + Q_5 + Q_p) + \frac{1}{2 R_z} N_1 N_5 (l^2 + n^2
- 1) \equiv M_{BPS} + \Delta M.
\ee
Thus one can see that if one fixes the three charges $(Q_1,Q_5,Q_p)$,
along with the moduli, then the solution is characterized by only one
other (integral) parameter. This parameter controls both the
non-extremality and the angular momenta, so the solution gives
precisely one microstate of a non-extremal D1-D5-P system with fixed
non-extremality.

It is useful to rewrite the six-dimensional metric as a fibration over
a four-dimensional base space, in order to facilitate comparison with
the general forms of supersymmetric solutions. Of course in the
non-supersymmetric case the base space does not have any special
character, but the hyper-K\"{a}hler structure is recovered in the
supersymmetric limit. The six-dimensional part of the metric can thus
be written as
\bea
ds^2 &=& \frac{1}{\sqrt{H_1 H_5}} \left ( - (f - m) (d \td{t} - (f -
m)^{-1} m c_1 c_5 (a_1 \cos^2 \q d \psi + a_2 \sin^2 \q d \phi))^2
\right . \nn \\
&& + \left . f (d \td{z} + f^{-1} m s_1 s_5 (a_2 \cos^2 \q d \psi + a_1 \sin^2 \q
d \phi))^2 \right )  \\
&& + \sqrt{H_1 H_5} \left ( \frac{r^2 dr^2}{ (r^2 + a_1^2)(r^2 + a_2^2) -
  m r^2} + d \q^2  \right . \nn \\
&& + (f (f-m))^{-1} [f (f-m) + f a_2^2 \sin^2 \q - (f-m) a_1^2 \sin^2 \q
d\phi^2 \nn \\
&& + 2 m a_1 a_2 \sin^2 \q \cos^2 \q d \psi d\phi \nn \\
&& + \left . (f (f-m) + f a_1^2 \cos^2 \q - (f-m) a_2^2 \cos^2 \q)
d\psi^2 ] \right ), \nn
\eea
where $\td{t} = t c_p - z s_p$ and $\td{z} = z c_p - t s_p$. In the
supersymmetric limit this six-dimensional metric can indeed be rewritten in
terms of an ambipolar hyper-K\"{a}hler base and harmonic functions on
this base \cite{Giusto:2004kj}.

The decoupling region of this geometry is obtained by
replacing $\td{H}_1 = Q_1$ and $\td{H}_5 = Q_5$ and also approximating
$m s_1 s_5 = m c_1 c_5 = \sqrt{Q_1 Q_5}$. This gives
\bea
ds^2 &=& \frac{1}{\sqrt{Q_1 Q_5}} \left ( - (f - m) (d \td{t} - (f -
m)^{-1} \sqrt{Q_1 Q_5} (a_1 \cos^2 \q d \psi + a_2 \sin^2 \q d \phi))^2
\right . \nn \\
&& + \left . f (d \td{z} + f^{-1} \sqrt{Q_1 Q_5} (a_2 \cos^2 \q d \psi + a_1 \sin^2 \q
d \phi))^2 \right )  \\
&&+ \sqrt{Q_1 Q_5} \left ( \frac{r^2 dr^2}{ (r^2 + a_1^2)(r^2 + a_2^2) -
  m r^2} + d \q^2  \right . \nn \\
&& + (f (f-m))^{-1} [f (f-m) + f a_2^2 \sin^2 \q - (f-m) a_1^2 \sin^2 \q
d\phi^2 \nn \\
&& + 2 m a_1 a_2 \sin^2 \q \cos^2 \q d \psi d\phi \nn \\
&& +  \left . (f (f-m) + f a_1^2 \cos^2 \q - (f-m) a_2^2 \cos^2 \q)
d\psi^2 ] \right ), \nn
\eea
This metric in turn can be rewritten as the twisted fibration of $S^3$
over BTZ as in (\ref{btzz}).
The regularity conditions of (\ref{con-1}) then translate to $J_3 =
0$ and $M_3 = -1$, in which case the BTZ part of the metric becomes
global $AdS_3$. Imposing both the regularity conditions and setting $m
\rightarrow 0$ restricts the fibration further, as we shall see below.

\bigskip

If one focuses on this decoupled region of the geometry it is
easy to understand why the non-extremality is pinned to the angular
momentum, and moreover what the corresponding CFT state is.
Consider the geometry
\bea
ds^2 &=& \sqrt{Q_1 Q_5} \left ( - (r^2 + \g_1^2 \mu^2) dt^2 + r^2 dy^2 +
\frac{dr^2}{r^2 + \g_1^2 \mu^2} \right ) \\
&& + \sqrt{Q_1 Q_5} \left (d\q^2 + \cos^2 \q (d\psi + j_{\psi})^2 +
\sin^2 \q  (d\phi + j_{\phi})^2 \right ) + (Q_1/Q_5)^{1/2} dz_4^2
\nn
\\
F &=& \sqrt{Q_1 Q_5} \left (dt \wedge dy \wedge dr + \cos \q \sin \q d \q
\wedge (d\phi + j_{\phi}) \wedge (d\psi + j_{\psi}) \right ), \nn
\eea
with constant dilaton $e^{-2 \Phi} = Q_5/Q_1$ and $y \sim y + 2 \pi
\mu^{-1}$, $\mu = \sqrt{Q_1 Q_5}/R$. Let
\be
j_{\phi} = j_{+} dx^{+} + j_- dx^{-}; \qquad
j_{\psi} = j_{+} dx^{+} - j_{-} dx^{-},
\ee
where $dx^{\pm} = \mu (dt \pm dy)$.
The geometry is regular if $\g_1^2 =1$, and has permissible conical
singularities if $\g_1 = 1/n$ with $n \in Z$. One can immediately
extract the holographic stress energy tensor and R symmetry currents
using the formulae for the holographic vevs given in (\ref{h-vevs2})
\be
\< T \> = \frac{N}{2\pi} \left (
( (j_{+})^2 - \g_1^2) (dx^{+})^2 + ( (j_{-})^2 - \g_1^2) (dx^{-})^2 \right
); \qquad
\< J^{\pm 3} \> = \frac{N}{2\pi} j_{\pm} dx^{\pm},
\ee
with $N = N_1 N_5$. The vevs of scalar operators vanish.
For the case of interest, one finds that $\g_1 = 1$ and
\be
j_{\pm} = \half (l \pm n),
\ee
and thus
\be
h = N (\qu (l+n)^2 ); \qquad
\bar{h} = N (\qu (l-n)^2); \qquad
j_+ = \half N (l+n); \qquad
{j}_- = \half N (l-n).
\ee
Note that the momentum charge at asymptotically flat infinity
is proportional to $(h - \bar{h})$
whilst the mass depends on $(h + \bar{h})$. The CFT state with
these charges corresponds to the spectral flow of the NS-NS
vacuum. One can see this by recalling that
under a spectral flow in the left sector by $\a_L = (l+n)$
units and in the right sector by $\a_R = (l-n)$ units,
\be
h \rightarrow h - \a_L j_+ + \frac{c}{24} \a_L^2; \qquad
j_+ \rightarrow j_+ + \frac{c}{12} \a_L,
\ee
with corresponding expressions for $(\bar{h},{j}_-)$, where $c = 6 N$.
Letting $\a_R = 1$ and $\a_L = 2 p + 1$ gives a BPS state in the RR
sector which is in the right moving vacuum and in a left moving
excited state. When $(\a_R,\a_L)$ are odd one obtains a
non-supersymmetric state in the RR sector which is a microstate of
the non-extremal black hole.

Since these states are related by spectral flow to the identity, it is
rather simple to compute properties of these states, for example,
decay rates of the non-supersymmetric states. (We will defer discussion
of this issue to section \ref{open}). At the same time, they are atypical black hole
microstates, and thus their properties cannot be assumed to be
representative. To be more precise, consider the microstates of the
BPS D1-D5-P (Strominger-Vafa) static black hole. Counting states with fixed
$(N_1,N_5,N_p)$ and either states with
$(j_+ = {j}_- = 0)$ or {\it all} states gives the same leading
contribution to the entropy, namely the famous result
\be
S = 2 \pi \sqrt{N_1 N_5 N_p}.
\ee
The point is that most states have $(j_+ = {j}_- = 0)$, so the
difference between the total degeneracy of states $d_t$ with fixed $(N_1,N_5,N_p)$
and arbitrary R-charges, and the degeneracy of states $d_0$ with
fixed $(N_1,N_5,N_p)$ and $(j_+ = {j}_- = 0)$ is subleading.
Fixing $j_+$ and ${j}_-$ corresponds to working in the canonical
ensemble, whilst allowing for all $(j_+,{j}_-)$ corresponds to working
in the grand canonical ensemble. Clearly the BPS state discussed
above, which has $j_+ \gg 1$, does not contribute to the canonical
ensemble for the Strominger-Vafa black hole, and it is a highly
non-representative state in the grand canonical ensemble, which is
peaked around states with zero R-charge. The BPS state above could be
interpreted as a microstate of a rotating BMPV black hole with this value of
$j_+$. However, recalling the BMPV entropy formula, and noting that
$j_{+}^2 = N_1 N_5 N_p$  one sees there are not enough microstates with this value of $j_+$
to give a black hole with macroscopic horizon area.

More generally, even if the state did not have atypical R-charges,
since it is obtained by spectral flow of the identity only
the stress energy tensor and R currents can acquire expectation values. By
contrast, a typical microstate will be characterized by the vevs of
operators within that state, and these vevs will determine the scale
at which the geometry starts to differ from the naive black hole
geometry.

Before moving to the correspondence between geometries and microstates
we should comment that the known non-extremal fuzzball solutions
have superradiant instabilities; see for example
\cite{Cardoso:2004nk,Cardoso:2005gj,Gimon:2007ps,Cardoso:2007ws}. This
is another indication that these solutions are not representative. As
we will discuss in section 6, one expects non-extremal fuzzballs to be
unstable, with their decay rate matching the decay rate of the
corresponding non-BPS microstates in the CFT. The radiation emitted by
a typical fuzzball should be very similar to that of the non-extremal
black hole. However, the known non-extremal fuzzball solutions decay
much faster; there is no contradication, as the decay rate matches
that of the corresponding CFT states, see
\cite{Chowdhury:2007jx,Dias:2007nj}, but these atypical fuzzballs are not
  representative.

\subsection{Correspondence between geometries and microstates} \label{corresp}

One of the most important outstanding issues is to derive the
correspondence between the candidate fuzzball geometries and the black
hole microstates. One would first like to match sufficient data to be sure
that the geometries do indeed correspond to black hole microstates,
rather than other states with the same charges. Then one would like to
explore what fraction of the black hole microstates are captured by
the known fuzzball geometries. No systematic analysis of the
correspondence has yet been carried
out, and a number of issues in matching geometries to microstates are rather
puzzling.

The bubbling geometries
\cite{Bena:2005va,Berglund:2005vb,Bena:2006is,Bena:2006kb,Cheng:2006yq,Bena:2007ju,Bena:2007qc,
Gimon:2007mha,Bena:2008wt} are viewed as the most promising candidates
for duals to typical black hole microstates. First of all, they have
the correct charges to correspond to three charge black holes and
in scaling solutions one
can obtain angular momenta comparable to that of a macroscopic BMPV black hole. By
replacing the asymptotically flat base space by an asymptotically
locally flat (Taub-NUT) space such that $V \rightarrow 1$ as $r
\rightarrow \infty$ one can also obtain candidate geometries for four charge
black holes in four dimensions, as  were discussed in \cite{Balasubramanian:2006gi},
again with the same charges as typical black hole microstates.

Secondly, the scaling solutions
\cite{Bena:2006kb,Bena:2007ju,Bena:2007qc} have throat regions, such that there
exists a decoupling region with the same asymptotics as in the black
hole geometry. In the context of the type IIB solutions, this implies
that bubbling solutions like the corresponding
black hole or black ring admit an $AdS_3 \times S^3 \times X_4$ decoupling
region. This is a necessary condition for the bubbling solution to
correspond to a black hole microstate: the AdS/CFT dictionary implies
that any microstate is dual to an asymptotically
$AdS_3 \times S^3 \times X_4$ geometry, with the information encoded
in the asymptotics determining the specific microstate.

Solutions of M theory compactified on a Calabi-Yau admit decoupling
$AdS_2 \times S^3$ or $AdS_3 \times S^2$ regions, depending on the
Taub-NUT charge. In the former case there has been
considerable progress on counting the relevant BPS states, via the counting of curves
in the Calabi-Yau, and on understanding the quiver quantum
mechanics of the D0-D2-D4-D6 branes which wrap these curves and
generate the black hole. At the same time, it is hard to use detailed AdS/CFT technology
to analyze these bubbling geometries as one is forced to
work within the rather poorly understood $AdS_2/CFT_1$
correspondence and there is no precise dictionary
developed between the decoupling $AdS_2$ regions and the dual theory.
In the other case, where one does have an $AdS_3$ region in an M theory
solution, one does not have a good description of the corresponding
2-dimensional CFT, given the limiting understanding of M-brane
theories.

Thus, to make full use of AdS/CFT technology to identify specific
bubbling solutions, it is natural to work instead with the type IIB solutions
which admit an $AdS_3 \times S^3 \times X_4$ decoupling
region and correspond to states in the relatively well
understood D1-D5 conformal field theory. A clear advantage of working
in this system is that one already has an identification of the geometries
dual to the microstates of the 2-charge black hole, namely the
Ramond ground states.

\bigskip

Let us turn to the correspondence between bubbling geometries in the
type IIB frame and D1-D5-P microstates in the conformal field
theory. Recall first the correspondence between two charge geometries
and Ramond ground states: the former are characterized by a curve
whose Fourier coefficients determine the corresponding dual
superposition of Ramond ground states. Restricting to ground states
built from the universal cohomology of $X_4$, the curve is a generic
closed curve in $R^4$. In the special case where the curve is a
circle, the corresponding dual is a specific R charge eigenstate
Ramond ground state.

Note that this dictionary determines the dual superposition of Ramond
ground states given the curve defining a smooth supergravity
solution. As discussed in \cite{Kanitscheider:2006zf, Kanitscheider:2007wq}
the geometric dual of a generic R charge eigenstate is not
known; most likely it is not well described in supergravity,
but one can also not exclude that the dual consists of multi-center
concentric circular curves. The latter is consistent with both the symmetry
of the CFT state and the charges, and cannot be excluded by comparing
vevs of operators, since these are just too small for reliable
comparisons \cite{Kanitscheider:2006zf}. A priori, however, it seems rather unnatural that the
geometric dual should be multi-center, as this generically signals
Coulomb branch, rather than Higgs branch, physics.

Moving on to the three charge bubbling geometries, recall that the defining data
for these geometries is the data at the centers of the Gibbons-Hawking
metric, namely $(\td{k}^a_i,q_i,\vec{x}_i)$. Here $q_i$ is the nut
charge of the $i$th center; $\vec{x}_i$ is its position in $R^3$, and
$\td{k}^a_i$ determines the associated contributions to the
charges. Lifting this data into four-dimensional language, to make
contact with the two charge geometries, the defining data is
associated with a set of circles on the four-dimensional base space,
with $\vec{x}_i$ determining their radii and relative orientation. In
particular, if the $\vec{x}_i$ are parallel, the circles are
concentric and $|\vec{x}_i|$ determines their radii. The latter
solutions preserve a $U(1)^2$ isometry group of the base space, and
should correspond to R charge eigenstates in the conformal field
theory.

The basic question is therefore: how does the data
$(\td{k}^a_i,q_i,\vec{x}_i)$ capture the dual D1-D5-P microstate?
Consider first the case where the $\vec{x}_i$ are parallel, so the
corresponding microstate must be an R charge eigenstate. Then as a
first guess one might think that the multi-center data corresponds to
fractionation in the CFT. Recall that the Ramond ground states in
the orbifold CFT language as
\be
\prod_{j} {\cal O}^{a_j \bar{a}_j}_{R n_j} | 0 \rangle, \qquad \sum_j
n_j = N,
\ee
where each twist $n_i$ operator ${\cal O}^{a_i \bar{a}_i}_{R n_i}$
is associated with the  $(a_i,\bar{a}_i)$ cohomology and
corresponds on spectral flow to an NS chiral primary.
The same operator can occur multiple times, with appropriate symmetrization.
The left moving excited states with total momentum $N_p$ are obtained
by exciting these ground states, with the momentum being distributed
between the different twist sectors. In the
string picture this corresponds to distributing the momentum
between each group of multiwound strings. Note that in such a
distribution the momenta $p_j$ given to each twist sector $n_j$ is non-negative.

It would be rather natural if this fractionation was also present in
the corresponding geometries. One might first try to identify each effective string with a
Gibbons-Hawking center, but this naive identification does not seem to be
correct. Recall that the total charges are given by
\be
Q_a = -2 {\cal N} |\ep_{abc}| \sum_{i} \frac{\td{k}^b_i \td{k}^c_i}{q_i} \equiv
\sum_i (Q_a)_i,
\ee
with $\sum_i \td{k}^a_i = 0$ and $\sum_i q_i = 1$. In particular,
the contributions to the three charges from centers with positive
$q_i$ cannot all be positive. At most, two out of the three $(Q_a)_i$
are positive, with the other negative, or only one of the three is
non-zero and positive. This does not however agree with the CFT
fractionation, where the total D1-D5-P charges are sums over
only positive contributions.

Actually this disagreement is not really surprising:
one should be rather careful about associating charge contributions to
each center, as there are no sources there. Whilst the solution is
built from harmonic functions sourced at the Gibbons-Hawking centers,
all the defining functions appearing in the solution are
finite at these centers. The distinguished hypersurfaces in the full
solution are clearly those where the Killing vector which is timelike
at infinity becomes null (\ref{sigma-beh}), which happens when the base space signature
changes. Thus it may be physically more natural to parameterize the
solutions in terms of functions which have poles at these
surfaces.

Indeed, this is explicitly demonstrated by the analysis of \cite{Giusto:2004kj} for the
specific 3-charge geometries obtained from limits of Cvetic-Youm
black hole solutions. Whilst one can rewrite these solutions in terms of six harmonic
functions on an ambipolar Gibbons-Hawking base space, the physical
interpretation is more manifest in the original Cvetic-Youm
coordinates, in which the defining functions are harmonic on $R^4$
with poles at the sources for the D-brane charges.
An important open question is thus whether
one can reparameterize the bubbling solutions in such a way
that the connection with fractionation in the CFT is manifest, and
moreover such that one can systematically solve the equations for
absence of closed timelike curves. For recent progress along related
lines see \cite{Bena:2008wt}: here coordinate transformations are used to
relate smooth three charge solutions with particular hyper-K\"{a}hler
bases to solutions in which several of the Gibbons-Hawking centers are
replaced by two charge (Lunin-Mathur) geometries. Such techniques may
help to understand which geometries correspond to
bound states and how the space of smooth three-charge solutions is
parameterized.

One should mention here that it is possible that the parameterization
in terms of the ambipolar Gibbons-Hawking space may still be natural
for the M theory geometries; perhaps the centers are related to the
nodes in the quiver quantum mechanics as has been
suggested in \cite{Bena:2007ju,Bena:2007qc}. Nonetheless, a systematic solution of the regularity
conditions in the M theory system is also still lacking.

Furthermore, without solving such conditions, it would be hard to
carry out geometric quantization for these fuzzball geometries; one
clearly needs to determine these constraints before quantization. Of
course, even if one could carry out the geometric quantization, it would be
hard to make further progress without determining which black hole
microstates the bubbling geometries sample, i.e. without carrying out the
matching.

Another related puzzle is the flight time calculation in the scaling
bubbling geometries: this gives a finite answer for symmetric
geometries, which can be interpreted in terms of the field theory mass
gap, but for non-symmetric geometries the time can become infinite
\cite{Bena:2007ju,Bena:2007qc} although one would still anticipate an
energy gap in the field theory. Thus either such geometries do not
correspond to microstates or the calculation needs to be interpreted
more carefully.

\section{Fuzzballs and black hole physics} \label{open}

The previous sections have focused on the explicit construction of
fuzzball solutions in supergravity for near supersymmetric black holes, and the
matching of these solutions with black hole microstates. Most of the
literature on the fuzzball proposal has concentrated on this
issue, since finding and matching candidate geometries for
black holes with macroscopic horizons has turned out to be difficult
even in the supersymmetric case. Results to date can be regarded as
evidence for the fuzzball proposal, but to make further detailed and
quantitative progress on issues such as coarse-graining one would
anticipate the need to match better the candidate bubbling geometries
with CFT data and to understand fuzzball solutions in the stringy regime.

At the same time, one can already envisage how many key conceptual
questions could be answered, and there are many interesting and
suggestive results in the current literature. Therefore in this section we will
consider how longstanding issues in black hole physics are addressed
by the fuzzball proposal.

\subsection{Quantizing fuzzball geometries}

Suppose one finds fuzzball geometries in supergravity with the correct
charges to correspond to microstates of a given black hole. Typically
the geometries will be characterized by continuous functions, and will
not be countable. Knowledge of the holographic map between these
solutions and the CFT microstates should determine how the continuous
functions are discretized, as in the case of 2-charge geometries.
Alternatively, one could consider quantizing the
geometries, as in the discussion in section (\ref{geo-quant}).
This would allow one to estimate the number of states accounted for by the
geometries, and, if it
reproduced the black hole entropy, this could be interpreted as evidence
that the geometries do describe black hole microstates.

At the same time, it is not clear why quantizing the fuzzball
geometries {\it visible in supergravity} should reproduce the full black hole
entropy in general. In the 2-charge case, quantizing the extrapolation
of the fuzzball solutions to supergravity (for the subset of solutions that this
was done) indeed reproduced the black
hole entropy, even though the average solution was string scale, but
one cannot assume that this agreement will
persist in other, less supersymmetric systems.
If geometric quantization of the fuzzball solutions found in
supergravity does reproduce the black hole entropy, it would suggest that
the extrapolation of the fuzzballs to supergravity is more
representative than one may have anticipated. Moreover, one would then
be able to address the important issue of coarse-graining, to
demonstrate explicitly how black hole properties emerge.

\subsection{Finite temperature} \label{temp}

Much of the discussion of previous sections has been restricted to
the case of supersymmetric (zero temperature) solutions, since one
can then use powerful tools to find supergravity solutions. Clearly
restricting to zero temperature removes an important feature of black
hole physics: extremal black holes do not Hawking radiate. So an
important question is how the fuzzball proposal works for a
non-extremal black hole.
Given the very few known candidate fuzzball geometries for non-extremal
black holes, the issue of Hawking radiation has not been discussed in
generality. However, in cases where AdS/CFT is applicable, one can
understand how the non-extremal fuzzballs will radiate, and moreover
use the specific known solutions as a testing ground.

First let us recall how Hawking radiation of asymptotically $AdS_{d+1}$
black holes is computed in the dual conformal field theory. Consider
radiation of a scalar field $\phi$, which corresponds to the dual
gauge invariant operator ${\cal O}_{\phi}$ of dimension
$\Delta_{\phi}$. Let the emitted radiation
have $d$-dimensional momentum $\vec{p}$.
Now consider a state $|i \rangle$ in the boundary field theory which
is dual to an asymptotically $AdS_{d+1}$ spacetime. The rate of
emission of the scalar field $\phi$ with momentum $\vec{p}$ in the
spacetime is computed from the decay rate of the corresponding field
theory state $| i \rangle$. The differential decay rate $d \Gamma(i)$
is given by
\be
d \Gamma(i) = \sum_{f} | M_{fi} |^2 d \Omega
\ee
where $| f \rangle$ are possible final states, $d \Omega$ is an
appropriate phase space factor, and the matrix element $M_{fi}$ is
given by
\be
{\cal N} \langle f | {\cal O}_{\phi} (\vec{p}) | i \rangle,
\ee
with ${\cal N}$ a normalization constant (fixed by two point functions
at the conformal point). In practice one computes this by using the
optical theorem to relate the sum of $|M_{fi}|^2$ over final states
into an appropriate discontinuity of the analytically continued
Euclidean two point function.

Let us now focus on the case of interest, a 2d CFT.
To compute the decay rate of the BTZ black hole corresponding to a thermal
state in a (grand) canonical ensemble, one must average over initial
states weighted by the appropriate Boltzmann factors.
The analytic continuation of the
Euclidean thermal two point function for the scalar operator of dimension
$\Delta_{\phi}$ is given by
\be
\Pi(x^+ x^{-}) \equiv \langle {\cal
  O}^{\dagger}_{\phi} (0,0) {\cal O}_{\phi} (x^+,x^-)
\rangle_{\rm thermal}  = {\cal C} \left [\frac{\pi T_+}{\sinh (\pi T_{+} x^+)} \right ] ^{\Delta_{\phi}}
 \left [ \frac{\pi T_-}{\sinh (\pi T_{-} x^-)} \right ]^{\Delta_{\phi}},
\ee
where ${\cal C}$ is a normalization factor, $x^{\pm} = t \pm z$ are lightcone coordinates and
$T_{\pm}$ are the left and right moving temperatures
respectively. Expressed in terms of the inverse temperature $\beta = T^{-1}$ conjugate to
the energy $\omega$ and the chemical potential $\mu_z$ conjugate to the
momentum $p_z$,
$2 T^{-1}_{\pm} = \beta \pm \mu_z$. In terms
of the BTZ temperature $T_{H}$ given in (\ref{btz-temp}) and the inner and outer horizon
locations $\r_{\pm}$ these quantities are
\be
T_{+}^{-1} = l T_{H}^{-1} \left ( 1 + \frac{\r_{-}}{\r_{+}} \right ); \qquad
T_{-}^{-1} = l T_{H}^{-1} \left (1 - \frac{\r_-}{\r_{+}} \right ).
\ee
Then the emission rate $\G_e$ is proportional to the discontinuity in the two
point function
\be
\G_e = {\cal F} \int dt dz e^{i \omega t - i p_z z} \Pi(t + i\ep,z)
\ee
where ${\cal F}$ is the flux of the emitted particle.
For example, in the case of $\Delta_{\phi} = 2$ and $p_z \rightarrow
0$ and inserting appropriate normalization factors one obtains
\be \label{decay}
\G_e = \frac{\pi^3 Q_1 Q_5 \omega} {(e^{\frac{\omega}{2T_+}} -1)
(e^{\frac{\omega}{2T_-}} -1)}
\ee
in agreement with the corresponding Hawking radiation rate computed in
the bulk. More details of this calculation may be found in
\cite{emission}. Note that detailed balance implies that the emission rate
$\G_e = e^{-\omega/T_H} \G$ where $\G$ is the absorption rate.

To compute instead the decay (or absorption) rate of a particular state $| i \rangle $
in an ensemble one needs to compute the appropriate discontinuity in the
two point function in that
state. For a typical member of a thermal ensemble one would
expect the variance from the thermal spectral density to be small. For
an atypical member of the ensemble the decay rate can however
differ substantially.

Now suppose one has a candidate asymptotically $AdS_{d+1}$ fuzzball
geometry dual to the specific state $| i \rangle$: how should the
instability of the state to decay by the operator ${\cal O}_{\phi}$ be
reflected in the bulk, which has no horizon so does not Hawking
radiate? The answer is that modes of the bulk field
$\phi$ with real momentum $\vec{p}$ should be tachyonic, with
the imaginary part of the frequency reflecting the decay rate.

This is precisely what has been found for the specific known examples of
non-extremal fuzzball solutions, such as that discussed in the
previous section. The instability of the bulk solutions
precisely matches the computed decay rate of the corresponding CFT
microstate \cite {Chowdhury:2007jx,Dias:2007nj}. However, the CFT
microstate under consideration
is a highly atypical member of the thermal ensemble, and thus the
emitted radiation is not even approximately thermal: $\G_e \approx
\w^2$, compared to (\ref{decay}). This rapid decay rate is responsible
for the superradiant behavior in the bulk solution.

\subsection{Path integral approach}

Despite its computational and conceptual difficulties, the Euclidean
path integral approach to gravity rather naturally accounts for black
hole thermodynamics. The Euclidean path integral is
\be
{\cal Z} = \int D[g] D [\Phi] e^{- I_{E} (g,\Phi)},
\ee
where $g$ is the metric, $\Phi$ collectively denotes matter fields and
$I_{E}$ is the Euclidean action. Whilst making sense of the
integration over geometries remains an open problem, it is interesting
that in the saddle
point approximation the onshell Euclidean action for a solution of the
Einstein (supergravity) equations behaves as the free energy $F$, i.e.
\be
I_{E} = \beta F = \beta (E - \mu_{i} Q_i) - S,
\ee
with $\beta$ the inverse temperature, $\mu_i$ the conjugate
potentials to conserved charges $Q_i$ (angular momentum, electric
charge etc) and $S$ the Bekenstein-Hawking entropy. This
identification is consistent with standard thermodynamic relations,
for example
\be
S = - \left ( \frac{\pa F}{\pa T} \right )_{Q_i} = \left (\beta
\frac{\pa I_{E}}{\pa \beta} - I_E \right ).
\ee
Note that extremal black holes should be treated as a limiting case of
non-extremal black holes, in which case their entropy is non-zero. If
one works directly at extremality ($T=0$) the
Euclidean action is linear in $\beta$, and thus the entropy would seem
to vanish.

The fuzzball solutions are however solutions of the
gravitational field equations with the same asymptotics as the black
hole, and are thus also saddle points of the path integral. Fuzzball
solutions which are stationary in the Lorentzian correspond to
topologically trivial Euclidean solutions, since the imaginary time
Killing vector has no fixed point sets. There is a natural
thermodynamic interpretation of the fuzzball solutions: each solution
has the same mass $E$ and conserved charges $Q_i$ as the black hole, and
quantization of the solutions should give a total degeneracy
${\cal D} = e^{S}$. Therefore the Euclidean action $I_{f_a}$  associated with each fuzzball
$f_a$ should be
\be
I_{f_a} = \beta F = \beta (E - \mu_{i} Q_i).
\ee
Clearly by construction one finds that
\be
\int_{f_a} D [g_a] D [\Phi_a]  e^{-I_{f_a}} = e^{-I_{f_a}
  + S} = e^{-I_{BH}},
\ee
where the functional integral is over all fuzzball geometries $f_a$,
with metric $g_a$ and matter fields $\Phi_a$, and
$I_{BH}$ is the action for the black hole of the same mass and
charges.

Evidently one would like to derive these formulae explicitly in a
specific example, by quantizing the set of fuzzball solutions. However
the argument above suggests that in the integration over geometries
one should not include both black holes and fuzzballs simultaneously,
as this is over counting. One should either include all topologically
trivial configurations with the same charges (namely the fuzzballs) or
one should include the topologically non-trivial configuration (namely
the black hole). In cases where AdS/CFT is applicable, this fits
exactly with field theory expectations: one either sums over
pure states or in equilibrium configurations typical states may be
represented by thermal states, but one does not simultaneously
include both pure and mixed states when computing a partition
function.

Hawking's proposed resolution of the information loss paradox
\cite{Hawking:2005kf} also makes use of the path integral over
asymptotically AdS metrics. However both topologically trivial and
non-trivial configurations are included simultaneously in the path
integral. The former are unitary and the latter are not, but it is
argued that contributions to correlation functions from topologically
non-trivial configurations fall off rapidly, and do not contribute at
late times. Thus just as in the fuzzball proposal one effectively only
includes the topologically trivial configurations in the path
integral although of course the justification is rather different.

\subsection{Distinguishing between a black hole and a fuzzball}

Given that the black hole curvature is small at the horizon scale, it
may initially seem surprising that the fuzzball solutions start to
differ from the black hole already at this scale. One might have
argued that deviations from the black hole spacetime which resolve its
singularity should be localized near the curvature singularity.
At the same time, standard results in black
hole physics already indicate that the horizon itself is deeply
connected with quantum gravity effects. Firstly, the existence of a
closed trapped surface is an input into the singularity theorems
relevant to black holes. As we review below, it is hard to evade these
theorems within supergravity when a horizon is present. Secondly,
Hawking radiation is associated with the existence of a horizon, and
thus with the well known trans-Planckian and backreaction issues
indicating the breakdown of semiclassical approximations.

One might worry that the small differences between the fuzzball geometry
and that of the black hole at the horizon scale manifest as
substantial differences in the measurements of an observer at
infinity. In asymptotically AdS examples, however, we have discussed in detail
how a black hole microstate is characterized by vevs and higher point
functions; the deviation of these from the corresponding ensemble
average is small for a typical microstate, and thus distinguishing
between black hole and fuzzball requires a large sequence of
measurements over a long timescale. Essentially the same point was
made by Hawking in \cite{Hawking:2005kf}: an observer at spatial infinity would
need to make measurements over an infinite time to be certain that a black hole
formed and then evaporated.

Inside the outer future event horizon of a non-extremal black hole the Killing
vector $\pa_{t}$ is spacelike and the radius necessarily decreases along
future directed timelike geodesics. In a corresponding fuzzball
solution the geometry at sub-horizon scales may be radically
different, with for example $\pa_t$ remaining timelike, so cannot an
observer at infinity detect this difference?  Again this question
needs to be phrased more precisely in terms of measurements accessible
to the observer: scattering of particles and measurement of emitted radiation.
The former corresponds to excitation and subsequent decay by radiation
of the black hole with the latter corresponding to spontaneous radiation.
A typical non-extremal fuzzball should emit radiation
with only small deviations from the thermal spectrum of the black
hole, which would take the observer a long time to measure.
Discussions of the small scale of deviations of correlation functions in a typical
state from that of the thermal state can be found in
\cite{Vijay,Balasubramanian:2007qv}; see also \cite{Balasubramanian:2005mg}.

Note that atypical non-extremal fuzzballs can contain ergoregions, in which a
Killing vector which is timelike at infinity becomes spacelike. Such
ergoregions are responsible both for superradiance processes
(mirroring excitation and decay of the black hole) and for
spontaneous emission (corresponding to Hawking radiation). However
these atypical fuzzballs decay much faster than the corresponding
black hole, as we saw in section (\ref{temp}). One can also understand
Hawking radiation from static black holes in terms of spontaneous emission
from an ergoregion bounded by the horizon. The key
difference between the black hole and the atypical fuzzball in this respect is
that in former case the ergoregion straddles the horizon, whereas for
the fuzzball there is no horizon to cloak it.

\subsection{Gravitational collapse}

Whilst eternal black holes represent a simplified system for explicit
analysis, one would also like to apply the fuzzball proposal to
astrophysical gravitational collapse. In astrophysical collapse it is
believed that given sufficiently dense and generic initial data the formation
of first a closed trapped surface and then a (cloaked) singularity is
inevitable. The fuzzball proposal in this context would be that supergravity
solutions exist which describe the collapse of initial data,
avoiding horizon and singularity formation. The late-time
quasi-stationary solution should differ from the corresponding black hole spacetime
only at sub-horizon scales.

To date only stationary fuzzball solutions have been found, and
it would be difficult to find the time dependent solutions
needed to describe gravitational collapse. However, in cases where one
can use AdS/CFT, the initial data corresponding to a pure state in the
field theory should collapse to form a fuzzball. Recall that the
fuzzball solutions generically involve all supergravity fields (and beyond)
reflecting the fact that in a generic state all primary  (and other) operators
acquire a vev. If one specifies initial data only for a subset
of fields then one is restricting to a subspace of the phase space,
and this could be interpreted as tracing over the complementary part of the
phase space, so this initial data should correspond to mixed states.
Such initial data should collapse to form black holes.
It would clearly be interesting to investigate this further, and
understand how this connects to results from
numerical simulations of gravitational collapse. In particular, it is
known that even for spherically symmetric collapse horizon formation
can be avoided for non-generic initial data
\cite{Choptuik:1992jv}. For the non spherically symmetric collapse
relevant to fuzzballs fewer results are known (see the review \cite{Gundlach:2002sx})
but perhaps the conditions on initial data needed
to avoid horizon formation can be related to conditions on the dual
theory being in a pure state.

It is also interesting to note here that many of the results of
numerical relativity support the fuzzball proposal: by adding just one
scalar field to the action one can find numerically solutions of the
Einstein equations which are horizonless, non-singular and asymptote
to the black hole solution. In particular, in a recent paper
\cite{Krueger:2008nq} numerical solutions were found which could
be interpreted as fuzzballs for the Schwarzschild black hole.

\subsection{Singularity theorems}

The fuzzball proposal states that each black hole microstate
corresponds to a non-singular horizon-free geometry. In this section
we will consider how the proposal is reconciled with the general
proofs of singularity theorems.

We have seen in explicit examples that fuzzball solutions for
spherically symmetric black holes break the spherical symmetry, and
involve many more of the supergravity fields than the black hole solution.
This at first sight seems somewhat reminiscent of proposals for
singularity resolution in general relativity suggested in the
1960s. The first singularity
theorems showed that there would be singularities in symmetric
solutions under certain reasonable conditions, but the results depended
on the symmetry being exact. It was thus suggested by a number of
authors (in particular, the Russian school of Lifshitz, Khalatanikov and
co-workers \cite{Lifshitz:1963ps}) that singularities were the result of symmetries and would not
occur in more general solutions. Indeed this idea seemingly fits with the fuzzball
proposal, as the non-singular solutions are less symmetric and involve
more fields.

However, the idea that singularities were non-generic was soon
disfavored, in part because the Russian school found that explicit
families of non-symmetric solutions did contain singularities \cite{Khalatnikov:1969eg}, but
principally because of the Penrose-Hawking
proofs \cite{Hawking:1969sw} of singularity theorems without assumptions of symmetry. So how
is the fuzzball proposal reconciled with the singularity theorems?

To understand this it is useful to recall the three main conditions
required to prove the existence of a singularity, or to be more precise,
geodesic incompleteness of the spacetime:
\begin{enumerate}
\item{An energy condition;}
\item{Condition on global structure;}
\item{Gravity strong enough to trap a region.}
\end{enumerate}
The exact conditions depend on the theorem under consideration, see
\cite{Hawking:1973uf}; for
example, the second condition can be the existence of a non-compact
Cauchy surface or that the chronology condition holds throughout the
spacetime. More importantly for our purposes, the third condition can
be that the spatial cross-section of the universe is closed or that
there is a closed trapped surface in the spacetime. The proofs input
these conditions into the
Raychaudhuri equation to show that timelike or null geodesics converge
within finite affine length, and thus that the spacetime is
geodesically incomplete.

The fuzzball solutions which are non-singular in supergravity evade
the singularity theorems via the third condition: the supergravity
fields do satisfy the weak and/or strong energy condition, and there is an
appropriate non-compact Cauchy surface, but there is no closed trapped
surface or horizon in the spacetime!

Of course, as we have emphasized throughout this report, a generic
fuzzball solution should be described as a solution of the
full string theory equations of motion, not just as a supergravity
background. In such cases the singularity theorems are simply not applicable.
For the same reason, whilst the singularity theorems imply that any black object in
supergravity has a singularity behind its horizon, stringy corrections
are expected to resolve these singularities.

\subsection{Emergence of horizons}

The main objective of the fuzzball proposal is to show that black hole
properties emerge upon coarse-graining over fuzzball geometries. In
particular, one would like to understand the emergence of the
defining property of a macroscopic black hole, its horizon. Given the
incomplete understanding of fuzzball geometries for even
supersymmetric macroscopic black holes, such a coarse-graining has not
yet been carried out.

One can however use rather general arguments to infer the scale of the
coarse-grained geometry. One first needs to estimate how much a typical
fuzzball geometry differs from the naive black hole solution, but this
is a question which is easy to answer, having set up the AdS/CFT
dictionary. Near the conformal boundary, in the far UV of the field
theory, all fuzzball solutions should look like the naive solution,
since in the UV we see only the conformal behavior. The solutions
should start differing at energies comparable to the scale set by the
vev of the most relevant gauge invariant operator. Using AdS/CFT this
translates into a certain radial distance scale.

Suppose one estimates the vev of the lowest dimension operator in a
typical microstate of the black hole, and then translates this value
into a characteristic radial scale $r_c$. Now compute the area of a
spatial surface of constant radius $r_c$: the area of this so-called
stretched horizon (in the naive metric)
should approximately reproduce the black hole horizon area. A version of this
argument was given in \cite{Lunin:2001jy,Lunin:2002qf,
Mathur:2007jn,Mathur:2007sc}, although here we use the AdS/CFT
dictionary and the vevs to determine more systematically the relevant
radial scale.

It is straightforward to carry out this calculation for the 2-charge
black hole. In this case the vev of the most relevant gauge invariant
operator is expressed in terms of the curve characterizing the
fuzzball solution. Thus the radial scale $r_c$ is determined by the
scale of a typical curve. Now recall that the defining curve $F^a(v)$
in $k$-dimensional space, with $a=1,\cdots k$, behaves
as
\be
F^a(v) = \mu \sum_{n > 0} (f^a_n \exp(2 \pi i nv/L) + (f^a_n)^{\ast} \exp(-2 \pi i nv/L)),
\ee
with $\mu = \sqrt{Q_1 Q_5}/R_z$, $L =2 \pi Q_5/R_z$  and
\be
Q_1 = \frac{Q_5}{L} \int^L_0 (\pa_v F^a)^2 dv \qquad \rightarrow
\qquad 2 \sum_{a, n >0} n (f_n^a (f_n^{a})^{\ast}) = 1.
\ee
A typical curve $F^a(v)$ satisfying this constraint will have a scale
of the order of
\be \label{r-c}
r_c \sim \frac{\mu}{\sqrt{N}}.
\ee
Now recall that the naive 2-charge metric (in the string frame) in the
decoupling region is
\be
ds^2 = \frac{r^2}{\sqrt{Q_1 Q_5}} (-dt^2 +dz^2) + \sqrt{Q_1 Q_5}
(\frac{dr^2}{r^2 } + d\Omega_2^2) + \frac{\sqrt{Q_1}}{\sqrt{Q_5}} ds^2(X_4),
\ee
with $e^{2 \Phi} = Q_1/Q_5$. Then the area of a spatial surface at
constant $r_c$ in the Einstein frame metric ($g_{ein} = e^{-\Phi/2}
g$) is ${\cal A}$ where
\be
{\cal A} = (2 \pi)^6 \pi V R_z \sqrt{Q_1 Q_5} r_c \sim (2\pi)^6 \pi
(\a')^4 \sqrt{N} .
\ee
Putting the area into the Bekenstein-Hawking entropy formula gives
\be
S =  \frac{2 {\cal A}}{(2 \pi)^6 (\a')^4} \sim 2 \pi \sqrt{N}.
\ee
Hence the area of the stretched horizon reproduces the microscopic
entropy of the 2-charge black hole. Of course one should be a little
cautious of this computation, as the typical scale implied by
(\ref{r-c}) is
\be
r_c \sim \frac{(\a')^2}{R_z \sqrt{V}},
\ee
which is clearly substring scale for the compactification radii of
interest, as indeed it must be for the 2-charge black hole which
has no horizon in supergravity.

\bigskip

Whilst these general arguments can be used to infer the characteristic
scale of a typical fuzzball geometry, explicitly demonstrating that a
horizon emerges upon coarse-graining remains an open problem. One
first needs to construct a representative basis of fuzzball geometries
for a black hole with a macroscopic horizon before one can explore
quantization and coarse-graining. The three charge system seems a
promising test case for exploring such issues.

Note that if the fuzzball proposal is correct then coarse-graining
over horizonless backgrounds of trivial topology should result in the
black hole background, whose topology is non-trivial. The black hole
should have a global feature, a horizon, which none of the individual
fuzzballs possess. On the Euclidean section this translates into
periodicity in imaginary time, and non-trivial topology.

It may at first seem surprising that the black hole has a property which is not
shared by any of the fuzzballs. Of course, given that this property is
associated with entropy, this must be the case if the fuzzball
proposal is to be correct. Moreover, the emergence of non-trivial
topology under coarse-graining precisely mirrors the corresponding
field theory behavior: thermal states in the field theory are
characterized by periodicity in imaginary time, which is not a
property of any state in the thermal ensemble.

\subsection{Stringy fuzzballs}

Throughout this report we have emphasized that many of the fuzzballs
for any given black hole will not be well-described by
supergravity. To develop the fuzzball proposal, and indeed to
understand more rigorously small black holes, one will most likely
need to work with backgrounds of the full string theory.
Clearly there are many technical obstacles to overcome in this area:
string theory in backgrounds with Ramond-Ramond fluxes is still rather
poorly understood, even given the progress with the pure spinor
formalism, and solving the worldsheet string theory in curved
backgrounds is generically very hard.

At the same time, not all issues relating to stringy fuzzballs are
likely to be intractable. Suppose one considers a worldsheet theory
in an asymptotically $AdS_3 \times S^3 \times X_4$ fuzzball
background, which is sufficiently strongly curved that the
supergravity approximation does not hold. Just as in the
case of $AdS_3 \times S^3 \times X_4$, the information that one
extracts from the worldsheet theory is the (holographic) correlation
functions.

Throughout this report, we have expressed the correspondence between
fuzzballs and black hole microstates in terms of these correlation
functions: the one-point functions are used to identify the specific
microstate whilst the two-point functions characterize the decay rate
of non-BPS fuzzballs etc. Even when a geometric description is
accessible, the correspondence is not expressed in terms of the local
geometry, but rather in terms of measurements made in the asymptotic
region. Thus the absence of a geometric description does not
necessitate a complete reformulation of the fuzzball proposal:
the information one computes from the worldsheet theory is
exactly the information needed to probe the fuzzball
proposal.

In the supergravity regime, one could compute holographic one-point
functions from a given supergravity solution from algebraic
manipulations even when solving fluctuation equations to extract
higher correlation functions is intractable.
The same should be true for the worldsheet theory: given the
string spectrum in the $AdS_3 \times S^3$ background one would expect
that one can extract holographic one-point functions from the worldsheet theory
in asymptotically $AdS_3 \times S^3$ backgrounds
even if one cannot fully solve the
worldsheet theory. Moreover, one would anticipate being able to
determine whether a given worldsheet theory corresponds to a black
hole (i.e. there is entropy) or to a fuzzball. One should
be able to address in general terms
how one point functions and entropy are encoded in
the worldsheet theory, and this would be interesting in itself.

\section{Conclusions}

The fuzzball proposal is a promising idea which has the potential to
resolve longstanding black hole related puzzles. In the few years since
this proposal was suggested a body of evidence
in support of the proposal has been found. In particular, families of
fuzzball solutions visible within supergravity have been found for
various black hole systems in string theory, and
holographic technology has been used to match their properties with
those of black hole microstates. All computations, be they of
holographic one point functions, scattering or geometric quantization,
support the interpretation of these geometries as black hole
microstates. The aim of this report was to collect in one place
all current evidence, emphasizing interconnections and pointing out
open problems and avenues for further research. We further sketched how
key issues in black hole physics could be addressed by the fuzzball proposal.

A recurrent theme in this report was the use of the AdS/CFT correspondence
to test, motivate and perhaps even explain why (a form) of the fuzzball
proposal should hold, at least for black holes which admit AdS near horizon
regions. In our view, precision holography would be needed if this interesting
idea is to become a physical model that would either be falsified or explain
black holes at a quantitative level and for this reason we placed
particular emphasis on this topic.

Whilst the fuzzball proposal has the potential to address black hole
issues, substantial work and technical progress
is still needed to demonstrate explicitly how the key issues are
resolved. One would like to show in concrete examples
how black hole properties, such as
the horizon and Hawking radiation, emerge upon coarse-graining over
fuzzballs. To address these issues one will need to understand better
the holographic map between solutions and microstates
and how to coarse-grain geometries. Most likely
one will also be forced to work with backgrounds of the full string
theory, rather than just supergravity.

The fuzzball proposal is a concrete idea which could explain black
hole physics and is very natural from a holographic perspective.
Given its potential and the increasing body of supporting
evidence, it merits further development and investigation.

\section*{Acknowledgments}

The authors acknowledge support from NWO, KS via the Vernieuwingsimplus grant
``Quantum gravity and particle physics'' and
MMT via the Vidi grant ``Holography, duality and time dependence in
string theory''. We would like
to thank the 2007 Simons Workshop and the 2007 Paris Summer
Workshop for their hospitality and the stimulating environment they
provided. We have
benefited from interesting discussions with many colleagues, including
S. Mathur, I. Bena, N. Warner, E. Rabinovici, R. Brustein, D. Marolf,
I. Kanitscheider, P. Kraus, F. Larsen, M. Berkooz and G. Horowitz.

\appendix

\section{Conventions for field equations and supersymmetry} \label{conv}
%\subsection{Field equations} \la{fe}

The equations of motion for IIA supergravity are:
\bea
\label{conv_IIA}
   && e^{-2\Phi}(R_{mn} +2\na_m \na_n \Phi - \frac{1}{4}
H^{(3)}_{mpq}
H_n^{(3)pq}) -\hp F^{(2)}_{m p} {F^{(2)p}_{n}} -\frac{1}{2\cdot3!} F^{(4)}_{mpqr}
F_n^{(4)pqr} \nono \\
   && \qquad +\frac{1}{4} G_{m n}(\hp (F^{(2)})^2 + \frac{1}{4!} (F^{(4)})^2) = 0, \\
   && 4\na^2 \Phi - 4(\na \Phi)^2 +R - \frac{1}{12} (H^{(3)})^2 =0, \nono \\
   && dH^{(3)} = 0, \qquad dF^{(2)} =0, \qquad \na_m F^{(2)m n} - \frac{1}{6} H^{(3)}_{pqr}
F^{(4) n pqr} = 0, \nono \\
&&  \na_m(e^{-2\Phi}H^{(3)mnp}) -\hp
F^{(2)}_{qr}F^{(4) qr np}
-\frac{1}{2\cdot (4!)^2} \e^{n p m_1 \cdots m_4 n_1 \cdots n_4} F^{(4)}_{m_1
  \cdots m_4} F^{(4)}_{n_1 \cdots n_4} = 0, \nono \\
   && dF^{(4)} = H^{(3)} \wedge F^{(2)}, \qquad \na_m F^{(4)mnpq} - \frac{1}{3!
  \cdot 4!}
\e^{n pq m_1 \cdots m_3 n_1 \cdots n_4} H^{(3)}_{m_1 \cdots m_3} F^{(4)}_{n_1
  \cdots n_4} = 0. \nn
\eea
The corresponding equations for type IIB are:
\bea
\label{conv_IIB}
   && e^{-2\Phi}(R_{m n} +2 \na_m \na_n \Phi - \frac{1}{4}
H^{(3)}_{m pq} H_n^{(3)pq}) - \hp F^{(1)}_{m} F^{(1)}_{n} -\frac{1}{4} F^{(3)}_{m pq}
F_n^{(3)pq} - \frac{1}{4\cdot4!} F^{(5)}_{m pqrs} {F_n}^{(5)pqrs} \nono \\
   && \qquad +\frac{1}{4} G_{m n}( (F^{(1)})^2 + \frac{1}{3!} (F^{(3)})^2 ) =0,
\nono \\
   && 4\na^2 \Phi - 4(\na \Phi)^2 +R - \frac{1}{12} (H^{(3)})^2 =0, \nono \\
   && dH^{(3)} = 0, \qquad \na_m(e^{-2\Phi}H^{(3)m n p}) -
F^{(1)}_{m}F^{(3) m n p } -\frac{1}{3!}
F^{(3)}_{m qr } F^{(5) m q r n p} = 0,  \\
   && dF^{(1)} =0, \qquad \na_m F^{(1) m} + \frac{1}{6} H^{(3)}_{pqr}
F^{(3) pqr} = 0,
\nono \\
   && dF^{(3)} = H^{(3)} \wedge F^{(1)}, \qquad \na_m F^{(3) m n p}
+ \frac{1}{6} H^{(3)}_{m q r } F^{(5) m q r n p} = 0, \nono \\
   && dF^{(5)} = d(\ast F^{(5)}) = H^{(3)} \wedge F^{(3)}, \nn
\eea
where the Hodge dual of a $p$-form $\omega_p$ in $d$ dimensions is
given by
\be
   (\ast \, \omega_p)_{i_1 \cdots i_{d-p}} = \frac{1}{p!}\e_{i_1
  \cdots i_{d-p} j_1 \cdots j_p} \omega_p^{j_1 \cdots j_p},
\ee
with $\e_{01 \cdots d-1} = \sqrt{-g}$. The RR field strengths are defined as
\be
   F^{(p+1)} = dC^{(p)} - H^{(3)} \wedge C^{(p-2)}.
\ee
The equations of motion for the heterotic theory are:
\bea
    &&4\na^2\Phi - 4\left(\na\Phi\right)^2 + R - \frac{1}{12}(H^{(3)})^2
          - \a' (F^{(c)})^2 = 0, \nono \\
    && \na_m\left(e^{-2\Phi}H^{(3) mnr}\right) = 0, \nono \\
    &&R^{mn} + 2\na^m \na^n \Phi
          - \frac{1}{4}H^{(3) mrs}H^{(3)n}_{rs}
          - 2\a' F^{(c) mr} F^{(c) n}_{r} = 0, \nono \\
   &&{\na_m}\left(e^{-2\Phi}F^{(c) mn}\right)
          + \half e^{-2\Phi}H^{(3) nrs}F^{(c)}_{rs} = 0. \nn
\eea
$F^{(c)}_{mn}$ with $(c) = 1, \cdots 16$ are the field strengths
of Abelian gauge fields $V^{(c)}_m$; we consider here only supergravity
backgrounds with Abelian gauge fields. This restriction means that the
gauge field part of the Chern-Simons form in $H_3$,
\be
H^{(3)} = d B^{(2)} - 2 \a' \w_3 (V) + \cdots,
\ee
does not play a role in the supergravity solutions, nor does the
Lorentz Chern-Simons term denoted by the ellipses.

The action for eleven-dimensional supergravity is
\be
S = \frac{1}{2 \k_{11}^2} \int d^{11}x \sqrt{-g_{11}} \left (
R - \frac{1}{12} F^2 - \frac{1}{432} \ep^{M_1 \cdots M_{11}} F_{M_1
  \cdots M_4} F_{M_5 \cdots M_8} A_{M_9 \cdots M_{11}} \right ),
\ee
whilst the supersymmetry variation of the gravitino is
\be
\delta \Psi_{M} = \left (D_{M} + \frac{1}{144} (\G_{M}^{\; \; NPQR} -
8 \d_{M}^{N} \G^{PQR}) F_{NPQR} \right ) \ep
\ee
with $\ep$ a 32-component Majorana spinor.

Our conventions for the (truncated) type IIB action are
\be
S = \frac{1}{2 \k_{10}^2} \int d^{10}x \sqrt{-g_{10}} \left(e^{-2 \Phi}
(R_{10} + 4 (\pa \Phi)^2) - \frac{1}{12} (F^{(3)})^2 + \cdots\right),
\ee
where $2 \k_{10}^2 = (2
\pi)^7 g_s^2 (\a')^4$; for the most part we set $g_s=1$ since it plays no role in our
discussion. The part of the supersymmetry variations relevant here are then
\bea
\delta \lambda &=& \G^{m} \pa_{m} \Phi \ep + \frac{i}{12}
e^{\Phi} \G^{mnp} F_{mnp} \ep^{\ast}; \label{IIbs} \\
\delta \Psi_{m} &=& D_{m} \ep
- \frac{i}{48} e^{\Phi} F_{npq} \G^{npq} \G_{m} \ep^{\ast}. \nn
\eea
where $\ep = \ep_1 + i \ep_2$ is a complex Majorana-Weyl spinor.

\section{Killing spinors} \label{spinors}

In this appendix we will review the derivation of the Killing spinors
of the bubbling solutions (\ref{m-bub}) and (\ref{gen1}).
The proof of the Killing spinors for the M theory solutions
(\ref{m-bub}) runs as follows. Introduce a vielbein
\be
e^{\hat{0}} = f^{-1/3} (dt + k); \qquad
e^{\hat{m}} = f^{1/6} \tilde{e}^{\hat{m}}; \qquad
e^{\hat{5}} = \left ( \frac{Z_2 Z_3}{Z_1^2} \right )^{1/6}
dy_{1}, \cdots
\ee
where $f = Z_1 Z_2 Z_3$, $\hat{m} =1 \cdots 4$ and $\tilde{e}^{\hat{m}}$ a vielbein for
the hyper-K\"{a}hler metric $h_{mn}$. The ellipses denote
corresponding expressions for the $e^{\hat{\mu}}$ with $\hat{\mu} = 5
\cdots 10$. The spin connection is
\bea
\omega^{\hat{0} \hat{m}} &=& \frac{1}{3} f^{-3/2} \tilde{e}^{m \hat{m}} \pa_m
f (dt + k) + f^{-1/2} (dk)_{lm} \tilde{e}^{m \hat{m}} dx^l; \\
\omega^{\hat{m} \hat{n}} &=& \frac{1}{6} f^{-1} \pa_n f \tilde{e}^{n
  \hat{n}}  \tilde{e}^{\hat{m}} + \tilde{\omega}^{\hat{m} \hat{n}} \nn
+ \half f^{-1} (dk)_{nm} \tilde{e}^{m \hat{m}} \tilde{e}^{n \hat{n}}
(dt+k); \nn \\
\omega^{\hat{5} \hat{m}} &=& \frac{1}{6} \left ( \frac{ \pa_m(Z_2
  Z_3)}{Z_2 Z_3 \sqrt{Z_1}} - 2 \frac{ \pa_m Z_1}{\sqrt{Z_1^3}}
  \right ) \tilde{e}^{m \hat{m}} dy_1, \nn
\eea
with corresponding expressions for the other $\omega^{\hat{\mu} \hat{m}}$. Here
$\tilde{\omega}$ is the spin connection for the hyper-K\"{a}hler
space. Now the components of the Killing spinor equations along the $T^6$ directions
reduce to
\bea
\left (\pa_{y_1} + \frac{1}{12}
{\omega}^{\hat{5} \hat{m}}_{y_1} \G_{\hat{5}
  \hat{m}} - \frac{1}{12} \G_{\hat{5} \hat{m}}  \tilde{e}^{m \hat{m}}
( \frac{ \pa_m Z_2} {Z_2 \sqrt{Z_1}}  \G^{\hat{0} \hat{7}
  \hat{8}}
+ \frac{ \pa_m Z_3} {Z_3 \sqrt{Z_1}}  \G^{\hat{0} \hat{9}
  (\hat{10})}
- 2 \frac{ \pa_m Z_1}{\sqrt{Z_1^3}} \G^{\hat{0} \hat{5}
  \hat{6}} ) \right )  \ep  \nn \\
+ \frac{1}{6 Z_1 \sqrt{Z_2 Z_3}} \G^{\hat{5} \hat{m} \hat{n}} \left ( Z_2
f^2_{\hat{m}\hat{n}} \G^{\hat{7} \hat{8}} +
Z_3 f^3_{\hat{m}\hat{n}} \G^{\hat{9} (\hat{10})} -
2 Z_1 f^1_{\hat{m}\hat{n}} \G^{\hat{5} \hat{6}} \right ) \ep = 0, \nn
\eea
where the two forms $f^a$ are defined as
\be
f^a_{\hat{m}\hat{n}} \tilde{e}^{\hat{m}} \wedge  \tilde{e}^{\hat{m}}
= \theta^{a} - \half Z_{a}^{-1} dk.
\ee
The projection conditions
\be \label{proj-m}
\G^{056} \ep = \G^{078} \ep = \G^{09 (10)} \ep = - \ep
\ee
%(\ref{proj-m})
along with the duality of
the two forms $\theta^{a}$ are sufficient to ensure these equations
are satisfied for spinors which are independent of the torus
coordinates. The time component of the Killing spinor equation is also
satisfied given the same conditions. The components along the
hyper-K\"{a}hler space reduce using the projections to
\be
\left (\pa_m + \qu \td{\w}_m^{\hat{m} \hat{n}} \G_{\hat{m} \hat{n}} + \sum_{a=1}^{3}
\frac{\pa_m Z_a}{6 Z_a} \right ) \ep = 0,
\ee
and are thus satisfied provided that the explicit form of the spinor is
\be
\ep = (Z_1 Z_2 Z_3)^{-1/6} \ep_0
\ee
with $\ep_0$ covariantly constant on the base
hyper-K\"{a}hler space. It is useful to extract the explicit
form of these spinors on a Gibbons-Hawking space, namely a
hyper-K\"{a}hler space with a $U(1)$ isometry. The metric on this
space can be written in the form
\be \label{gh}
ds_4^2 = V^{-1} (d \psi + A)^2 + V dx^{i} dx^{i}
\ee
where $x^i$ with $i=2,3,4$ are coordinates on $R^3$ and the connection $A$ satisfies
$\ast_3 dA = dV$. Introducing the vierbein
\be
\tilde{e}^{1} = V^{-1/2} (d \psi + A); \qquad
\tilde{e}^{i} = V^{1/2} dx^i,
\ee
the spin connection is
\bea
\tilde{\omega}^{1i} &=& - \half V^{-2} \pa_i V (d \psi + A) + \half
V^{-1} (dA)_{ij} dy^j; \\
\tilde{\omega}^{ij} &=& \half V^{-2} (dA)_{ij} (d \psi + A) + \half
V^{-1} \pa_j V dx^i. \nn
\eea
Thus the Killing spinor equations become
\bea
(\partial_{\psi} - \qu V^{-2} \pa_i V (\G^{1i} + \ep_{ijk} \G^{jk}))
\ep = 0; \\
(\partial_{i} - \qu (V^{-2} A_i \pa_k V  - V^{-1} \pa_j V \G^{i1})
(\G^{1j} + \ep_{jkl} \G^{kl})) \ep = 0. \nn
\eea
These equations are solved for $\G^{1234} \ep = \ep$ for {\it
  constant} $\ep_0$.

\bigskip

Next let us consider the type IIB bubbling solutions (\ref{gen1}).
To prove the spinors for the type IIB solution, introduce a vielbein such that
\bea
e^{\hat{0}} &=& (Z_1 Z_2)^{-1/4} Z_{3}^{-1/2} (dt + k); \qquad
e^{\hat{1}} = Z_3^{1/2} (Z_1 Z_2)^{-1/4} (dz + {\cal A}_3); \\
e^{\hat{m}} &=& (Z_1 Z_2)^{1/4} \td{e}^{\hat{m}}; \qquad
e^{\hat{a}} = (Z_2/Z_1)^{1/4} dy^a, \nn
\eea
where $\hat{m} = 2,3,4,5$, $\hat{a} = 6,7,8,9$ and $\td{e}^{\hat{m}}$ is a vierbein on the
hyper-K\"{a}hler space. Note that the three form can conveniently
be written in terms of this vielbein as
\bea
F^{(3)} &=& (Z_1 Z_2)^{1/4} Z_{3}^{-1/2} (d \eta_2 - Z_2^{-1} d k)
\wedge e^{\hat{1}} + Z_1^{1/2} Z_2^{-3/2} d Z_2 \wedge e^{\hat{0}} \wedge e^{\hat{1}} \nn
\\
&& + \ast_{4} dZ_1 + Z_1^{5/4} Z_2^{-3/4} Z_3^{-1/2} e^{\hat{0}} \wedge (d
\eta_1 - Z_1^{-1} \ast_4 dk),
\eea
where $\ast_4$ refers to the Hodge dual on the hyper-K\"{a}hler space.

The associated spin connection takes the form
\bea
\omega^{01} &=& - \half (Z_1 Z_2)^{-1/4}
Z_3^{-3/2} \pa_{\hat{m}} Z_3 \td{e}^{\hat{m}}; \\
\omega^{0 \hat{m}} &=& - \qu (Z_1 Z_2)^{1/4} \left ( \frac{\pa^{\hat{m}} Z_1}{Z_1} +
\frac{\pa^{\hat{m}} Z_2}{Z_2} \right ) e^0 \nn \\
&& - (Z_1 Z_2)^{-1/2}( \half Z_3^{-1/2} \pa^{\hat{m}} Z_3
(dz + \eta_3) - \half Z_3^{-1/2} (dk)^{\hat{n}\hat{m}} \td{e}_{\hat{n}}); \nn \\
\omega^{1 \hat{m}} &=& - \qu (Z_1 Z_2)^{1/4} \left ( \frac{\pa^{\hat{m}} Z_1}{Z_1} +
\frac{\pa^{\hat{m}} Z_2}{Z_2} \right ) e^1 \nn \\
&& + (Z_1 Z_2)^{-1/2} ( \half Z_3^{-1/2} \pa^{\hat{m}} Z_3
(dz + \eta_3) - \half (Z_3^{1/2} (d\eta_3)^{\hat{n}\hat{m}} + Z_3^{-1/2}
(dk)^{\hat{n}\hat{m}}) \td{e}_{\hat{n}}); \nn \\
\omega^{\hat{m} \hat{n}} &=& \tilde{\omega}^{\hat{m} \hat{m}} + \qu
\td{e}^{\hat{m}} \left ( \frac{\pa^{\hat{n}} Z_1}{Z_1} +
\frac{\pa^{\hat{n}} Z_2}{Z_2} \right ) \nn \\
&& -\half  (d\eta_3)^{\hat{m} \hat{n}} (Z_3 (dz + \eta_3) + (dt + k))
  - \half (dk)^{\hat{m} \hat{n}} (dz + \eta_3); \nn \\
\omega^{\hat{a} \hat{m}} &=& \qu (Z_1 Z_2)^{-1/4} \left (
\frac{\pa^{\hat{m}} Z_2}{Z_2} - \frac{\pa^{\hat{m}} Z_1}{Z_1} \right ) e^{\hat{a}}. \nn
\eea
where $\pa_{\hat{m}}Y = \td{e}^{m}_{\hat{m}} \pa_m Y$. Noticing that
\be
(dt + k) = Z_3^{1/2} (Z_1 Z_2)^{1/4} e^{0}; \qquad
(dz + \eta_3) = Z_3^{-1/2} (Z_1 Z_2)^{1/4} (e^0 - e^1),
\ee
all terms in the spin connection can be expressed in terms of the
vielbein.

The spin connection together with the conveniently expressed three
form allow us to check the type IIB supersymmetric equations
(\ref{IIbs}). The dilatino equation reduces to
\bea
&& i \G^{\hat{m}} \tilde{e}^{m}_{\hat{m}} \left ( \frac{\pa_m Z_2}{Z_2} - \frac{\pa_m Z_1}{Z_1}
\right ) \ep^{\ast} + (Z_1 Z_2 Z_3)^{-1/2} (dk)_{\hat{m} \hat{n}}
\G^{\hat{m} \hat{n}} (\G^0 + \G^1) \ep \nn \\
&& + (Z_1 Z_2 Z_3)^{-1/2} \left (Z_2 (d \eta_2)_{\hat{m} \hat{n}} \G^1 + Z_1
(d \eta_1)_{\hat{m} \hat{n}} \G^0 \right ) \G^{\hat{m} \hat{n}} \ep \\
&& + \G^{\hat{m} 01} \tilde{e}^{m}_{\hat{m}} \frac{\pa_m
  Z_2}{Z_2} \ep
- \frac{\pa_m Z_1}{Z_1} \tilde{e}^{m}_{\hat{m}} \ep^{\hat{m} \hat{n}
  \hat{p} \hat{q}} \G_{\hat{n} \hat{p} \hat{q}} \ep = 0. \nn
\eea
This equation is satisfied by the spinors (\ref{spiniib}), using the
projection conditions
\be
\ep = \G^{01} \ep = \G^{0126789} \ep = - i \ep,
\ee
along with the self-duality of the forms $d
\eta_a$. Next consider the components of the gravitino equation along
the compact directions, which reduces to:
\be
\w^{\hat{a} \hat{m}}_{a} \G_{\hat{m}} \ep = - \frac{i}{24} F^{(3)}_{MNP}
\G^{MNP} \ep^{\ast}.
\ee
This is also satisfied using the projection conditions (\ref{spiniib})
along with the self-duality of the forms $d \eta_a$, as are the
components of the gravitino equation
along the string directions. Finally the gravitino equation
along the hyper-K\"{a}hler space reduces to:
\be
\left (\pa_m + \qu \td{\w}^{\hat{m} \hat{n}}_{m} \G_{\hat{m} \hat{n}} +
\frac{\pa_m Z_3}{4 Z_3} + \frac{\pa_m Z_1}{8Z_1} + \frac{\pa_m Z_2}{8Z_2}
\right ) \ep = 0.
\ee
This is satisfied provided that the spinor is
\be
\ep = (Z_1 Z_2)^{-1/8} Z_3^{-1/4} \ep_0,
\ee
where the spinor $\ep_0$ is covariantly constant on the
hyper-K\"{a}hler space.

\section{Properties of spherical harmonics} \label{sphere}

Scalar, vector and tensor spherical harmonics on the unit radius $S^3$
satisfy the following equations
\bea \label{sph_Y}
\Box Y^{I} &=& - \Lambda_{k} Y^{I}, \\
\Box Y_{a}^{I_v} &=& (1 - \Lambda_{k}) Y_{a}^{I_v}, \hsp D^{a}
Y_{a}^{I_v} = 0, \nn \\
\Box Y_{(ab)}^{I_t} &=& (2 - \Lambda_{k}) Y_{(ab)}^{I_t}, \hsp D^{a}
Y_{k(ab)}^{I_t} = 0, \nn
\eea
where $\Lambda_k = k (k+2)$ and the tensor harmonic is traceless. It
will often be useful to explicitly indicate the degree $k$ of the harmonic;
we will do this by an additional subscript $k$, e.g. degree $k$ spherical
harmonics will also be denoted by $Y_k^I$, etc.
$\Box$ denotes the d'Alambertian along the three sphere. The
vector spherical harmonics are the direct sum of two irreducible
representations of $SU(2)_L \times SU(2)_R$ which are characterized by
\be
\ep_{abc} D^b Y^{c I_v \pm} = \pm (k+1) Y_{a}^{I_v \pm}
\equiv \l_k Y_{a}^{I_v \pm}. \label{vec-dual}
\ee
The degeneracy of the degree $k$ representation is
\be
d_{k,\ep} = (k+1)^2 - \ep,
\ee
where $\ep = 0,1,2$ respectively for scalar, vector and tensor harmonics.
For degree one vector harmonics $I_v$ is an adjoint
index of $SU(2)$ and will be denoted by $\a$.
We use normalized spherical harmonics such that
\be
\int Y^{I_1} Y^{J_1} = \Omega_3 \d^{I_1 J_1}; \hsp
\int Y^{a I_v} Y_a^{J_v} = \Omega_3 \d^{I_v J_v}; \hsp
\int Y^{(ab) I_t} Y_{(ab)}^{J_t} = \Omega_3 \d^{I_t J_t},
\ee
where $\Omega_3 = 2 \pi^2$ is the volume of a unit 3-sphere.
We define the following triple integrals as
\bea
\int Y^{I} Y^{J} Y^K &=& {\Omega_3} a_{IJK}; \label{ap-ov0} \\
\int (Y^{\a \pm}_{1})^a Y^{j}_{1} D_a Y^{i}_1 &=& \Omega_3 e^{\pm}_{\a ij};
\label{ap-ov3}
\eea

\end{document}